\documentclass[%
 aps,
 prd,
twocolumn,
nofootinbib,
 amsmath,amssymb,
 aps,
]{revtex4-2}
\usepackage{graphicx}
\usepackage{adjustbox}
\usepackage{dcolumn}
\usepackage{bm}
\usepackage[dvipsnames]{xcolor}
\usepackage{color, colortbl}
\usepackage[mathlines]{lineno}

\def\vecsign{\mathchar"017E}
\def\dvecsign{\smash{\stackon[-1.95pt]{\vecsign}{\rotatebox{180}{$\vecsign$}}}}
\def\dvec#1{\def\useanchorwidth{T}\stackon[-4.2pt]{#1}{\,\dvecsign}}
\usepackage{stackengine}
\stackMath
\usepackage{xcolor}
\usepackage{dcolumn}
\usepackage{bm}
\usepackage{physics}
\usepackage{caption}
\usepackage{comment}
\usepackage{subcaption}
\usepackage{hyperref}
\usepackage[compat=1.1.0]{tikz-feynhand}
\usepackage{tikz-feynman}
\tikzfeynmanset{compat=1.1.0}
\usepackage{feynmp}
\usepackage{tikzsymbols}
\usepackage{float}

\usepackage{soul}

\definecolor{Gray}{gray}{0.9}
\definecolor{LightCyan}{rgb}{0.88,1,1}
\definecolor{maroon}{cmyk}{0,0.87,0.68,0.32}
\newcommand{\LIR}{\Lambda_{\text{IR}}}
\newcommand{\LUV}{\Lambda_{\text{UV}}}
\newcommand{\Lir}{\Lambda_{\text{IR}}}
\newcommand{\Luv}{\Lambda_{\text{UV}}}
\newcommand{\JJ}{J_{\mu}^{\text{SM}} J_{\text{DS}}^{\mu}}
\newcommand{\pds}{p_{\mathrm{DS}}}
\newcommand{\DS}{\text{DS}}
\newcommand{\QCD}{\text{QCD}}
\newcommand{\EW}{\text{EW}}
\newcommand{\egeo}{\epsilon_{\mathrm{geo}}}
\newcommand{\ldsp}{\text{LDSP}}
\newcommand{\eds}{E_{\text{DS}}}

\newcommand{\npot}{N_\text{POT}}

\def\sv#1{{\bf 
\textcolor{blue}{[SV: {#1}]}}}

\begin{document}

\preprint{APS/123-QED}

\title{Model Agnostic Probes of Dark Sectors at Neutrino Experiments}

\author{Marco Costa$^{a,b}$}

\author{Rashmish K.~Mishra$^{c}$}%

\author{Sonali Verma$^{a,b}$}%
\affiliation{${}^a$Scuola Normale Superiore, Piazza dei Cavalieri 7, 56126 Pisa, Italy\\
${}^b$INFN Sezione di Pisa, Largo Bruno Pontecorvo 3, 56127 Pisa, Italy \\
${}^c$Harvard University, 17 Oxford Street, Cambridge, MA, 02138, USA}

\begin{abstract}
Present and upcoming neutrino experiments can have considerable sensitivity to dark sectors that interact feebly with the Standard Model. We consider dark sectors interacting with the SM through irrelevant portals that are motivated on general principles. We derive bounds on such scenarios by considering decays of dark sector excitations inside the neutrino detector, placed downstream from the target. Our approach is model agnostic and applies to a wide range of dark sector models, both strongly and weakly coupled. In this approach, the dark sector is characterized by two scales: $\Lambda_{\text{UV}}$ (mass of mediators generating the portals) and $\Lambda_{\text{IR}}$ (mass gap of the dark sector). At intermediate energies, far away from these scales, the theory is approximately scale-invariant. This allows the calculation of production rates independent of the threshold corrections, although some mild model-dependent assumptions are needed. We look at various dark sector production processes relevant at neutrino experiments such as meson decays, direct partonic production, and dark bremsstrahlung. We consider representative experiments from past (CHARM), present (ICARUS, NOvA, MicroBooNE), and upcoming future (DUNE-MPD), and compare their reach to existing bounds from high energy experiments (LHC and LEP) and dedicated future LLP experiments (SHiP). We find that the upcoming DUNE-MPD can probe $\Luv$ in the TeV range, and $\Lir$ in the 0.1-1 GeV range, covering parts of parameter space currently inaccessible in high energy experiments and fixed-target/beam-dump experiments, and is comparable to future LLP experiments. In general, future neutrino experiments can be an efficient probe of dark sectors, providing complementary as well as new reach in parameter space.
\end{abstract}

\maketitle


\section{Introduction}
\label{sec:intro}
Secluded sectors with their own particle content and dynamics, that interact with the Standard Model (SM) feebly, are well motivated for a variety of reasons. If such sectors contain or interact with dark matter, they are a natural scenario to probe in terrestrial and cosmological studies. Such sectors can also arise naturally from bottom-up BSM considerations motivated to address various issues in SM, as well as in top-down string constructions. The hidden valley scenario~\cite{Strassler:2006im} initiated various model-building and phenomenological aspects of such dark sectors (DS), and this remains a focus of much present research activity to date.

Dark sectors with sufficiently low production threshold are generally probed by high-intensity experiments with centre-of-mass energies much lower than typical high energy collider machines. Such high-intensity experiments involve a proton or electron beam
(or even muon beam, see ref.~\cite{Kahn:2018cqs,Sieber:2021fue,Cesarotti:2022ttv} for future proposals of fixed target experiments with a muon beam) 
hitting a fixed target, producing a high flux of SM particles alongside a beam of putative DS particles (see ref.~\cite{Batell:2022dpx, Batell:2022xau, Krnjaic:2022ozp, Berger:2022cab, Gori:2022vri, Coloma:2022dng} for general reviews on the capabilities of high intensity experiments). Due to weak couplings to the SM, lightest DS particles once produced tend to have long lifetimes, allowing them to travel macroscopic distances before decaying back to visible particles. Suitable detectors placed downstream from the fixed target can be used to discriminate a possible DS signal against SM background. 

Short and long-baseline neutrino experiments happen to be placed behind some of the most powerful proton beams up to date. Thus, they provide an ideal and pre-existing infrastructure for probing low scale dark sectors. In fact, neutrinos themselves are a prototype for a DS, so it is not a surprise that a facility for studying them is useful more generally.  Further, a rich short-baseline experimental program for neutrinos has been planned at Fermilab (e.g. DUNE, SBNP, see ref.~\cite{MicroBooNE:2015bmn,MicroBooNE:2016pwy,Machado:2019oxb} for experiment details, ref.~\cite{Berryman:2019dme,Chauhan:2022iuh, Batell:2019nwo} for recent studies on DS search at these experiments). These new proposals will improve upon the current neutrino experiments, using a higher number of protons on target (POT), and better detectors that can help in reducing SM background~\cite{Ballett:2018uuc}, therefore leading to an increased sensitivity for DS searches. 

Apart from neutrino experiments, a natural setup to probe DS is at future experiments proposed for long-lived particle (LLP) searches (e.g. see~\cite{Curtin:2018mvb,Beacham:2019nyx} for a review). These experiments are in the approval phase and may require a longer timescale. Existence of current data from neutrino experiments, and a fairly short timescale for future ones to come online make neutrino experiments a powerful and efficient probe of dark sectors.

Any broad enough search for dark sectors must explore all axes of ignorance of such scenarios. The dimensionality $D$ of the portal interaction between the DS and SM is one such axis. While searches for DS interacting with the SM via relevant portals ($D\leq4$) have been well studied (see refs.~\cite{Essig:2013lka,Alexander:2016aln,Beacham:2019nyx} and references therein), the case of irrelevant portals ($D>4$) is equally well motivated. Among these, the \emph{axion} portal ($D=5$) has been studied the most, especially at the high intensity frontier \cite{Dobrich:2015jyk, Dobrich:2019dxc, Altmannshofer:2019yji, Asai:2021ehn, Bertuzzo:2022fcm, Harland-Lang:2019zur, deNiverville:2018hrc} (see~\cite{Co:2022bqq} for a study specific to neutrino experiments). Regarding other irrelevant portals ($D\geq5$) that might connect the DS and SM, some recent progress has been made~\cite{Contino:2020tix,Arina:2021nqi,Cheng:2021kjg,Darme:2020ral, Bertuzzo:2020rzo, Berlin:2020uwy, Carmona:2021seb, Barducci:2022gdv}. In this work, we probe SM-neutral dark sectors which interact via irrelevant portals with the SM at neutrino oscillation experiments. On general principles we are led to $D\sim5$ and $D\sim6$ portals. Such dark sectors are in general \textit{very elusive} due to the irrelevant nature of the portal. The results presented here are complementary to the constraints from current high energy terrestrial experiments and astrophysical data, presented in~\cite{Contino:2020tix}, and in specific cases, much stronger, as we point out in the relevant sections. 

In this work we consider various (inclusive) DS production processes. Depending on the DS 4-momentum $\pds$, a different production process can be relevant: meson decays (for $\sqrt{\pds^2} \lesssim$ M, the parent meson mass), direct partonic production (for $\sqrt{\pds^2} \gtrsim \Lambda_{\text{QCD}}$) or dark bremsstrahlung (for $\sqrt{\pds^2} < \Lambda_{\text{QCD}}$). We require the DS states to be produced away from any mass thresholds, which allows estimating the rates based on general principles. DS particles once produced are required to decay to SM particles inside the neutrino near-detectors placed generally $\sim \mathcal{O}(100\text{m})$ downstream from the target\footnote{Far detectors are less constraining due to a very small angle subtended to the interaction point.}.

Compared to previous attempts at studying irrelevant portals, our work is more comprehensive, as we point out now. Compared to~\cite{Darme:2020ral, Bertuzzo:2022fcm}, we study a more complete set of operators, both in production and decay of DS particles. We perform a detailed study of production modes through irrelevant portals, such as dark bremsstrahlung and partonic production, that have either been neglected or only considered partially~\cite{Darme:2020ral, Bertuzzo:2020rzo, Berlin:2020uwy}. We find that the bremsstrahlung mode can be comparable to other modes and is necessary for a complete analysis. Compared to~\cite{Cheng:2021kjg, Contino:2020tix} we focus on high intensity experiments, particularly on neutrino and other proton dump experiments that were not considered previously in this framework. We do so by adopting the model agnostic strategy outlined in~\cite{Contino:2020tix}. More importantly, this allows us to put bounds on \emph{strongly coupled} light dark sectors through irrelevant portals, which as far as we know is not a thoroughly studied scenario at high intensity experiments (except in~\cite{Contino:2020tix}. See~\cite{Bernreuther:2022jlj, Schwaller:2015gea, Carmona:2021seb, Cheng:2021kjg} and references therein for searches at collider experiments). Strongly coupled GeV-scale DS are relevant in frameworks containing composite resonances from a new gauge group, such as composite versions of Asymmetric Dark Matter~\cite{Kaplan:2009ag, Zurek:2013wia, Petraki:2013wwa}, Mirror world models~\cite{Foot:1991bp, Foot:2014mia}, some incarnations of the Twin Higgs paradigm~\cite{Chacko:2005pe} and are a natural realization of the Hidden Valley scenario~\cite{Strassler:2006im}.

The outline of the paper is as follows: in section~\ref{sec:darkcft} we describe the dark sector portals and the relevant model agnostic framework for estimating inclusive rates, lifetime and multiplicity of DS particles, also pointing out the mild model dependent assumptions we have to make to proceed. In section~\ref{sec:DSprod} we describe various DS production processes relevant at neutrino experiments, while in section~\ref{sec:experiments}, we give an overview of the neutrino experiments we use to constrain the parameter space, and describe our strategy for estimating signal events from DS decaying inside the neutrino detector. Our results and bounds can be found in Section~\ref{sec:results} with a discussion and summary in section~\ref{sec:summary}. Appendices~\ref{app:FormFactors},\ref{app:MesonDecay},\ref{app:DecayProb},\ref{app:egeo} and~\ref{app:factorization} contain technical details.

\section{A Model Agnostic Strategy (And its Limitations)}
\label{sec:darkcft}
In this section we discuss the relevant theoretical details for studying dark sectors with irrelevant portal to the SM. The emphasis is towards being as model agnostic as possible, and only allowing for minimal model dependence where necessary. We point out the assumptions we have to make at various stages for this. Our work builds upon the model agnostic approach first undertaken in~\cite{Contino:2020tix}, wherein more details can be found.

Dark sectors with portal interactions to the SM from irrelevant operators can be generated in a large class of models, generically by exchange of heavy mediators charged under both the SM and the DS. The general form of such a portal is
\begin{align}
    \frac{\kappa}{\LUV^{D-4}}\mathcal{O}_\text{SM}\, \mathcal{O}_\text{DS}\:,
\end{align}
where $\Luv$ is the mass scale of the heavy mediator, $\kappa$ is a dimensionless coupling, and $\mathcal{O}_\text{SM}$ ($\mathcal{O}_\text{DS}$) are local operators made of SM (DS) degrees of freedom. The dimensionality of portal $D=\left[\mathcal{O}_\text{SM}\right]+\left[\mathcal{O}_\text{DS}\right]$ is greater than 4 for irrelevant portals. States in the DS are further characterized by a mass gap $\Lir$, and the dynamics between the scales $\Luv$ and $\Lir$ is approximately scale invariant. A large hierarchy between these scales is a working assumption of this scenario.
In order to avoid strong constraints, we also assume that the portal preserves both CP and flavor symmetries of the SM.

The most constraining portals are expected to be those with lowest dimension $D$. In concrete examples of such DS, both weakly and strongly coupled, the two lowest dimension DS operators are a scalar operator $\mathcal{O}$ (of dimension $\Delta_\mathcal{O} \lesssim 4$) and a conserved current operator $J_\mu^\text{DS}$ (of dimension 3). While the dimension of a conserved current operator is fixed to be 3 in 4D, the reason to take the scalar slightly marginal is to ensure that the condition $\Lir/\Luv \ll 1$ is realized naturally. We will consider $\Delta_\mathcal{O} = 3,4$ in this work. Specific to a given model there can be other operators that generate portals to the SM. However in the absence of a symmetry, their dimension is either unprotected or requires additional assumptions about the DS. We will therefore limit ourselves with only a current and a scalar operator on the DS side. The gauge invariant operators on the SM side that can be used to make a portal operator, with increasing scaling dimension, are $H^\dagger H$ and $J_\mu^\text{SM} = \bar{f}\gamma_\mu f,\,\bar{f}\gamma_\mu \gamma^5 f,\: f = l, q$ or $J_\mu^\text{SM} = H^\dagger i\overleftrightarrow{D}_\mu H$ (see table 1 in ref.~\cite{Contino:2020tix} for a complete list of scenarios). The lowest dimension Lorentz invariant combinations are then $J_\mu^\DS\,J^\mu_\text{SM}$ and $\mathcal{O}\,H^\dagger H$. In the unitary gauge, $H^\dagger i\overleftrightarrow{D}_\mu H \sim Z_\mu$ so that for this portal the interactions with the DS proceed through a Z-boson. We will refer to this as the \textit{Z portal}. At the energy scales relevant at neutrino experiments, $Z$ is never produced on-shell, so we can integrate it out and generate an effective $JJ$ operator where now the SM current is the one that couples to $Z$. Therefore, considering $JJ$ portals where the SM current is either generic or the one for $Z$, we cover all possibilities. We will refer to these as the \textit{generic} $JJ$ portal and the \textit{Z-aligned} $JJ$ portal respectively.

Hence the lowest dimension portals that can be formed are
\begin{align}
    &\mathcal{L_{\text{portal}}} \nonumber \\
    &= 
    \frac{\kappa_{\mathcal{O}}}{\LUV^{\Delta_{\mathcal{O}}-2}}  
    \mathcal{O} H^{\dagger} H 
    + 
    \frac{\kappa_J}{\LUV^2} 
    J_{\mu}^{\text{DS}} 
    J^{\mu}_{\text{SM}}
    +
    \frac{\kappa_Z}{\Luv^2}
    J_{\mu}^{\text{DS}} 
    H^\dagger i\overleftrightarrow{D}^\mu H
    \nonumber \\
    &= 
    \frac{\kappa_{\mathcal{O}}}{\LUV^{\Delta_{\mathcal{O}}-2}}  
    \mathcal{O} H^{\dagger} H 
    + 
    \frac{\kappa_J}{\LUV^2} 
    J_{\mu}^{\text{DS}} 
    J^{\mu}_{\text{SM}}
    +
    \frac{\kappa_Z}{\Luv^2}\frac{v_\text{EW}}{m_Z}
    J_{\mu}^{\text{DS}} 
    J^{\mu}_{\text{SM, Z}}
    \:,
    \label{eq:portals}
\end{align}
where $\kappa_{\mathcal{O}},\kappa_J, \kappa_Z$ are dimensionless coefficients, $v_\text{EW}$ is the electroweak VEV and in the second line we have integrated out $Z$, which couples the DS current to $J^{\mu}_{\text{SM, Z}}$, the SM current that couples to Z. The three terms in eq.~\eqref{eq:portals} are the Higgs portal, the generic $JJ$ portal and the $Z$ portal respectively. It is clear that a Z-aligned $JJ$ portal can be obtained from the Z-portal with a rescaling: $\kappa_J = \kappa_Z (v_\text{EW}/m_Z)$. For $\Delta \lesssim 4$, all these portals are of dimension $D \sim 6$. In principle, a DS described by a local QFT also possesses a stress-energy tensor $T_{\mathrm{DS}}^{\mu\nu}$ of dimension 4, that can be used to build dimension 8 operators with SM dimension 4 operators. However, given a larger suppression compared to the dimension 6 portals in~\eqref{eq:portals}, the bounds on them are too weak to be of any interest.

If the energy $\sqrt{s}$ of an experiment that probes the DS is such that $\LIR \ll \sqrt{s} \ll \LUV$, the DS states are produced directly in the conformal regime. Inclusive DS production rates can be estimated using only the scaling dimension of the DS operator, along with the optical theorem. The optical theorem allows to sum over the DS phase space in an inclusive manner and relates it to the imaginary part of the two point function of the DS operator, which in turn is fixed by the scaling dimension of the operator. In particular, the optical theorem gives
\begin{align}
   \sum_{n} 
   \int \dd\Phi_{\text{DS}} 
   \left|\bra{\Omega}\mathcal{O}_{\text{DS}}\ket{n}\right|^2 
   &= 
   2\,\text{Im}
   \left(
   i\bra{\Omega} \mathcal{O}_{\text{DS}} \mathcal{O}_{\text{DS}} \ket{\Omega}
   \right)\:,
\end{align}
where the DS operator $\mathcal{O}_{\text{DS}}$ interpolates a DS state $\ket{n}$ from vacuum $\ket{\Omega}$ and the integration is over the entire dark sector phase space $\dd\Phi_{\text{DS}}$. 

This approach allows calculating the cross section for DS production without specifying the fields that make the composite operator $\mathcal{O}_{\text{DS}}$. While for irrelevant portals, the matrix element does not decrease with energy (specific behavior depends on the production mode), this needs to be convolved with the structure functions (e.g. the pdfs/form-factors/splitting functions, depending on the production channel), and this changes where the bulk of events come from. As long as the involved $\pds$ values are away from $\Luv, \Lir$, one can ignore the events near the thresholds in a self-consistent manner. Relatedly, the two point function of $\mathcal{O}_\DS$ will also depend on the ratio $\Lir^2/\pds^2$ and $\pds^2/\Luv^2$. For self-consistency, we again need both these ratios to be small. In particular, the condition $\pds^2/\Luv^2 < 1$ effectively ensures the mediators of mass $\Luv$ are not directly produced and the effective local operator for the portal is a good description. In ref.~\cite{Contino:2020tix}, this was enforced by ensuring that the obtained bound on $\Luv$ always satisfies this condition for the highest $\pds^2$ used in the calculation. In practice, this effectively resulted in a lower limit on the parts of $\Luv$ ruled out, or completely invalidated certain bounds. As we will see, for neutrino experiments, where the involved energy is much smaller than LHC or LEP, this condition is less detrimental. By restricting to $\Luv\gtrsim 50$ GeV, we are able to get useful bounds as well as be consistent with the EFT condition. The condition on $\Lir$ on the other hand needs to be imposed, which we do for each production mode.

After production, the DS states will interact and decay among each other, and eventually all the DS degrees of freedom would decay to the \emph{Lightest Dark Sector Particle} (LDSP), which we denote by $\psi$. We will take the mass of $\psi$ to be of order $\Lir$ and this can be taken as our definition for the mass gap $\Lir$. In the absence of additional symmetry, $\psi$ will decay back to SM states from the portal interactions itself. Since the portal interactions are weak, the typical time for DS states to decay among each other is much smaller than the typical lifetime of $\psi$, and can be safely ignored. Note that the LDSP is not the DM candidate in the scenario under consideration---a DM candidate would need to be much more long lived, and will have a missing energy signal. In this work we will assume that the DS relaxes entirely to LDSPs, and leave the question of considering a fraction of events to be missing energy, for future work.

The signatures of $\psi$ depend on its lifetime, and this is the first place where some assumptions have to be made, which bring some model dependence. At high energy colliders, depending on the lifetime of $\psi$, one can get missing energy events, displaced vertices, or prompt decays, ordered by decreasing lifetimes. Missing energy events, being most inclusive, need minimal information about the underlying dynamics of DS, while displaced vertices and prompt decays being exclusive, need some information. Note that the requirement $\Lir/\Luv\ll1$ puts us away from the prompt decay regime, since in this limit, the lifetime increases. Focusing on neutrino experiments, since the detectors are placed some distance from the interaction point, we are in the displaced vertex scenario. It is possible to detect the decay of $\psi$ inside the detector, or its scattering against electrons or nucleons of the detector~\cite{Batell:2021ooj,deNiverville:2018dbu,Buonocore:2019esg,deNiverville:2011it,Batell_2014,Batell:2009di,deNiverville:2016rqh,deNiverville:2012ij}.
Both signatures need some knowledge of the IR behaviour of the underlying theory, and are model dependent, however with varying degrees. For these two signatures, the relevant DS matrix elements are:
\begin{align}
    \text{Decay}:&
    \:\:\langle \Omega | \mathcal{O}_\DS | \psi \rangle 
    = a\, f\, \Lir^{\Delta_{\mathcal{O}}-2}\:, 
    \nonumber \\
    \text{Scattering}:&
    \:\:\langle \psi (p_f) | \mathcal{O}_\DS | \psi (p_i) \rangle 
    = a\, F(p_i,p_f,\Lir) \;,
    \label{eq:generic_ff}
\end{align}
where $\left|\Omega\right>$ is the vacuum, $f$ is a decay constant, $F$ is a form factor and $a$ is an $\mathcal{O}(1)$ number, all of which are model dependent. 
In this work, we will only focus on the decay mode, as we explain this choice now.

The model dependence that enters in the decay case comes in the combination $a\,f$. Scattering process, on the other hand, requires knowing the form factor which can be a complicated function of the momenta (especially for strongly coupled sectors). The functional dependence also influences who $\psi$ recoils against most efficiently. The spin of $\psi$ does not fix the portal, since one can make multiple total spin states using two $\psi$. 
Further, depending on the spectrum, an LDSP might up scatter to a close by state, making the scattering inelastic (similar to what happens in inelastic Dark Matter scenarios~\cite{Tucker-Smith:2001myb}), leading to a different parametric dependence for the scattering cross-section. 
These aspects make it clear that scattering processes require additional model dependent assumptions, and we will not consider them here.
A further reason to choose decays over scattering is that they have a larger signal-to-noise ratio, and we will have more to say about it in sec~\ref{sec:experiments}. Note that
while there are weakly coupled models in which all the LDSPs are stable under some accidental symmetry, and therefore can only be studied through scatterings in the experiments under scrutiny (and therefore our analysis will not apply to such scenarios), in strongly coupled models unstable resonances are expected generically. 

The lifetime of $\psi$ to decay to SM states, via the portal itself, can be estimated in a straightforward manner. However there are differences when the decay is from mixing with a SM state or a direct decay. For a direct decay from a portal of dimension $D$, the lifetime can be estimated to be
\begin{align}\label{eq:ldsp_lifetime}
    \frac{1}{\tau_{\psi}} \sim 
    \LIR \frac{\kappa^2}{8 \pi} \frac{f^2}{\LIR^2} \bigg(\frac{\LIR^2}{\LUV^2}\bigg)^{D - 4}\:,
\end{align}
where the decay constant $f$ is defined by the matrix element $\bra{\Omega} \mathcal{O} \ket{\psi} = a\,f\,\LIR^{\Delta_{\mathcal{O}} - 2}$ and $a$ is an $\mathcal{O}(1)$ number taken to be 1. Further, $f$ can be estimated to be $f = \sqrt{c}\,\LIR/ 4\pi$, where $c$ is the number of degrees of freedom of the DS. On the other hand, if the LDSP decays through mixing with a SM particle such as Higgs, if the LDSP is spin 0, or $Z$, if the LDSP is spin 1 (e.g. through $\mathcal{O}H^\dagger H$ or $J_\mu^\text{DS} H^\dagger i\overleftrightarrow{D}^\mu H$ respectively), the lifetime in the limit $\Lir \ll m_{Z/h}$ is given as
\begin{align}
    \frac{1}{\tau_\psi} = \Gamma_i \sin^2\theta_i\:,
    \:\: \tan 2\theta_i = \frac{2\delta_i}{m_i^2}\:,\:\:
    i = Z, h\:,
\end{align}
where $\Gamma_{Z/h}$ is the decay width of $Z/h$ evaluated at $m_{Z/h}=\Lir$, and the mixing parameter $\delta_i$ is
\begin{align}
    \delta_h &= \kappa_{\scriptscriptstyle \mathcal{O}}\, v_{\scriptscriptstyle\text{EW}} f
    \left(\frac{\Lir}{\Luv}\right)^{\Delta_{\mathcal{O}}-2}\:,
    \nonumber\\
    \delta_Z &= \kappa_{\scriptscriptstyle J} \,v_{\scriptscriptstyle\text{EW}} f 
    \frac{m_Z \Lir}{\Luv^2}\:.
\end{align}

To model the hadronic decay of the scalar LDSP (that mixes with the Higgs), we use the spectator quark model for $\LIR> 2$ GeV and the dispersive analysis for $\Lir < 2$ GeV, following~\cite{Winkler:2018qyg}. For a spin-1 LDSP (mixing with the Z) we again use the spectator quark model for $\LIR>2$ GeV, and a data-driven approach for $\Lir < 2$ GeV, following~\cite{Ilten:2018crw, Baruch:2022esd} for the vector and axial vector component respectively.

The next model dependent assumption needed in order to evaluate the reach at high intensity experiments is how many LDSPs are produced per DS shell, or equivalently how many are excited by the DS operator acting on the vacuum.
We will take two benchmark values, $n_\ldsp=2$ for weakly coupled dark sectors and $n_\ldsp = n(\pds^2)$ a function of the invariant mass squared $\pds^2$ of the DS system, similar to the case of QCD~\cite{Webber:1984jp}:
\begin{align}
    n(\pds^2) &= A \left(\log x\right)^B 
    \exp{C \left(\log x\right)^D}\:,
\end{align}
where $x = \pds^2/\bar{\Lambda}^2$, $\bar{\Lambda} = 0.1\Lir$, $A=0.06,B=-0.5,C=1.8,D=0.5$, and $\pds^2$ is the invariant mass squared of the DS system. Our results are not very sensitive to small changes in $n_\ldsp$.

Finally, we need to know the directional distribution of the produced LDSPs, to estimate if they interact with the detector. In the strongly coupled benchmark where typically $n_\ldsp > 2$, we assume that the LDSPs have a uniform angular distribution in the rest frame of DS (i.e. the frame in which $\pds$ only has a time component), and we can boost it to the lab frame to know its relevant distribution. A uniform distribution in the rest frame is a simplifying choice, and is well motivated, at least for a certain class of strongly coupled theories (e.g. see ref.~\cite{Cesarotti:2020uod} for such a scenario). Further, even if the distribution is not uniform per event, it can be uniform when all the events are considered.
For light enough LDSPs, which are very boosted in the lab frame, small deviations from this assumption do not change our results significantly. 

The weakly coupled case is in principle different, and the angular distribution depends on the production mode, spin of produced DS particles and the specific form of the portal.
In general we expect $\mathcal{O}(1)$ differences among the possible LDSP angular distributions in the DS rest frame.
For example, in DY production the typical LDSP distribution is either proportional to $\sin^2\theta$ or $1+\cos^2\theta$ for scalar and light fermion LDSPs respectively. The difference between the two distributions is that the scalar distribution is more peaked around the most probable LDSP lab angle $\sim 1/\gamma_\mathrm{DS}$. 
However, since the LDSP is produced with a high boost, any differences in the distribution are washed out, and we can assume an isotropic distribution in the DS rest frame as before. We have checked this by an explicit computation.

\section{Production Modes at Proton Beam based Experiments}
\label{sec:DSprod}
\begin{figure}[ht!]
\includegraphics[width=0.48\textwidth]{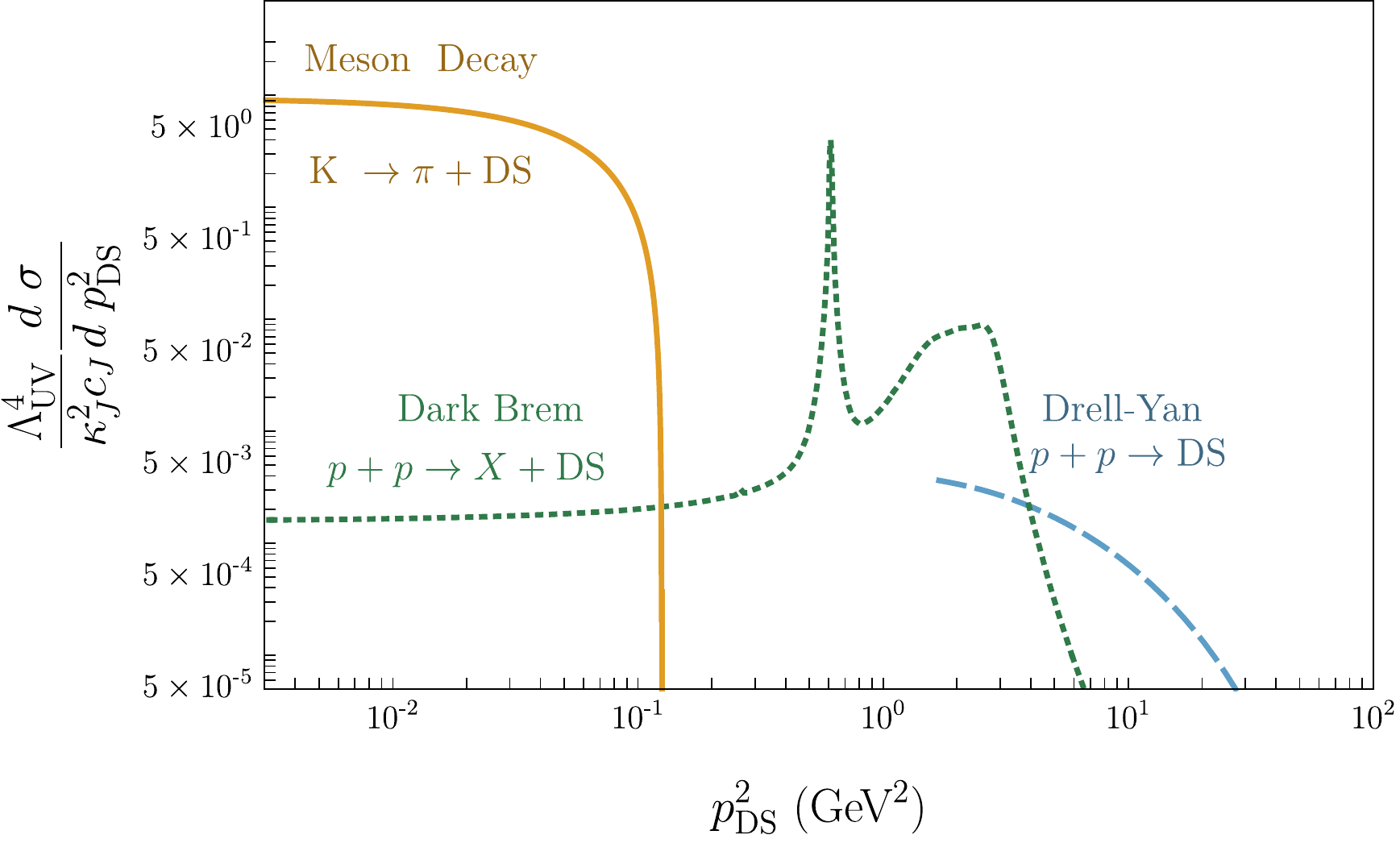}
\caption{\small{Relative importance of various production modes: the scaled differential cross-section for DS production at DUNE-MPD ($E_\text{beam} = 120$ GeV) as a function of $p_\DS^2$ for various DS production modes (for $Z$ portal). Solid yellow shows meson decay: $K \to \pi + \text{DS}$
, dotted green line shows dark bremsstrahlung mode ($p+p\to X +\DS$), and dashed blue line shows Drell-Yan mode ($p+p\to\DS$). The reported cross-section is per proton-on-target, and is without the geometric acceptance factor $\epsilon_\text{geom}$ (which at DUNE is approximately $10^{-3}$ for DY and meson modes and around $10^{-2}$ for DB mode).}}
\label{fig:prod modes}
\end{figure}

Even though at neutrino experiments the primary process is a proton interacting with a nucleus, depending on the energy scale of the process, there are various production modes to consider. In this work, we consider experiments based on 120 and 400 GeV beam energies. For such energies there are three relevant production modes. First of all, the proton nucleus interaction creates mesons, which may decay into lighter mesons and DS states, or completely annihilate into DS states. Denoting the 4-momentum carried by the DS state as $\pds$, this requires $p_\DS^2 \leq \left(\mathbf{M}_\text{heavy} - \mathbf{M}_\text{light}\right)^2$ for the first scenario and $p_\DS^2 = \mathbf{M}_\text{heavy}^2$ for the second. We will refer to these as radiative and annihilation meson decays (MD) respectively. For $p_\DS^2 \gtrsim \Lambda_\QCD^2$, the incoming proton is at high enough energies so one has to consider partonic process involving constituents from the incoming proton and the nucleons in the target, and we refer to this as Drell-Yan (DY) production mode. For $p_\DS^2 \lesssim \Lambda_\QCD^2$, DS states can be produced from initial state emission, which we will refer to as Dark Bremsstrahlung (DB) mode. For each of these processes, the production cross section has a different differential distribution in $p_\DS^2$. Fig.~\ref{fig:prod modes} shows a comparison of the differential DS production cross-section for DY, DB and radiative MD mode, for $Z$ portal, at $120 \text{ GeV}$ beam energy. The radiative MD mode is flat in $\pds$, switching off when the phase space for DS production closes, which in turn is set by the parent meson mass. The DB mode switches off around $\Lambda_\text{QCD}$ beyond which it is not a valid description of the scattering process. The sharp peak in the DB mode is due to meson resonance, as seen in the form factors (see App.~\ref{app:FormFactors}).
The switch off of DY mode comes from the drop in the PDFs of constituents of the proton at higher $\pds^2$, given $\sqrt{s}$ of the experiment, and is a slower drop.

We remark that dependence of the production cross section on the center of mass energy $\sqrt{s}$ of the experiment is not the one given by naive power counting (\emph{i.e.} $\sigma \sim s^{D-5}/\LUV^{2D-8}$), and is general dependent on the production mode. For MD and DB modes, the typical scale of the process is not set by the center of mass energy of the experiment, but rather by the hadronic resonances. There is a residual dependence on $\sqrt{s}$ in the meson production cross section and in the proton-nucleon cross section respectively, but typically this dependence is much weaker than the expected one above the specific hadronic production threshold. DY production instead is more sensitive on $\sqrt{s}$. These features are easily seen in fig.~\ref{fig:prod modes}.

Independent of the production mode, we need to estimate the number of DS signal events $S$ produced. This is generically given as
\begin{align}\label{eq:factorized_xsec}
   S= N_\text{signal} &= N_\text{DS}\,P_\text{decay}\,\egeo
   \:,
\end{align}
where $P_\text{decay}$ is the probability for at least one LDSP to decay inside the radial location of the detector, $\egeo$ is the geometric acceptance for the LDSP direction to intersect with the detector and $N_\text{DS}$ is the number of DS states produced. For more than one production mode, a sum is implied. Note that we have defined a signal event as one in which \text{at least one} LDSP decays inside the detector. The case of more than one LDSPs can be accounted for by multiplying the single LDSP probability with the number of LDSPs produced, and it's included in the definition of  $P_\text{decay}$ (see App.~\ref{app:DecayProb} for a detailed discussion of this). As the final step to get the number of signal events $S$, we have to express $N_\DS$ in eq.~\eqref{eq:factorized_xsec} in terms of the (inclusive) signal cross section $\sigma_\text{S}$ as
\begin{equation}\label{eq:luminosity}
    N_\DS = \frac{N_\mathrm{POT}}{\sigma_{\mathrm{pN}}} \sigma_\text{S} \;,
\end{equation}
where $\sigma_\mathrm{S}$ is the cross section for DS production, $N_\mathrm{POT}$ is the total number of proton delivered on target during the duration of the experiment (projected years for future experiments) and $\sigma_\mathrm{pN}$ is the typical proton-nucleus cross section for the proton beam hitting the target, taken constant for the center of mass energies of the experiments we consider~\cite{Blinov:2021say}:
\begin{equation}
    \sigma_\mathrm{pN} = A^{0.77}\, 49.2 \text{ mb} \;,
\end{equation}
with $A$ the target nucleus' atomic weight. In eq.~\eqref{eq:luminosity}, we are considering only DS production in the first interaction length of the target (or the dump for beam-dump experiments), neglecting production happening at later lengths with a degraded beam. Our computations are therefore conservative. 

Specific to the case of meson decays, for a given meson $\mathbf{M}$ and in a given decay channel $C$, $N_\DS$ is given as 
\begin{align}\label{eq:NDS_meson}
    N_\text{DS} &= N_\text{POT} 
    \,N_\text{M}
    \text{Br}_\text{C}\Big(\textbf{M} \rightarrow \text{DS } (+\mathfrak{m})\Big)
    \:,
\end{align}
where $N_\text{M}$ is the number of mesons produced per collision and $\text{Br}_\text{C}$ is the branching ratio of the meson $\mathbf{M}$ to the DS (which may be in association with other mesons $\mathfrak{m}$).

Strictly speaking, the various factors that go into the estimation of the number of signal events $N_\text{signal}$ depend on the kinematic information, the production mode, and the details of the detector (\emph{e.g.} on- vs off-axis). For example, depending on $\pds^2$, the boost of the DS states and therefore its decay probability is different. Further, depending on whether the DS is produced with a non-zero transverse momentum or not, the angle subtended at the detector can be different. The correct procedure would be to consider differential quantities and integrate over the allowed range.\footnote{Note that for meson annihilation decay $\mathbf{M} \to \DS$, $\pds^2$ is fixed to $M^2$.} This however can obscure the relation between a given experiment and the probed parameter space. As a way out, we use the average value of boost factor for estimating the probability, and compute the average geometric acceptance. In app.~\ref{app:factorization} we compare this procedure, referred to as \emph{factorized} approach, with the exact procedure, called the \emph{full} approach, and show that the difference between the two is small. 

This simplified strategy to compute bounds is useful for the following reason. While production quantities such as the cross section depend in a trivial way on $\LUV$ and very weakly on $\LIR$ via the kinematic  condition $\pds^2 \geq n_\ldsp^2 \LIR^2$, the decay probability depends on both parameters. By using averages in the production quantities allows factorizing them from the decay probability. 
This procedure therefore allows an analytic understanding of $\LUV$ dependence on the number of signal events. Given the vast array of cases, coming from different experiments, different production channels (which can depend also on extra parameters like the dimension $\Delta$), different decay channels, and the strongly vs weakly coupled scenario, this factorization allows to track the $\LUV$ dependence clearly, and also speeds up the computations.

We next briefly outline the details of the three production modes discussed earlier.

\subsection{Meson Decays}\label{sec:meson_prod}
The considered portals between the SM and the DS can cause mesons to decay into DS states. Once the mesons are produced by the incoming proton hitting the target, they can decay in two ways. The first possibility is a heavier meson $\textbf{M}$ decaying into a lighter SM state (such as another meson $\mathfrak{m}$) along with DS states. This is to be contrasted with the case when the mesons decay just into the DS states and nothing else. These two are the radiative decay and annihilation decay modes respectively.
The differential production cross section for radiative decay, as shown in fig.~\ref{fig:prod modes} for the $Z$ portal, is flat in $\pds$ up to kinematic threshold. The decay width for both modes can be approximately estimated, keeping the portal generic: 
\begin{align}
      \Gamma &\sim \kappa^2\,g_{\mathrm{SM}}^2\,\Phi(\Delta)
      \left(\frac{f_M}{M}\right)^a\,
      \begin{cases}
      \frac{v^2}{m_h^4}\,
      \frac{M^{2 \Delta_\mathcal{O}-1}}{\Luv^{2\Delta_\mathcal{O}-4}}\:, & \mathcal{O}H^\dagger H\text{ portal,} \\
      & \\
      \frac{M^5}{\Luv^4}\:, & JJ \text{ portal.}
      \end{cases}
\end{align}
where $\kappa$ is the portal coupling, $M$ is the mass of the parent meson, $f_M$ is the decay constant, $\Delta_{\mathcal{O}}$ is the dimension of $\mathcal{O}$, $g_\mathrm{SM}$ is a dimensionless coupling built out of dimensionless SM couplings (like the gauge couplings, loop factors, extra SM particles' phase space, and relevant spurions), and $\Phi(\Delta)$ is the phase space factor coming from the integration over the DS degrees of freedoms (e.g. see eq.~\eqref{eq:BR-Z-H-ratio}). The exponent of the dimensionless ratio $(f_M/M)$ depends on the process, and is $-2$ for processes coming from the axial anomaly, $+2$ for tree level processes from the chiral Lagrangian and $0$ for processes directly proceeding through the portal (without going through the chiral Lagrangian).  In this estimate we have ignored the lighter meson mass for radiative decay, and have not included the meson form factors for simplicity. In our full analysis we include all these effect. We next discuss specific details of the radiative and annihilation decays as DS production modes.

\subsubsection*{Radiative Decays}
For the radiative decay of the form $\mathbf{M} \to \mathfrak{m} + \text{DS}$ proceeding via a flavour violating loop, the DS state is produced either by the quark line, and/or by the internal $W$ loop (which is necessary to change the quark flavor). This depends on the portal operator. For $J_{\mu}^{\text{SM}} J_{\text{DS}}^{\mu}$ portal where $J_{\mu}^{\text{SM}}$ is the quark current, the DS is produced just by the quark lines, whereas for $J_{\mu}^{\text{SM}} = i H^{\dagger} \dvec{D}_{\mu} H \sim Z_\mu$, the DS states can be produced by attaching a $Z$ to the quarks, or to the $W$ in the loop\footnote{We are working in the unitary gauge.}. The DS states can also be produced by the Higgs portal $\mathcal{O} H^{\dagger} H$. To understand their relative importance, let's consider the ratio of the branching ratios of the two different portals for DS production:
\begin{align}
    \frac{\text{BR}_{OHH}}{\text{BR}_{J\,HDH}}\sim\frac{{96} \ \Gamma(\Delta_{\mathcal{O}} +1/2)}{\pi^{1/2}\Gamma(2\Delta_{\mathcal{O}}) \Gamma(\Delta_{\mathcal{O}}-1)}\frac{m_t^2}{m_h^4} \frac{(M - m)^{2\Delta_{\mathcal{O}}-6}}{\LUV^{2\Delta_{\mathcal{O}} -8}}
    \label{eq:BR-Z-H-ratio}
\end{align}
with $M$ $(m)$ being the mass of the heavy (light) meson, $\Delta_{\mathcal{O}}$ the dimensionality of the operator $\mathcal{O}$, and the factor of $m_t$ comes from top Yukawa in the Higgs portal case.
It is clear that for $\Delta_{\mathcal{O}} \geq 4$, production through the Higgs portal is suppressed with respect to the $Z$ portal and will give weaker bounds. For production through a $\Delta_{\mathcal{O}} = 3$ Higgs portal, even though the $\LUV$ scale probed is higher than $Z$ portal production case (close to 900 $\mathrm{GeV}$ for 120 GeV DUNE-MPD and around 3 TeV for 400 GeV SHiP, discussed in more detail in sec.~\ref{sec:result_higgs}), as can be estimated by a rescaling of the branching ratio using eq.~\ref{eq:BR-Z-H-ratio}, the bound is still at most only marginally stronger compared to  missing energy searches at LHC~\cite{Contino:2020tix}. 
In this subsection we will mostly focus only on the $Z$ portal production for mesons, but will make some comments about the Higgs portal case in sec.~\ref{sec:results}.

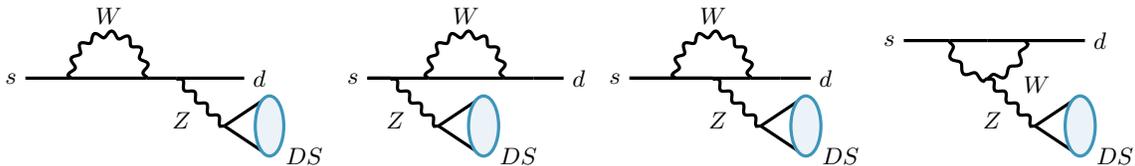
\begin{figure*}[t]
\centering
    \begin{tikzpicture}[/tikzfeynman/medium]
       \begin{feynman}
        \vertex (a1){\(s\)};
        \vertex[right=0.8cm of a1] (a2);
        \vertex[right=1cm of a2] (a3);
        \vertex[right=0.4cm of a3] (a4);
        \vertex[right=0.9cm of a4] (a5){\(d\)};
        \vertex[below right=0.9cm of a4] (c1);
        \vertex[right = 0.6 cm of c1](b1);
        \vertex[above=0.4 cm of b1](b2);
        \vertex[below=0.4 cm of b1](b3);
       \vertex[below=1.5cm of a3] (c2);
       \vertex[right=0.1 cm of b3](c5) {\(DS\)};
       \diagram* {
       {
       (a5) --[very thick] (a4) --[very thick] (a3) -- [very thick](a2) --[very thick] (a1),
       },
       (a3) -- [boson, very thick, out=90, in=90, looseness=2.0, edge label'=\(W\)] (a2),
       (a4) -- [boson, very thick,edge label'=\(Z\)](c1),
       (c1) --[very thick](b2),
       (c1) --[very thick](b3)
        };
        \end{feynman}
        \filldraw[color=MidnightBlue!60, fill=MidnightBlue!5, very thick](b1) ellipse (0.19 and 0.4);
       \end{tikzpicture}
        \begin{tikzpicture}[/tikzfeynman/medium]
       \begin{feynman}
        \vertex (a1){\(s\)};
        \vertex[right=0.5cm of a1](a0);
        \vertex[right=1.0cm of a1] (a2);
        \vertex[right=1cm of a2] (a3);
        \vertex[right=0.4cm of a3] (a4);
        \vertex[right=0.4cm of a4] (a5){\(d\)};
        \vertex[below right=0.9cm of a0](c1);
        \vertex[right=0.6 cm of c1](b1);
         \vertex[above=0.4 cm of b1](b2);
        \vertex[below=0.4 cm of b1](b3);
       \vertex[right=0.1 cm of b3](c5) {\(DS\)};
       \diagram* {
       {
       (a5) --[very thick] (a4) --[very thick] (a3) --[very thick] (a2) -- [very thick](a1) --[very thick] (a0),
       },
       (a3) -- [boson, very thick, out=90, in=90, looseness=2.0, edge label'=\(W\)] (a2),
       (a0) -- [boson, very thick, edge label'=\(Z\)](c1),
       (c1) --[very thick] (b2),
       (c1) -- [very thick](b3)
        };
        \end{feynman}
        \filldraw[color=MidnightBlue!60, fill=MidnightBlue!5, very thick](b1) ellipse (0.2 and 0.4);
       \end{tikzpicture}
        \begin{tikzpicture}[/tikzfeynman/medium]
       \begin{feynman}
        \vertex (a1){\(s\)};
        \vertex[right=0.8cm of a1] (a2);
        \vertex[right=0.5cm of a2](a0);
        \vertex[right=1cm of a2] (a3);
        \vertex[right=0.4cm of a3] (a4);
        \vertex[right=0.4cm of a4] (a5){\(d\)};
       \vertex[below right=0.9cm of a0](c1);
        \vertex[right=0.6 cm of c1](b1);
         \vertex[above=0.4 cm of b1](b2);
        \vertex[below=0.4 cm of b1](b3);
       \vertex[right=0.1 cm of b3](c5) {\(DS\)};
       \diagram* {
       {
       (a5) --[very thick] (a4) --[very thick] (a3) --[very thick] (a2) -- [very thick](a1) --[very thick] (a0),
       },
       (a3) -- [boson,very thick, out=90, in=90, looseness=2.0, edge label'=\(W\)] (a2),
       (a0) -- [boson,very thick, edge label'=\(Z\)](c1),
       (c1) -- [very thick](b2),
       (c1) -- [very thick](b3)
        };
        \end{feynman}
        \filldraw[color=MidnightBlue!60, fill=MidnightBlue!5, very thick](b1) ellipse (0.2 and 0.4);
       \end{tikzpicture}
        \begin{tikzpicture}[/tikzfeynman/medium]
       \begin{feynman}
        \vertex(a0){\(s\)};
        \vertex[right=0.8cm of a1] (a2);
        \vertex[right=0.5cm of a2] (a3);
        \vertex[right=0.5cm of a3] (a4);
         \vertex[right=0.8cm of a4] (a5){\(d\)};
        \vertex[below=0.5cm of a3](c1);
        \vertex[below right=0.9cm of c1](c2);
         \vertex[right=0.6 cm of c2](b1);
         \vertex[above=0.4 cm of b1](b2);
        \vertex[below=0.4 cm of b1](b3);
       \vertex[right=0.1 cm of b3](c5) {\(DS\)};
       \diagram* {
       {
       (a5) --[very thick] (a4) -- [very thick](a3) --[very thick] (a2) --[very thick] (a1),
       },
      (a2) -- [boson, very thick, quarter right] (c1) -- [boson,  very thick, quarter right, edge label'=\(W\)] (a4),
       (c1) -- [boson,  very thick, edge label'=\(Z\)](c2),
       (c2) -- [very thick] (b2),
       (c2) -- [very thick] (b3)
        };
        \end{feynman}
        \filldraw[color=MidnightBlue!60, fill=MidnightBlue!5, very thick](b1) ellipse (0.2 and 0.4);
       \end{tikzpicture}
       \caption{\label{fig:Ktopi_fcncloop} \small{The underlying quark level transition in DS production via $Z$ portal in flavour violating decays such as $K \to \pi + \text{DS}$.}}
\end{figure*}

Examples of radiative meson decay processes are $K^+ \to \pi^+ + \DS$, $B^{+} \to K^{+} + \DS$ and $D^{+} \to \pi^{+} + \DS$, and a prototypical diagram (for $K^+ \rightarrow \pi^+ + \DS$ ) is shown in fig.~\ref{fig:Ktopi_fcncloop}. 
In general, these processes proceed through insertion of two $\text{CKM}$ entries, so that for flavor $i$ going to $j$, the amplitude approximately scales as $\sum_k V^\text{CKM}_{ik} V^\text{CKM}_{kj} f(m_k/m_W)$, where $m_k$ is the quark mass of flavor $k$, $m_W$ is the W mass and $f(x)$ is a loop function~\cite{Buras:1998raa, Inami:1980fz}. For $D$ mesons, for which the underlying process is $c\rightarrow u$, there is no top quark in the loop, as opposed to $B, K$ decays, which makes the D-meson process suppressed. As a result, the $D$ decays are not very constraining---e.g. the large number of $D$ mesons expected at SHiP (enhancement by $\sim 10^4$ compared to B meson production, see ref.~\cite{SHiP:2018xqw}) is not enough to overcome the CKM suppression of $\sim10^{-12}$.

Due to the abundant number of $K$ mesons produced at neutrino  experiments, $K \to \pi + \DS$ decay is an important mode for DS production. For this process, and for the $Z$ portal case, the Feynman diagrams are shown in fig.~\ref{fig:Ktopi_fcncloop}. Note that one must include penguin diagrams as well as self-energy diagrams~\cite{Inami:1980fz,Buras:1998raa}. The DS production rates can be obtained from the SM calculation for $d\bar{s} \to \bar{\nu} \nu$, but with some modifications. We can use the SM results if we keep only the penguin diagrams and omit the box diagrams in the $d\bar{s} \to \bar{\nu} \nu$ process, since the latter are specific to the neutrino coupling (e.g. see~\cite{Inami:1980fz}). This however must be done in the unitary gauge since the box and the penguin diagrams are needed together to make the result gauge invariant in an arbitrary gauge, but their gauge dependent parts vanish individually in the unitary gauge\footnote{If we stayed in arbitrary gauge, the DS would also couple to the longitudinal modes of W and hence the box diagrams would also contribute. In the unitary gauge, $H^\dagger\,D_\mu\,H \sim Z_\mu$, the DS does not couple to W, and the box diagrams' contributions vanish.}~\cite{Inami:1980fz}. Once these subtleties are addressed, we can simply replace the neutrino current coupling to Z, $g_\EW/2 \cos \theta_W (\bar{\nu}_L \gamma_{\mu} \nu_L)$, with the DS current coupling to Z, $(\kappa_J v_\EW  m_Z /\Lambda_{\text{UV}}^2 ) J_{\mu}^{\text{DS}}$. This allows us to write the rate of decay of $K^+ \to \pi^+ + \DS$, using the optical theorem, as:
\begin{align}
\label{eq:gamma_Ktopi_Z}
& \Gamma_{K^+ \to \pi^{+} + \text{DS}}  
= \frac{1}{2 M_K} \left(\frac{G_F}{\sqrt{2}} 
\frac{g_\EW \cos \theta_W}{8 \pi^2 }\right)^2
\frac{m_Z^2 v_\EW^2 
\kappa_J^2}{\LUV^4}  
\nonumber \\
& \times 
\bigg( 
\sum_{j=c,t} V_{js}^{*} V_{jd} \bar{D}(x_j, x_u = 0) 
\bigg)^2 
\int{\frac{\dd^3 p_{\pi}}{(2 \pi)^3} \frac{1}{2 E_{\pi}}}
\mathcal{M}_\mu \mathcal{M}_{\nu}^*  
\nonumber \\
& \times 
2 \ \text{Im} 
\expval{J_{DS}^\mu(p_{\text{DS}}) J_{DS}^{\nu}(p_{\text{DS}})}\;,
\end{align}
where $G_F$ is the Fermi constant, $\theta_W$ is the weak mixing angle, $g_\EW$ is the electroweak gauge coupling, $M_K$ is the mass of the K meson, $v_\EW$ is the Higgs VEV, $m_Z$ is the $Z$ boson mass, $E_{\pi} = \sqrt{m_{\pi}^2 + \abs{\vec{p}_{\pi}^2}}$, $p_{\text{DS}} = p_K - p_{\pi}$, $\mathcal{M}_{\mu} = \bra{\pi^+} \bar{d} \gamma_{\mu} s \ket{K^+}$ is the SM QCD matrix element (see App.~\ref{appendix:mesontoP} for details), and $V_{ij}$ are CKM matrix elements. The loop functions $\bar{D}(x_j)$, where $x_j = m_j^2/m_W^2$, sum the contributions from various diagrams. The upper limit for $p_\pi$ integral is fixed by the kinematic requirement $\pds^2 \geq n_\ldsp^2\Lir^2$.

Apart from the decay $K^+ \to \pi^+ + \text{DS}$, one can also consider the decays of $K_L^0$ and $K_S^0$. We obtain the partial width of $K_L^0$ from that of $K^+$ using ref.~\cite{Buras:1998raa} by replacing $|V^*_{js} V_{jd}|^2$ in eq.~\ref{eq:gamma_Ktopi_Z} with $|\mathrm{Im} (V^*_{js} V_{jd})|^2$. The $K_S^0 \to \pi^0 + \DS$ decay is less constraining since it has a smaller branching ratio due to the large width of $K_S^0$ (see refs~\cite{Batell:2019nwo, Feng:2017vli}).

We next consider decays of B mesons to DS which is relevant at proton-beam experiments with higher beam energies (e.g. SHiP and CHARM, with $E_\text{beam} \sim 400 \text{ GeV}, \sqrt{s} \sim 27 \text{ GeV}$). These high energy proton beam experiments would also have a high K meson production rate but a large number of them get absorbed in the beam dump or target.  Unlike $B$ mesons, kaons have a decay length\footnote{The decay length of $K^{\pm}, K_{L}^0$ is $\sim3\text{ meters} \gg l_H \sim 15.3\text{ cm}$ for SHiP and CHARM target~\cite{Winkler:2018qyg}.} which largely exceeds the hadronic interaction length ($\l_H$) hence they tend to be absorbed in thick targets (for a target length of several $\l_H\,$s) and only a fraction of them then decay to $\DS$ before absorption~\cite{Winkler:2018qyg, Gorbunov:2020rjx}. For estimating this, we use ref.~\cite{Gorbunov:2020rjx} for SHiP, and ref.~\cite{Winkler:2018qyg} for CHARM. 

The B meson decays to lighter mesons like $K$ and $\pi$ take place via $Z$-penguin diagrams which we already encountered in the case of $K^+ \to \pi^+ + \DS$ (see fig.~\ref{fig:Ktopi_fcncloop}) except for the appropriate exchange of external quark flavors ($b \to s/d + \DS$ instead of $s \to d + \DS$). Of all the B decay modes, we find that the largest contribution to signal comes from the decays $B \to K + \DS$ and $B \to K^* + \DS$ ~\cite{Boiarska:2019jym}. For example, even though the partial width for the decay of $B_s \to \rho + DS$ is twice of $B \to K + \DS$, the number of signal events from $B_s$ decays are suppressed due to smaller number of $B_s$ mesons produced with respect to $B^{\pm}$ and $B_0$ mesons at SHiP~\cite{SHiP:2018xqw}. The contribution from $B \to \pi + \DS$ is suppressed with respect to $B \to K + \DS$ by a factor $\sim 20$ coming from $|V_{ts}|^2/|V_{td}|^2$ that enters in the respective decay widths~\cite{Buras:1998raa}. This same suppression applies when comparing B meson decays to vector mesons: $B \to \rho + \DS$ is suppressed with respect to $B \to K^* + \DS$. We calculate the partial decay widths for these B decays in the same way as in eq.~\eqref{eq:gamma_Ktopi_Z} using the appropriate QCD matrix elements from eq.~\eqref{eqn:BtoP_meSM} and eq.~\eqref{eq:M_decay_vec} in Appendices~\ref{appendix:mesontoP} and~\ref{appendix:mesontoV}. 

We do not consider DS production from radiative decays of pseudoscalar mesons like like $\pi, \ \eta, \ \eta'$. Their radiative decay into $\gamma + \mathrm{DS}$ through a generic $JJ$ portal is suppressed by the loop factor from the chiral anomaly triangle diagram, from the electromagnetic coupling and from the lightness of the meson in the $\pi$ case~\cite{Darme:2020ral}.\footnote{For the $Z$ portal case, one external leg of the triangle diagram would produce $Z$ which can couple with DS. This mode can give bounds at LSND due to the huge number of pions ($N_{\pi_0} \sim 10^{22}$), and we find that the $\LUV$ probed is comparable to CHARM in the Meson Production mode.}

Radiative decays of vector mesons like $\rho$ and $\omega$ can also produce DS via decay modes like $\rho^0 \to \pi^0 + DS$, etc. These decays would occur via flavour conserving transitions producing DS either through $Z$ portal or SM vector quark current. The number of DS events from this mode is sub-leading due to the large width of $\rho$ meson with respect to K meson width ($\Gamma_{\rho} \sim 10^{-1}$ GeV $\gg$ $\Gamma_K \sim 10^{-17}$ GeV). Moreover, the radiative decays of vector mesons $V \rightarrow \mathrm{DS} + P$ where $P$ is a generic pseudo scalar are anyway suppressed since the interaction mediating the process come from the same triangle diagram mediating pseudoscalar radiative decays like $\pi^0 \to \gamma + \DS$.

Recently~\cite{Altmannshofer:2022izm} considered three-body leptonic decays of mesons to put bounds on leptophilic ALPs. In our case too, DS can be produced from such leptonic charged meson decays such as from the decay $K^+ \to \mu^+ + \nu + DS$ via $Z$ portal. However we find this mode to be very suppressed with respect to $K \to \pi + \DS$, due to phase space suppression (see also \cite{Barger:2011mt}).

Eventually, to calculate the number of DS events from a meson decay, we use eq.~\eqref{eq:NDS_meson}. It is clear from eq.~\eqref{eq:NDS_meson} that the meson decay mode that gives the strongest bound would depend on $N_M$, the number of parent mesons produced per POT at a given neutrino experiment. In general, this can be estimated as the ratio of production cross section of the meson to the total cross section between proton beam and target: $N_\text{M} = \sigma_{pN \to M}/\sigma_{pN}$. We take these numbers for various experiments from ref.~\cite{SHiP:2018xqw} (also see references within) for 400 GeV beam energy and from ref.~\cite{Berryman:2019dme} for 120 GeV beam energy, which are obtained using PYTHIA simulations. 

\subsubsection*{Annihilation decays}
DS states can also be produced via annihilation decays of vector mesons through the $JJ$ portal: $V \to \DS$ where $V$ can be  $\rho, \phi, \omega, J/\psi$. We do not consider DS production from the annihilation decay of pseudoscalar mesons from this portal since it will not be model-independent under our approach~\cite{Contino:2020tix}: the pseudoscalar decay matrix element is proportional to its momentum $p_\mu$, which either vanishes when contracted to a conserved DS current, or gives a term proportional to a new, model dependent scale if the DS current is not conserved (corresponding to the internal DS symmetry breaking scale).

In principle the same topology can happen for the Higgs portal and scalar mesons (the matrix element for the spin 1 annihilation through this portal vanishes). However, given the uncertainties in the details of scalar meson production and their subdominance, we do not consider this possibility here. The leading contribution in this topology for the Higgs portal comes from FCNC CP-violating pseudoscalar annihilation decays such as $K \to \DS$~\cite{Hostert:2020gou}, and we will briefly discuss them together with radiative decays in Sec.\ref{sec:results}. 

For a general $V$, and for the case of $V \to \DS$ via $Z$ portal, we can compute the decay width as before:
\begin{equation}\label{eq:VtoDS}
\begin{split}
 \Gamma(V \to \DS) & = \frac{1}{2\,m_V} \frac{1}{3} g_Z^2 \frac{\kappa_J^2 v^2}{m_Z^2 \LUV^4}  f_V^2 m_V^2 \\
 \times  \sum \epsilon_{\mu}^{*}(p)& \epsilon_{\nu}(p) \ 2 \  \text{Im} \expval{J_{\DS}^\mu(p) J_{\DS}^{\nu}(p)}|_{p^2=m_V^2} \\
&= \frac{\kappa_J^2 c_J}{96 \pi} \frac{g_Z^2 v^2}{m_Z^2} \frac{m_V^3 f_V^2}{\LUV^4}\:,
\end{split}
\end{equation}
where $m_V$ is the mass of the vector meson, $f_V$ is the decay constant defined by $\left<V(p)\right|\bar{q}\gamma_\mu\,q\left|\Omega\right> = i f_V m_V \epsilon^*_\mu (p)$, $\epsilon^*_\mu(p)$ is the polarization vector for $V$ meson, and $g_Z$ is the coupling of $\bar{q}\gamma_\mu\,q$ to $Z$ boson. Here we have again used the optical theorem to do integration over DS phase space and used the expressions reported in \cite{Contino:2020tix} for the imaginary part of the correlators at $m_V \gg \LIR$. Out of $\phi$, $\omega$ and $J/\psi$, the largest branching ratio to DS would be that of $J/\psi$ because of the narrow total width, and a partial width which is enhanced by the mass. 
\begin{figure}
\centering
\begin{tikzpicture}
\begin{feynman}
   \vertex(a0){\(V\)};
   \vertex[right=0.2cm of a0](a1);
   \vertex[right=1.2cm of a1](a2);
  \vertex[right=0.2cm of a1](s1);
  \vertex[right=0.6cm of a1](s2);
  \vertex[right=1.0cm of a1](s3);
  \vertex[below=0.5cm of a1](a3);
  \vertex[below=0.5cm of a3](a5);
    \vertex[right=0.2cm of a5](v1);
    \vertex[right=0.6cm of a5](v2);
  \vertex[right=1.0cm of a5](v3);
   \vertex[right=1.2cm of a5](a6);
   \vertex[right=2cm of a3](a4);
   \vertex[right=0.8cm of a4](c1);
   \vertex[right=2.3cm of a2](c2);
   \vertex[below=0.5cm of c2](c0);
   \vertex[below=1cm of c2](c3);
   \vertex[below=1cm of c3](c4);
   \vertex[right=0.2cm of c0](label){\(DS\)};
    \diagram* {
       {
       (a1) --[very thick] (a2) --[anti fermion, edge label = \(\overline q\), very thick] (a4),
       (a6)--[fermion,  edge label = \(q\), very thick] (a4),
       },
      (a6) --[very thick] (a5),
      (s1)--[scalar](v1),
      (s2)--[scalar](v2),
      (s3)--[scalar](v3),
       (a4) -- [boson, very thick, edge label'=\(Z\)](c1),
       (c1) -- [very thick](c2),
       (c1) -- [very thick](c3),
        };
\end{feynman}
\filldraw[color=MidnightBlue!60, fill=MidnightBlue!5, very thick](c0) ellipse (0.2 and 0.5);
\end{tikzpicture}
\caption{\label{fig:V_DS_JJ}\small{DS produced via $Z$ portal in annihilation decays of vector mesons.}}
\end{figure}
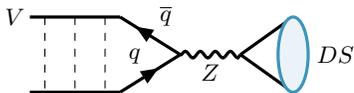    

We find that the bounds from $J/\psi \to \DS$ are comparable to those from $B \to K/K^* + \DS$ decays at SHiP. Despite $\text{BR}(J/\psi \to \DS)/\text{BR}(B \to K/K^* + \DS) \sim 10^{-2}$, the large number of $J/\psi$ mesons expected at SHiP as compared to B mesons, $N_{J/\psi}/N_B \sim \mathcal{O}(100)$ compensates for this.  

For neutrino experiments based on the 120 GeV NuMI beam line, annihilation decays of lighter vector mesons like $\rho, \ \omega, \ \phi$  
can give contribution to signal events. Out of the three vector mesons $\rho, \ \omega, \ \phi$, we find that the leading contribution to DS production is via $\phi$ meson decay to DS. We can compare the branching ratio for $\phi$ and $\omega$ decay to DS via $Z$ portal:
\begin{equation}
\begin{split}
\frac{\text{BR}(\phi \to \text{DS})}{\text{BR}(\omega \to \text{DS})} = \frac{\text{BR}(\phi \to e^+ e^-)}{\text{BR}(\omega \to e^+ e^-)} \frac{m_{\phi}^4}{m_{\omega}^4} 
\frac{\big(-\frac{1}{2} + \frac{2}{3} \sin^2 \theta_W \big)^2}{e_s^2 \ 4  \sin^4 \theta_W}\:,
\end{split}
\end{equation}
where $e_s=-1/3$ is the EM charge of strange quark. Using this we expect DS produced in $\phi$ decay to dominate over $\omega$ decay to DS by a factor given by: $N_{\phi}/N_{\omega} \times \text{BR}(\phi \to \text{DS})/\text{BR}(\omega \to \text{DS})$ $\sim$ $0.007/0.03 \times 50 \sim 10$. Here we have used numbers for $\phi$ meson production at 80 GeV from ref. \cite{NA61SHINE:2019gqe} and $\omega$ meson production at 120 GeV from \cite{Darme:2020ral}. A similar estimate shows that the case of $\rho$ is also subleading. Therefore we only focus on the $\phi$ decay and do not consider $\rho$ and $\omega$. Note that $\rho$ and $\omega$ annihilation decays overlap with the (vector) bremsstrahlung production mode when $\pds ^2$ hits the resonance peak~\cite{deNiverville:2016rqh}: not including them avoids over-counting such contributions.
We do not consider the annihilation decays of heavier mesons like  $\Upsilon$ since its production will be very suppressed at neutrino experiments due to its large mass.

Now we outline how we compute the LDSP boost entering the decay probability and geometric acceptance factors for the meson production mode. More details can be found in the appendix~\ref{appendix:ego_mesons}. In order to calculate the decay probability of the LDSP, we use the following estimate for the average boost factor for the LDSP produced from meson decays:
\begin{align}
    \expval{\gamma}_\text{LDSP} \approx \frac{\expval{E_{\text{DS}}^{\text{lab}}}}{\left<n_{\text{LDSP}}\right> \Lir}\:,
    \label{eq:boost_meson}
\end{align}
where $\expval{\eds^{\text{lab}}}$ is the average energy of the DS produced from parent meson decay in the lab frame. We have checked that an honest average of $\langle \gamma\rangle_{\ldsp}$ matches this estimate very well. To obtain $\expval{\eds^{\text{lab}}} $, the strategy is as follows:
for radiative decays of the form $\text{M} \to \text{m} + \DS$, in the parent meson rest frame, the DS 3-momentum $\vec{p}_{\text{DS}}^{\ 0} = \{ (\pds^0)_T, (\pds^0)_z \}$ can be written using energy conservation as: 
\begin{equation} \label{eq:pds0_rad}
    |\vec{p}_{\text{DS}}^{\ 0}|= \sqrt{ \frac{(M^2 - m^2 + \pds^2)^2}{4 M^2} - \pds^2}\:,
\end{equation}
and fixes $(p_\DS^0)_z = |\vec{p}_{\text{DS}}^0|\,\cos \theta^0_\DS$, where $\theta_\DS^0$ is the angle that the DS makes with the meson flight direction, in its rest frame. For annihilation decays, of the form $\text{M} \to \DS$, the DS 3-momentum in the meson rest frame is zero by momentum conservation. We further assume that 3-momentum of the mesons that decay to DS is perfectly aligned along the beam axis i.e. $\theta_\text{meson} = 0$.\footnote{A more refined analysis using~\cite{Berryman:2019dme} shows that the most probable value for the ratio between transverse and longitudinal components of 3-momentum of decaying K mesons $|p_\text{meson}^\text{T}/p_\text{meson}^\text{z}| \sim \theta_\text{meson} \sim 10^{-2} \ll 1$. Using a non-zero but small value of $\theta_\text{meson}$ does not change our final results.}

We next calculate ($E_\DS^\text{lab}$, $(p_\DS^\text{\, lab})_z$) from ($E_\DS^0$, $(p_\DS^{\, 0})_z$) using the boost and the velocity of the parent meson in the lab (which is along the z-axis), obtained from the average meson momentum values for various experiments from table 6 in ref.~\cite{Darme:2020ral}. From $(p_\DS^\text{\, lab})_z$, we can obtain $\left|\vec{p}_{\DS}^\text{ lab}\right|$ by noting that the transverse component is unaffected by the z direction boost, so that everything is a function of $\theta^0_\DS$ and $\pds^2$. Finally, to get the average value of DS 3-momentum $\expval{\left|\vec{p}_{\DS}^\text{ lab}\right|}$, we average over $\cos \theta_\DS^0$, since DS is isotropic in this variable and set $\pds^2$ to its average value for each radiative meson case. We have again checked that this matches a true average.

To get the final number of signal events as in eq.~\eqref{eq:factorized_xsec}, we also need the geometric acceptance, which we again compute as an average. See App.~\ref{appendix:ego_mesons} for details of these computations, and App.~\ref{app:factorization} for a comparison between using this average procedure with a more refined analysis. Some typical values of $\expval{\gamma}_\text{LDSP}$ are given in table~\ref{tab:avg_boosts}.

\begin{table*}[]
    \centering
    \begin{adjustbox}{width=2.1\columnwidth,center}
    \begin{tabular}{|l||c|c|c|c|c|c||c|c|c|c|c|c|}
    \hline
   \large{$E_\mathrm{beam}^\mathrm{lab}$} (GeV) & \multicolumn{6}{c||}{\large{Z portal production}}  &  \multicolumn{6}{c|}{\large{H portal production}} \\
    \hline
     & $\langle \gamma \rangle_\text{DS}^\text{weak}$  & $\langle \gamma \rangle_\text{DS}^\text{strong}$ & $\langle \gamma \rangle_\text{LDSP}^\text{weak}$ & $\langle \gamma \rangle_\text{LDSP}^\text{strong}$ & $\langle\egeo\rangle^\text{weak}$ &$\langle\egeo\rangle^\text{strong}$ & $\langle \gamma\rangle_\text{DS}^\text{weak}$  & $\langle \gamma\rangle_\text{DS}^\text{strong}$& $\langle \gamma \rangle_\text{LDSP}^\text{weak}$& $\langle \gamma \rangle_\text{LDSP}^\text{strong}$& $\langle\egeo\rangle^\text{weak}$& $\langle\egeo\rangle^\text{weak}$\\
       \hline 
        
        \rowcolor[gray]{.9}
         \multicolumn{13}{|l|}{Drell-Yan}\\
           120 (DUNE-MPD) & 12 & 12 & 1490 & 160 & 0.004 & 0.004 & 9 & 9 & 1600 & 150 & 0.002 & 0.002  \\
           400 (SHiP) & 25 & 25 & 4310 & 400 & 0.63 & 0.63 & 17 & 17 & 4830 & 364 & 0.54 & 0.54  \\
           \hline
            \rowcolor[gray]{.9}
        \multicolumn{13}{|l|}{Dark Bremsstrahlung}\\
         120 (DUNE-MPD) & 80 & 80 & 4500 & 655 & 0.040 & 0.040 & 74 & 74 & 4630 & 650 & 0.039 & 0.039  \\
         400 (SHiP) & 270 & 270 & 15000 & 2200 & 1 & 1 & 250 & 250 & 15700 & 2200 & 0.96 & 0.97   \\
         \hline
         \rowcolor[gray]{.9}
         \multicolumn{13}{|l|}{Meson Radiative Decay $K\to\pi+\DS$}\\
          120 (DUNE-MPD) & 31 & 26 & 215 & 48 & 0.003 & 0.004 & 31 & 26 & 215 & 48 & 0.003 & 0.004  \\
         400 (SHiP) & 55 & 45 & 375 & 84 & 0.79 & 0.89 & 55 & 45 & 375 & 84 & 0.79 & 0.89\\
         \hline
          \rowcolor[gray]{.9}
         \multicolumn{13}{|l|}{Meson Annihilation Decay $\phi \to \DS$ }\\
          120 (DUNE-MPD) & 8 & 8 & 403 & 61 & 0.001  & 0.001 & - & - & -& - & - & \\
         400 (SHiP) & 14 & 14 & 702 & 107 & 0.27  & 0.27 & - & - & -& - & - & - \\
         \hline
    \end{tabular}
    \end{adjustbox}
    \caption{\small{Average quantities $\langle\gamma\rangle_\text{DS} =\langle E_{\text{DS}}^{\text{lab}}/\pds\rangle$, $\langle\gamma\rangle_\text{LDSP}$ and $\langle\egeo\rangle$ for $Z$ portal and H portal production, for various production modes, and for weak/strong case. The reported numbers are for fixed $\Lir=10$ MeV. For Higgs portal, we have taken $\Delta_\mathcal{O} = 4$. The shown numbers are for DUNE-MPD (at 120 GeV) and SHiP (at 400 GeV) target materials (which sets the target atomic weight and number $A,Z$ respectively). The average DS boost $\langle\gamma\rangle_\text{DS}$ depends on weak/strong case through the kinematic condition $\sqrt{\pds^2} \geq n_\ldsp \Lir$ imposed when calculating the average, and is a weak dependence. Annihilation decays of vector mesons does not proceed through the Higgs portal due to mismatch in quantum numbers.
    }}
    \label{tab:avg_boosts}
\end{table*}

\subsection{Drell-Yan production}\label{sec:dy_prod}
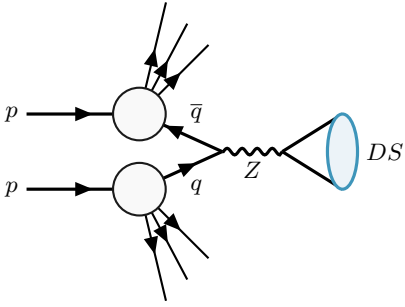
\begin{figure}
\centering
\begin{tikzpicture}
\begin{feynman}
   \vertex(a0);
   \vertex[right=0.2cm of a0](a1);
   \vertex[right=1.2cm of a1](a2);
   \vertex[left=1.5cm of a2](a10){\(p\)};
   \vertex[left=1.5cm of a6](a11){\(p\)};
   \vertex[above right=1.3cm of a2](p1);
   \vertex[above left=0.4cm of p1](p2);
   \vertex[above left=0.4cm of p2](p3);
   \vertex[below right=1.3cm of a6](b1);
   \vertex[below left=0.4cm of b1](b2);
   \vertex[below left=0.4cm of b2](b3);
   \vertex[below=0.5cm of a1](a3);
   \vertex[below=0.5cm of a3](a5);
    \vertex[right=0.2cm of a5](v1);
    \vertex[right=0.6cm of a5](v2);
   \vertex[right=1.0cm of a5](v3);
   \vertex[right=1.2cm of a5](a6);
   \vertex[right=2.3cm of a3](a4);
   \vertex[right=0.8cm of a4](c1);
   \vertex[right=0.8cm of c1](c0);
   \vertex[above=0.5cm of c0](c2);
   \vertex[below=0.5cm of c0](c3);
   \vertex[below=1cm of c3](c4);
   \vertex[right=0.2cm of c0](label){\(DS\)};
    \diagram* {
       {
        (a2) --[anti fermion, edge label = \(\overline q\), very thick] (a4) --[edge label = \(q\), very thick, anti fermion] (a6),
        (a10) --[very thick, fermion](a2),
        (a11) --[very thick, fermion](a6),
        (a2) --[thick, fermion](p1),
        (a2) --[thick, fermion](p2),
        (a2) --[thick, fermion](p3),
        (a6) --[thick, fermion](b1),
        (a6) --[thick, fermion](b2),
        (a6) --[thick, fermion](b3),
       },
       (a4) -- [boson, very thick, edge label'=\(Z\)](c1),
       (c1) -- [very thick](c2),
       (c1) -- [very thick](c3),
        };
\end{feynman}
\filldraw[color=black!90, fill=gray!5,  thick](a2) circle (0.35);
\filldraw[color=black!90, fill=gray!5,  thick](a6) circle (0.35);
\filldraw[color=MidnightBlue!60, fill=MidnightBlue!5, very thick](c0) ellipse (0.2 and 0.5);
\end{tikzpicture}
\caption{\label{fig:dy_z}\small{ Prototypical Drell-Yan process in $Z$ portal for the DS production.}}
\end{figure}

If the typical exchanged momentum from the protons to the DS is comparable or larger than $\Lambda_{\mathrm{QCD}}$, the process is able to probe the partonic constituents of the nucleon. Given the energy scales involved, the protons are ultra-relativistic, and using the parton distribution functions (PDF) language to model the interaction between the constituents is justified. Notice that in our case, the condition to probe the partonic structure of the nucleon is $\pds ^2\gtrsim 1$ GeV$^2$, which is a request on the total DS system, and not on the mass $\LIR$ of the DS constituents. This is unlike what happen in models in which the mediator is produced on-shell, such as in light dark photon models.
The production cross-section is in general dependent on the portal. 
A general estimate for the amplitude of DY through a given portal can be obtained on dimensional grounds, by assuming the typical momentum to be $\sqrt{\pds^2}$, and integrating over it to get the cross section. For Higgs portal, the partonic cross-section comes from Higgs exchange and is dominated by gluon initial states, while for $Z$ portal, there is a $Z$ exchange, and the initial states are the quarks. For the Z-aligned $JJ$ portal, the results of $Z$ portal apply, once appropriately rescaled, if the couplings are assumed to be Z-aligned (both in axial-vector and isospin space). The Feynman diagram for such a process is shown in fig.~\ref{fig:dy_z}. Due to the similarity with Drell-Yan (DY) annihilation process we dub this production channel DY. 

Consider first the Higgs portal. The leading interaction at the constituent level is due to \emph{gluon-gluon fusion} (ggF) processes: indeed light quarks, while abundant in the proton, have a suppressed coupling to the Higgs, while heavy quarks are rare in the proton.
Following~\cite{Boiarska:2019jym}, the effective ggF operator is, after integrating out the Higgs, 
\begin{align}
    \mathcal{L}\supseteq F(\hat{s}) \frac{\kappa_\mathcal{O}}{\LUV^{\Delta-2}}\frac{\alpha_s}{4 \pi m_h^2} G^{\mu\nu\; a}G_{\mu\nu}^a \mathcal{O} \;,
    \label{eq:ggF}
\end{align}
where $F(\hat{s})$ is a function of the center of mass energy $\hat{s}$ of the process that accounts for the loops of internal quarks. Given that our computation is valid only for $\pds$ above the QCD scale, we retain in $F$ only the contributions coming from the top, bottom, charm and strange quarks.
This expression holds for center of mass energies much smaller than the Higgs mass (true for typical neutrino and beam dump experiments).

The cross section to produce a DS shell of total momentum $\pds^2$ can be computed by integrating over the DS phase space using optical theorem:
\begin{align}
    &\sigma^\text{DY(Higgs)} = A\, \sigma_\text{pp}^\text{DY(Higgs)} = \frac{A\,\alpha_s^2}{128 \pi} \frac{2\kappa_{\mathcal{O}}^2c_{\mathcal{O}}}
    {m_h^4 \pi^{3/2}}  \frac{\Gamma(\Delta_{\mathcal{O}}+1/2)}{\Gamma(2\Delta_{\mathcal{O}}) \Gamma(\Delta_{\mathcal{O}}-1)} 
    \nonumber \\
    &\qquad\times
    \int_{Q_0^2}^s \frac{\dd \pds ^2}{2\pi}
    \left|F(\pds ^2)
    \right|^2
    \frac{\pds ^{2\left(\Delta_{\mathcal{O}} -1\right)}}{\LUV ^{2\Delta_{\mathcal{O}}-4}} 
    \nonumber \\
    &\qquad \times \int_{\pds ^2/s}^1 \frac{\dd x}{s x}f_g\left(\pds,x\right)f_g\left(\pds,\pds^2/(s x)\right)\:,
    \label{eq:ggfh}
\end{align}
where $A$ is the atomic number of the target nucleus, $f_g$ are the gluon PDFs\footnote{To compute the PDF integral, we used the nCTEQ15 PDF values \cite{Kovarik:2015cma}, included in the MANEPARSE Mathematica package \cite{Clark:2016jgm}.}, $x$ is the longitudinal momentum fraction of one of the initial gluons in the CM frame and $\sqrt{s}$ is the center of mass energy of the protons. 
The lower limit of the integral over $\pds ^2$ is cutoff at $Q_0^2 = 1.3 \;\mathrm{GeV}^2$, the lowest for which the PDFs have been fitted, and below which the process does not probe a single parton and the DY picture breaks down. For consistency, $Q_0^2$ must be more than the minimum invariant mass of the DS system, $(n_\text{LDSP} \LIR)^2$, which we impose internally. 

Next consider the $Z$ portal. Since the quark-Z coupling depends only on the up or down type of the quark, the process is dominated by light quark-antiquark annihilations. We will consider only contributions coming from up and down quarks, for which the couplings are given as
\begin{align}
    g_u^2 &=\frac{g_{\mathrm{EW}}^2}{\cos^2 \theta_W}\left( \frac{1}{8}+\frac{4}{9}\sin ^4 \theta_W-\frac{1}{3}\sin ^2 \theta_W\right) \;,
    \nonumber\\
    g_d^2&=\frac{g_{\mathrm{EW}}^2}{\cos^2 \theta_W}\left( \frac{1}{8}+\frac{1}{9}\sin ^4 \theta_W-\frac{1}{6}\sin ^2 \theta_W\right) \;.
\end{align}

Unlike the ggF case, the relevant PDFs depend on whether the target nucleon is a proton or a neutron. We approximate the neutron PDFs $f^n_i$ to be the isospin-rotated PDFs of the proton $f^p_i$:
\begin{align}\label{eq:pdfpn}
    f^p_u=f^n_d \,,\:\: f^p_d=f^n_u \,,\:\: f^p_{\bar{u}}=f^n_{\bar{d}} \,, \:\:f^p_{\bar{d}} = f^n_{\bar{u}} \;.
\end{align}
The partonic cross section for the pp/pn interaction in the limit of massless quarks reads:
\begin{align}
    & \sigma_{pp/pn}^{\mathrm{DY(Z)}}
    =\frac{1}{1152 \pi}\frac{g_\mathrm{EW}^2 v^4 \kappa_J^2 c_J}{m_Z^4\cos^2 \theta_W\LUV ^4}
    \int_{Q_0^2}^s \dd \pds^2 \pds^2
    \nonumber \\
    &\qquad \times
    \int_{\pds^2/s}^1 \frac{\dd x}{sx}\:\:X_{pp/pn}(s, x, \pds^2)\:,
    \label{eq:qdyz_pp}
\end{align}
where
\begin{align}
    X_{pp}(s, x, \pds^2) 
    &=  
    2\sum_{i = u , d}
    \left(g_i^2 f^p_i(x) f^p_{\bar{i}}(\pds^2/(s x))\right) \;,
    \nonumber \\
    X_{pn}(s, x, \pds^2) 
    &=  
    g_u^2\,f^p_u(x) f^p_{\bar{d}}(\pds^2/(s x)) 
    \nonumber \\
    &\:\:+ g_u^2\,f^p_{\bar{u}}(x)f^p_{d}(\pds^2/(s x))
    \nonumber \\
    &\:\:+ g_d^2\,f^p_d(x) f^p_{\bar{u}}(\pds^2/(s x))
    \nonumber \\
    &\:\:+ g_d^2\,f^p_{\bar{d}}(x)f^p_{u}(\pds^2/(s x))\:.
\end{align}    
In the PDFs used, we have taken the factorization scale to be the exchanged momentum $\pds^2$ and not indicated it explicitly to keep the expressions simpler. 
Putting the contributions from the protons and neutrons together, the total DY cross section for the $Z$ portal is:
\begin{equation}\label{eq:total_qdyz}
    \sigma^{\mathrm{DY(Z)}}=Z \sigma^{\mathrm{DY(Z)}}_{pp}+(A-Z)\sigma^{\mathrm{DY(Z)}}_{pn}\;,
\end{equation}
where $Z,A$ are respectively the atomic and weight number of the target nuclei.
Notice that in both Higgs and $Z$ portal scenarios the cross section increase with $\pds$, as expected on dimensional grounds. The drop at high $\pds$ seen in fig.~\ref{fig:prod modes} is due to the PDF convolutions.

In order to estimate the decay probability, we estimate the average boost of the LDSP in the lab frame (not to be confused with the boost of the total DS system) as given in eq.~\eqref{eq:boost_meson}.
The value for these averaged quantities is given in table~\ref{tab:avg_boosts}. 

In principle, the boost should take into account the angle in the DS frame: while particles in the DS frame have roughly the same energy, in the lab frame particles emitted along the beam are more boosted with respect to particles emitted in the opposite direction. We have checked that this effect is negligible, when restricting to particles hitting the detector.
For the geometric acceptance, we notice that in the DY production mode, the DS system has no transverse momentum and is collinear to the beam axis. After boosting the LDSP momentum in the DS frame we compute the angles that corresponds to the detector. To estimate $\egeo ^{\mathrm{DY}}$, we follow the prescription given in App.~\ref{app:egeo}.

In general, in the DY production mode, events are produced with larger $\pds^2$ compared to other modes (see fig.~\ref{fig:prod modes}). We also find that the average energy of the DS system in the lab frame is not as high as in bremsstrahlung. These lead to lower $\gamma$ of the DS system, a larger LDSP angular spread and therefore a slightly smaller $\egeo$ for on-axis detectors.

\subsection{Dark Bremsstrahlung ($pp \to \DS + X$)}\label{sec:brem_prod}
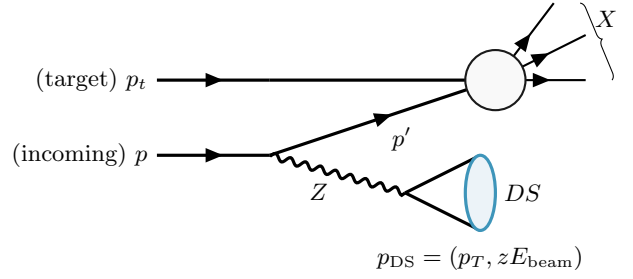
\begin{figure}
\centering
\tikzfeynmanset{
  every blob/.style={draw=green!40!black, pattern color=green!40!black},
  }
\begin{tikzpicture}
\begin{feynman}
   \vertex(a0);
   \vertex[left=0.01 cm of a0](t2){\(\text{(target)} \ p_t\)};
   \vertex[blob, right=4.5cm of a0](a1);
   \vertex[below=1.0cm of a0](a2);
   \vertex[left=0.01cm of a2](t1){\(\text{(incoming)} \ p\)};
   \vertex[below=1.5cm of a0](b1);
   \vertex[right=3.3 cm of b1](b2);
   \vertex[right=1cm of b2](b3);
   \vertex[below=0.5cm of b3](b4);
   \vertex[above=0.5cm of b3](b5);
    \vertex[below=0.1cm of b4](b6){\( p_\text{DS}=(p_T, z E_\text{beam}) \)};
   \vertex[right=1.5cm of a0](a3);
   \vertex[below=1.0cm of a3](a4);
   \vertex[right=2.0cm of a4](a5);
   \vertex[right=1cm of a5](a6);
   \vertex[below=0.5cm of a6](a7);
    \vertex[above=0.5cm of a6](a8);
    \vertex[right=1.2cm of a1](a9);
    \vertex[above=0.6cm of a9](a10);
    \vertex[above left=0.6cm of a10](a11);
    \vertex[right=0.35cm of a11](a12);
    \vertex[right=0.35cm of a9](a13);
    \vertex[right=0.2cm of b3]{\(DS\)};
    \diagram*{
       {
       (a0) --[very thick, fermion] (a3) -- [very thick] (a1),
      (a2)  --[very thick, fermion](a4) -- [very thick, fermion, edge label'=\(p'\) ] (a1),
      (a1) --[thick, fermion](a9),
      (a1) --[thick, fermion](a10),
      (a1) --[thick, fermion](a11),
       },
       (a4) -- [boson, very thick, edge label'=\(Z\)](b2),
       (b2) -- [very thick](b5),
       (b2) -- [very thick](b4),
        };
       \draw [decoration={brace}, decorate] (a12.north east) -- (a13.south east)
       node [pos=0.2, right]{\(X\)};
\end{feynman}
\filldraw[color=black!90, fill=gray!5,  thick](a1) circle (0.4);
\filldraw[color=MidnightBlue!60, fill=MidnightBlue!5, very thick](b3) ellipse (0.2 and 0.5);
\end{tikzpicture}
\caption{\label{fig:V_DS_JJ}\small{ DS produced via $Z$ portal in the Bremsstrahlung production mode.}}
\end{figure}    

Another possibility is for the DS states to be produced directly from proton as an initial state radiation. In this case, the exchanged momentum $\pds^2$ is not hard enough to probe the partonic structure. Following \cite{Foroughi-Abari:2021zbm}, we model the process using the initial state radiation (ISR) splitting function formalism. The idea is to treat the DS as incoming from the leg of the initial beam proton, which then becomes slightly virtual---an almost on-shell particle participating in the rest of the process. 
In the following, we will use the standard jargon: $p_T$ for the transverse momentum of the DS system  (in the plane orthogonal to the beam direction) and $z\,E_\text{beam}$ for its longitudinal momentum, where $E_\text{beam}$ is the beam longitudinal momentum (in the lab frame).

The splitting function formalism works well when the virtual particle is almost on shell. This means that in order to get reasonable cross sections, we must integrate the variables $p_T$ and $z$ in a sub-region of their kinematically allowed values, in which the virtual proton is not too off-shell.
Denoting the proton after DS emission as $p'$,  concretely, we will consider the region in which the virtuality is small:
\begin{align}\label{eq:brem_thresh}
    \frac{p^{\prime \,2}-m_p^2}{E_{p^{\prime}}^2}=\frac{z^2m_p^2+(1-z)\pds^2+p_T^2}{z(1-z)^2E_{\mathrm{beam}}^2}<0.1\:.
\end{align}
The choice of $0.1$ is arbitrary and our results are not sensitive to small changes in this.
In order to compute the splitting functions, we need to compute the vertex between the proton and the $Z$ or Higgs.
For the Higgs, by using low energy theorems~\cite{SHIFMAN1978443} we can compute the coupling between the Higgs and nucleon at zero momentum to be $g_{hNN} h\bar{N}N$, where $g_{hNN}=1.2\times 10^{-3}$.
To model the momentum dependence of the form factor, we employ a generalization of the \emph{extended Vector Meson Dominance} (eVMD) model, in which the DS state interacts with the hadron by mixing with the scalar, CP even hadronic resonances. The resonances' propagators are taken to be Breit-Wigners (BW), and the mixing coefficients are fixed by using sum rules and by fitting the zero momentum values. The form factor is taken from \cite{Foroughi-Abari:2021zbm} and the specific values used are reported in appendix \ref{app:ff_h}.

We also need  to take into account the fact that for too high virtuality the quasi-real proton stops interacting with the target proton as a coherent object, and the bremsstrahlung computation breaks down. To do this we multiply the previous form factor by a smooth cutoff~\cite{1999}:
\begin{align}\label{eq:cutoff_FF}
    F_D(Q^2)=\frac{\Lambda _p^4}{\Lambda_p ^4+Q^4}\;,
\end{align}
where $\Lambda_p=1.5$ GeV is the cutoff, taken to be near the proton mass, and $Q^2$ is the virtuality of the intermediate proton:
\begin{align}
    Q^2 = \frac{z^2 m_p^2 + (1-z)\pds ^2 + p_T^2}{z^2} \;.
\end{align}
Finally, the cross section for the process is calculated by factorizing the total cross section into the Bremsstrahlung part and a proton-nucleus (after Bremsstrahlung) part. The proton-nucleus cross section $\sigma_{pN}^{\mathrm{nTSD}}$ is calculated using the difference between the total inelastic proton-nucleus cross section and the \textit{target single diffractive} (TSD) contribution, in which the target nucleus is diffracted but not disintegrated. This choice allows neglecting possible interference between the initial state and the final state radiation \cite{Blinov:2021say, Foroughi-Abari:2021zbm}. According to \cite{Blinov:2021say, Carroll:1978hc}, $\sigma_{pN}^{\mathrm{nTSD}}$ is a slowly varying function of energy, and for the energies involved, we can approximate it as a constant
\begin{align}
    \sigma_{pN}^{\mathrm{nTSD}}&=762 \left(A/56\right)^{0.71}\left(1-0.021\left(56/A\right)^{0.36} \right)\:,
\end{align}
where $A$ is the target atomic weight. 
We will now generalize the results of \cite{Boiarska:2019jym, Foroughi-Abari:2021zbm} to higher dimensional portals. 

So far the discussion applies to any of the portals. However, once a portal is specified, the involved form factors change. Consider first the Higgs portal. Putting everything together, the inclusive production cross section is given as
\begin{align}
\label{eq:h_brem}
&\sigma^\text{Brem(Higgs)}_{pN}
\nonumber\\
&=
\sigma_{pN}^\text{nTSD}g_{hNN}^2
\frac{v^2}{m_h^4 16 \pi^{7/2}}
\frac{\Gamma (\Delta_{\mathcal{O}} +1/2)}{\Gamma(\Delta_{\mathcal{O}}-1) \Gamma(2 \Delta_{\mathcal{O}})}
\frac{\kappa_{\mathcal{O}}^2 c_{\mathcal{O}}}{\LUV^{2\Delta_{\mathcal{O}}-4}}
\nonumber \\
& \times \:\: 
\int\dd \pds ^2\dd p_T^2\dd z
\left|F_H\right|^2
\nonumber \\
& \times \:\: 
z\left( \left(2-z \right)^2m_p^2+p_T^2\right)
\nonumber\\
&\times \:\:
\left( \frac{1}{m_p^2z^2+(1-z)\pds ^2+p_T^2}\right)^2 \pds^{2\Delta_{\mathcal{O}}-4}\:,
\end{align}
where $F_H$ is the Higgs bremsstrahlung form factor built as outlined before, and can be found in App.~\ref{app:ff_h}. The limits of integration are chosen to respect the kinematic condition $\pds^2\geq n_\text{LDSP}^2\Lir^2$.

Next consider the $Z$ portal case. The only difference is in the form factors of the axial and vector current of the proton. For the vector case, the production cross section is given as
\begin{align}
& \sigma^\text{Brem(Z, Vector)}_{pN}
\nonumber\\
&= 
\sigma_{pN}^\text{nTSD}
\frac{\kappa_J^2 c_J}{2^{11}\pi^4}
\frac{v^4}{m_Z^4\LUV ^4}
\frac{g_{\mathrm{EW}}^4}{\cos \theta_W^4}
\nonumber\\
&\times\:\: 
\int \dd \pds^2 \dd p_T^2 \dd z 
\left|F^\text{V}_Z\right|^2 
\nonumber\\
&\times\:\:
\left(\frac{2}{z}
+\frac{4\pds^2\,z\,\left(p_T^2+m_p^2(z^2+2z-2)\right)}{\left(m_p^2z^2+(1-z)\pds^2+p_T^2\right)^2}
\right)\;.
\label{eq:z_brem_vec}
\end{align}
The vector form factor $F^{V}_Z(\pds^2)$ is modeled by $\rho$ (iso-triplet) and $\omega$ (iso-singlet) exchange. We take three states for each tower. Details are given in App~\ref{app:FormFactors}. For the axial case, the cross section is given as
\begin{align}
& \sigma^\text{Brem(Z, Axial)}_{pN}
\nonumber\\
&= 
\sigma_{pN}^\text{nTSD}
\frac{\kappa_J^2 c_J}{2^{11}\pi^4}
\frac{v^4}{m_Z^4\LUV ^4}
\frac{g_{\mathrm{EW}}^4}{\cos \theta_W^4}
\nonumber\\
&\times\:\: 
\int \dd \pds^2 \dd p_T^2 \dd z 
\left|F^\text{A}_Z\right|^2 
\nonumber\\
&\times\:\:
\frac{2}{z}
\left( \frac{1}{m_p^2z^2+(1-z)\pds^2+p_T^2}\right)^2
\nonumber\\
&\times\:\:
\left( 
\pds^2(1-z)^2+\left(p_T^2+z^2m_p^2 \right)^2+2\pds^2 \right.
\nonumber\\
&\times\left.\left(m_p^2z^2\left(5+(z-5)z+p_T^2(1+z^2-z) \right) \right)\right)\;.
\label{eq:z_brem_ax}
\end{align}
Similar to the vector case, for the axial form factor we take the respective iso-triplet axial vector exchange (there is no contribution from the axial iso-singlet resonances). Details about the axial form factor $F^{A}_Z(\pds^2)$ are in App~\ref{app:FormFactors}.
Combining the vector and the axial pieces we get
\begin{align}
    \sigma^\text{Brem, (Z)}_{pN}
    &=
    \sigma^\text{Brem, (Z, Vector)}_{pN}
    +
    \sigma^\text{Brem, (Z, Axial)}_{pN}\:.
    \label{eq:z_brem_tot}
\end{align}
Notice that in eq.~\ref{eq:z_brem_tot} the interference term between the vector and axial piece vanishes, due to the different quantum numbers under parity of the two possible states.

The vector contribution to the cross section is subdominant with respect to the axial one, due to an accidental cancellation in the vectorial quark coupling. The vector contribution has a more narrow distribution in $\pds^2$ than the axial one, and it's peaked at $m_\omega ^2$: this is because $\omega$ is much more narrow than the iso-triplet vectors and axial vectors resonances mixing with the Z. Notice that the same exact computation holds for a $JJ$ portal aligned (in both Lorentz and flavor space) to the $Z$ quantum numbers. For different coupling structure of $J_\mathrm{SM}$ for a generic $JJ$ portal, we can decompose the proton vector and axial form factors in their iso-singlet and iso-triplet components to get the correct form factor, shown in App.~\ref{app:FormFactors}.

An estimate for the cross section can be given by exploiting the fact that the cross section is dominated by the Breit-Wigner (BW) peaks in the form factors. The $\pds^2$ integral of the BW associated with an intermediate meson $\mathfrak{m}$ can be estimated as $\pi f_\mathfrak{m}^2 m_\mathfrak{m}^3/(2 \Gamma_\mathfrak{m})$, where we have used the notation of App.~\ref{app:FormFactors}.
Therefore, using the splitting function to give the correct momentum scaling, upto an $\mathcal{O}(1)$ factor, the cross section for radiating DS particles through a dimension $D$ portal made of a SM operator and a dimension $\Delta_{\mathcal{O}}$ DS operator can be estimated as:
\begin{align}
    \sigma \sim \sigma_{pN}g_\mathrm{SM}^2\Phi(\Delta)f_\mathfrak{m}^2 
    \begin{cases}
      \frac{v^2}{m_h^4}\,
      \frac{m_\mathfrak{m}^{2\Delta_{\mathcal{O}}-1}}{\Luv^{2\Delta_{\mathcal{O}}-4}\Gamma_\mathfrak{m}}\: &:\: \mathcal{O}H^\dagger H\text{ portal} \:,\\
      & \\
      \frac{m_\mathfrak{m}^5}{\Luv^4 \Gamma_\mathfrak{m}}\: &:\: JJ \text{ portal}\: ,
      \end{cases}
\end{align}
where, as in the meson case, $g_\mathrm{SM}$ is a dimensionless SM factor built out of dimensionless couplings (like the gauge couplings), $\Phi(\Delta)$ is the phase space factor coming from the integration over the DS degrees of freedoms, $f_\mathfrak{m}$ is the coupling of proton to the meson $\mathfrak{m}$ and $\Gamma_\mathfrak{m},m_\mathfrak{m}$, is its decay width and mass respectively (see App~\ref{app:FormFactors}).
In presence of multiple resonances, the estimate can be done by restricting to the leading contribution of the BW.
For both the Higgs and $Z$ portals the DS is produced as a collimated state forming an angle with the beam $\theta_{\mathrm{DS}} \sim p_T/E_{\mathrm{DS}}^{\mathrm{lab}}=p_T/( z E_{\mathrm{beam}})$. The average acceptance $\egeo^{\mathrm{brem}}$ is computed by averaging over all the kinematic variables, and it doesn't differ much from the one obtained by replacing $\theta_\mathrm{DS}$ with its average.
Details of the computation of the geometric acceptance are given in App~\ref{app:egeo}.
To estimate the average decay probability, we use the average LDSP boost, as defined in eq.~\eqref{eq:boost_meson} (see also table~\ref{tab:avg_boosts} for typical values), with the only difference being that the probability distribution is given by the splitting function.
The typical energy of the DS system in the lab frame is roughly $(3/4) E_\mathrm{beam}$ for all the portals considered, larger than in DY case.

\section{Experimental Setups, Signal and Background Estimation}
\label{sec:experiments}
In this section we briefly discuss the experiments we consider for obtaining the bounds on the $\LUV$ and $\LIR$ scales, and the assumptions we make when obtaining these bounds.

\begin{figure*}[t]
\includegraphics[width=0.8\textwidth]{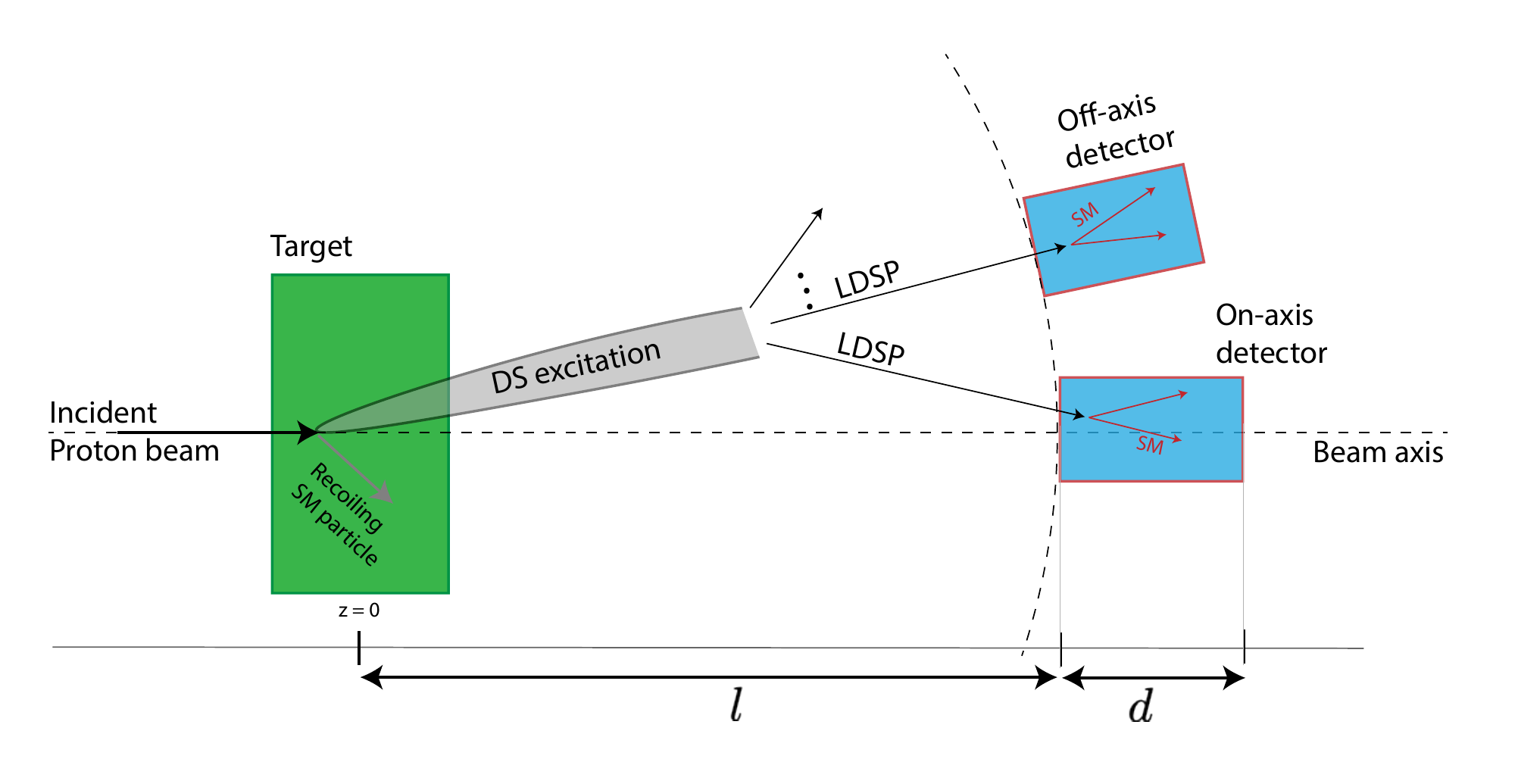}
\caption{\small{A cartoon of DS events produced at a typical neutrino detector (drawn not to scale): both on-axis and off-axis cases are shown. In general the DS state may not be produced along the beam-axis, as shown, though this is dependent on the production mode: annihilation decays of Mesons and DY production modes produce DS along the beam-axis, while radiative decays of Mesons and DB production mode produce DS at a small angle from the beam-direction.}}
\label{fig:typical_exp}
\end{figure*}

As explained in Sec.~\ref{sec:darkcft}, the relevant characteristics of high intensity experiments are their beam energy $E_\mathrm{beam}$, their integrated luminosity (reported as the total protons on target $N_{\text{POT}}$) and geometric details of the experimental setup. Further, the detectors in these experiments can be placed on-axis (i.e. along the line of the incoming beam) or off-axis (see fig.~\ref{fig:typical_exp} for a cartoon of the experimental setup), which can change the geometric acceptance if certain production modes are forward peaked. Specific to the MD production mode, the number of meson $N_\text{M}$ produced at a given experiment is an additional input, as seen from eq.~\eqref{eq:NDS_meson}. This depends on the target details as well as the energy in the centre of mass frame, $\sqrt{s} \approx \sqrt{2 E_{\text{beam}} m_p}$.

In this work we consider a representative set of past present and future experiments. Table~\ref{tab:nu_experiments} gives the list of considered experiments with the relevant parameters. 

We show bounds from recasts of BSM search results from past and current experiments, and projections from current and upcoming experiments, considering a few representatives from each category. We will also consider a future dedicated LLP search experiment, SHiP, for comparison, since it has a broad reach in typical DS models (e.g. see ref.~\cite{SHiP:2015vad}). We will take the SHiP parameters to be very optimistic, to have a conservative comparison with the neutrino experiments. A more comprehensive analysis that also considers other dedicated LLP experiments can be useful, and will be done in future.

Out of past experiments, we show CHARM, a beam dump experiment that ran on the CERN SPS (400 GeV) beamline in the 1980s. CHARM searched for the decay of axion-like particles (ALPs) into a pair of photons, electrons and muons and found no events \cite{CHARM:1985anb}, and we will recast this search for our bounds. We do not recast Heavy Neutral Lepton decay searches of CHARM or other past experiments like BEBC, or dark particle scatterings as done for example in~\cite{Marocco:2020dqu, Barouki:2022bkt}, given the different final state topology.

From existing experiments, we choose the MicroBooNE and ICARUS experiments, based on the 120 GeV NuMI beamline. These are two of the three detectors of Fermilab's  Short-Baseline Neutrino program (SBNP)~\cite{MicroBooNE:2015bmn}.\footnote{The third detector, SBND, is too far off-axis with respect to the NuMI beamline and therefore its geometric acceptance is too low to give meaningful constraints. These three detectors also run on the 8 GeV Booster Neutrino Beam (BNB), which is at a lower energy than the NuMI energy, 120 GeV. We find the bounds to be subleading compared to DUNE, and do not consider it.} For MicroBooNE, we will use the analysis in ref.~\cite{ MicroBooNE:2021usw} for dark scalars decaying into electron-positron pairs. 
For ICARUS, we will use the results in~\cite{Batell:2019nwo}, which studied DS coupled through the renormalizable Higgs portal.\footnote{Since the target specifications for NuMI beamline experiments are the same as that of proposed DUNE-LNBF beamline, we recycle the meson production numbers for DUNE-LNBF~\cite{Berryman:2019dme} also for the NuMI beamline experiments ICARUS.}
DM searches at another current experiment, MiniBooNE, based on 8 GeV BNB beamline, use scattering~\cite{MiniBooNEDM:2018cxm,deNiverville:2018dbu}, and as explained in Sec.~\ref{sec:darkcft}, they require additional model dependent assumptions, so we will not consider them here.
Another currently running NuMI-based experiment is NO$\nu$A.
We are not aware of any search for DM decays done at this experiment (for a scattering analysis, see \cite{deNiverville:2018dbu, Filip:2020eat}, based on \cite{Bian:2017axs} ).
Since NO$\nu$A is currently running, we show a possible prospect of such a search. We assume that it will be possible to reduce the backgrounds to negligible amounts, given the good angular resolution of the detector.

For future experiments we look at the Deep Underground Neutrino Experiment (DUNE). Ref.~\cite{Berryman:2019dme} has proposed the use of the multipurpose, high pressure gaseous chamber- the Multi-Purpose Detector (MPD) present in DUNE near detector complex for DS searches. We show projections for our DS scenario for the future DUNE-MPD as well.

\begin{table*}[t]
\begin{ruledtabular}
\begin{tabular}{l l l l l c c}
Experiment  & $N_\text{POT}$ (total) & $E_{\text{beam}}$ (GeV) & $l$ (m) & $d$ (m) & Off-axis angle, $\theta_{\mathrm{det}}$ (rad) & $\theta_{\text{acc}}$ (rad)\\[0.1cm] 
 \hline
 \hline\\[0.1cm]
CHARM \cite{CHARM:1985anb, DORENBOSCH1986473, Gninenko:2012eq} & $2.4 \times 10^{18}$ & 400 & 480 & 35 & 0.01 & 0.003\\[0.1cm] 
NO$\nu$A-ND \cite{deNiverville:2018dbu, Bian:2017axs} & $3 \times 10^{20}$ & 120 & 990  & 14.3 & 0.015 & 0.002\\[0.1cm]
MicroBooNE (KDAR) \cite{MicroBooNE:2021usw} & $1.93 \times 10^{20}$ & 120 & 100  & 10.4 & - & 0.013\\[0.1cm]
ICARUS-NuMI \cite{Batell:2019nwo,MicroBooNE:2015bmn} & $3 \times 10^{21}$ & 120 & 803 & $19.6$ & 0.097 & 0.005 \\[0.1cm]
DUNE-MPD\cite{Berryman:2019dme, DUNE:2020ypp} & $1.47 \times 10^{22}$  & 120 & 579 & 5 & 0 & 0.004\\[0.1cm]
SHiP \cite{SHiP:2015vad,Gorbunov:2020rjx} &  $2 \times 10^{20}$  & 400 & 64 &  50 & 0 & 0.078\\
\\[0.1cm]
\end{tabular}
\end{ruledtabular}
\caption{
\small{The relevant parameters for the experiments considered in this work. The quantities $l$ and $d$ are defined in fig.~\ref{fig:typical_exp}. $\theta_\mathrm{det}$ stands for the position of the detector centre with respect to the beam line, with the origin taken at the interaction point. Entries with zero $\theta_\mathrm{det}$ indicate that the detector is placed along the beam axis. $\theta_\mathrm{acc}$ stands for the detector half angular opening. Note that for MicroBoone KDAR analysis, the K mesons are produced at rest (in the lab frame) so that $\theta_\text{det}$ is irrelevant. The angle $\theta_\text{acc}$ for this case is measured with the origin at the NuMI hadron absorber, placed $\mathcal{O}$(600) m from the interaction point~\cite{MicroBooNE:2021usw}.}}
\label{tab:nu_experiments}
\end{table*}

In order to estimate the sensitivity of the selected current and future experiments, an assessment of the background is needed.
We assume that beam dump experiments can be made background free by imposing cuts with $\mathcal{O}(1)$ signal efficiencies, as seen in past searches, e.g. at CHARM~\cite{CHARM:1985anb}.
On the other hand, at neutrino experiments, the neutrino beam itself can be a source of background events.
At these experiments the typical mass of the LDSPs probed is $\mathcal{O}$(10-100) MeV, therefore the available channels for the LDSP decay to the SM are mostly photons and electron-positron pairs, which produce electromagnetic showers in the detector. Heavier decay products, such as muons, will be reconstructed as tracks (but for all practical purposes, we will treat them similar to the showers in this section). The following discussions hold for any of the decay products. 

In principle, the two shower signature has no irreducible background. Reducible background events come from hard radiation of a single photon, or from neutral-current $\nu$ scattering against a nucleus producing a $\pi^0$, which then decays into $\gamma \gamma$. The produced photons then can convert into $e^+e^-$ pairs, that mimic the signal. However, it's not guaranteed that the two daughter particles will be reconstructed as separate showers. 

The typical condition in order to reconstruct the two particles involves an isolation cuts between the decay products, or in other words an angular separation cut. The specific implementation depends on the specific detector and analysis strategy. We will briefly review what has been suggested in previous works. However many of the relevant aspects can be understood more generally, which we will elaborate with a relevant prototypical experiment in mind. 

For the ICARUS experiment, as suggested in~\cite{Batell:2019nwo}, an angular separation of $10^\circ$ is enough to be able to separate the two showers. Background events instead have a narrow separation between the charged particles, or potentially two showers that do not originate from the same vertex. The angular cut reduces the background events to a negligible amount. In \cite{Batell:2021ooj} the authors elaborate on an analysis with less stringent cuts but with $\mathcal{O}$(100) background events.
Indeed a strong isolation cut has low efficiency for lighter, and therefore more boosted, LDSPs. This is especially true for the models under consideration here, in which DB and DY production modes are non-negligible and generate LDSPs more boosted than the ones coming from meson decays. For example, in DB, for the weakly coupled case $n_\mathrm{LDSP}=2, \Lir=100$ MeV, we expect the daughter particles to be separated by an angle of $1/\gamma_{\mathrm{LDSP}}\approx 0.15^\circ$, which is smaller or comparable to the angular resolutions of some of the detectors. This highlights a potential problem in our framework, when reconstructing the signal.
For this reason, we suggest that at ICARUS, it might be better to avoid a stringent cut in angular separation and work with $\mathcal{O}(100)$ background events~\cite{Batell:2021ooj}, possibly reduced with an energy cut and a cut on the direction of the DS system with respect to the beam.
Interestingly, for strongly coupled DS we expect the angular separation condition to be less stringent on the signal. Because of a larger $n_\ldsp$, the energy is split among more LDSPs, leading to a suppression of the single LDSP boost factor. For such sectors, assuming an average $n_\ldsp$ of $\mathcal{O} (10)$, the average separation angle is typically $\mathcal{O}(1^\circ)$ for DB and $100$ MeV masses, and less for other production modes. Since this is of the order of the angular resolution of ICARUS, it should be feasible to reconstruct the signal events as separated tracks for masses not too light.

Specific to the DUNE-MPD detector, in~\cite{Brdar:2020dpr, Kelly:2020dda,Berryman:2019dme} it has been shown that boosted signal events have a narrower angular separation compared to the more isotropic background distributions. Due to this difference the search can effectively be rendered background free. Even in this case the two decaying particles must be reconstructed as separate particles, which these references claim to achieve. In these studies, the typical opening angle between the decay products is comparable to the weakly coupled scenarios we consider in this work. Therefore we take this search to be background free.

If instead the two decay products are not separated, the event will be reconstructed as a single electron event. The background to this kind of event comes from $\nu_e$ charged-current scattering, from $\nu$-$e$ elastic scattering events or $\nu$ neutral-current quasi-elastic events.
The idea of decaying particle hiding behind the single electron signature has been explored in~\cite{Essig_2010,Foroughi_Abari_2020}. In these works it is shown how to recast the analysis of the LSND experiment (with a beam energy of 0.8 GeV) that looked for $\nu_e$ charged-current scattering~\cite{LSND:1997vqj,LSND:2001aii} to put bounds on BSM particles decaying into $e^-e^+$, hiding as single electron events. In particular, the decaying particle would present as an excess of high energy electron events near the maximum value of the energy analyzed (200 MeV)\footnote{We do not recast LSND bounds in the DB and DY modes, even if the intensity is one of the highest. For DY, due to a very low beam energy of $0.8$ GeV, the integration range of the partonic center of mass energy is very small. For DB, the condition on the integration domain of eq.~\eqref{eq:brem_thresh} is very constraining. Relaxing the condition, by setting the RHS of eq.~\eqref{eq:brem_thresh} to 1, the bounds are still worse than DUNE.} 
 
Considering now a higher beam energy experiment, searches for charged-current at NuMI based experiment (with beam energy 120 GeV) typically look for neutrino with GeV energies, a bit lower than the typical LDSP energy (see for example \cite{NOvA:2022see}). 
Other scattering analysis typically look for $\nu$-$e$ elastic scattering. The cuts imposed require a low energy recoil, and a very forward electron. It's unclear whether or not they can be used to put stringent bounds on misidentified $e^-e^+$ pairs. 
It will be very interesting to explore this signature of single electron hiding in the high energy tail of scattering events at high beam energy experiments, but we will not study this signature here.

We would like to point out that it should be possible to run the neutrino experiments in a beam dump mode, essentially removing all the background, while keeping almost all the signal (except the one coming from charged meson decays). 
A beam dump proposal for DUNE has been studied in~\cite{Bhattarai:2022mue}, showing that indeed running in the beam dump mode allows neglecting all the SM backgrounds in the DUNE detector, albeit at a reduced luminosity of roughly two orders of magnitude (one order of magnitude, for the optimistic scenario). The idea to suppress neutrino background by steering the beam off the target (as in the beam dump mode) has been already implemented at MiniBooNE~\cite{MiniBooNEDM:2018cxm} (although looking for DM scattering events) and MicroBooNE experiments. We will recast the MicroBooNe search, which however is quite different in spirit from the typical beam dump search, as it is optimized to look for Kaon Decay At Rest (KDAR). The idea is to look for the decay products of Kaons decaying at rest in the NuMI hadron absorber, which have a very peculiar directionality: in usual cases, the decay products of produced kaons are collected by a detector placed further down the beamline, whereas here the MicroBooNE detector is placed on the back side of the NuMI hadron absorber (e.g. see fig.~1 in ref.~\cite{MicroBooNE:2021usw}). This peculiarity allows the signal events to be easily distinguished from background events, with an estimated efficiency of $0.14$ on signal selection.

In addition to the backgrounds discussed so far, there is an extra component coming from \emph{neutrino trident events}, in which a neutrino scatters against a nucleus in a purely electroweak process, to produce a lepton-antilepton pair. As argued in~\cite{Berryman:2019dme, Altmannshofer:2019zhy}, the expected number of events is $\mathcal{O}(10)$ events at DUNE-MPD, while it is lower in other liquid Argon detectors like ICARUS. In more conventional detectors like NO$\nu$A, $\mathcal{O}(10)$ $e^-e^+$ trident events are expected~\cite{Ballett:2018uuc}.
Given the rather peculiar kinematics, it's possible to bring down these background events and neglect them in the analysis~\cite{Berryman:2019dme}.

For all these reasons, we will compute the signal yield contours for 10 and 100 event lines when discussing prospects for DUNE-MPD. Indeed 10 events represent a reasonable proxy for an almost background free search in the presence of $\mathcal{O}(1)$ experimental efficiency, although this number could be brought down in specific experiments by a more careful analysis of the reducible backgrounds. The 100 event lines instead can be representative of some signal loss due to selection cuts (which could be present for example in the weakly coupled case due to a small angular separation) or for a reduction in $N_\mathrm{POT}$, for example due to running in dump mode for a limited amount of time.

From the experimental analysis we recast, we use $95\%$ confidence level, including the signal efficiencies reported. For CHARM~\cite{CHARM:1985anb} which observed 0 event, we set a bound at $95 \%$ confidence level of $N_\text{signal} < 3$, using efficiency of 0.51 and 0.85 for the $e^+ e^{-}$ and $\mu^+ \mu^-$ modes respectively. For MicroBooNE KDAR~\cite{MicroBooNE:2021usw} which observed 1 event compared to a background expectation of 1.9 events, we require $N_\text{signal} < 3.8$ at $95\%$ confidence level, with a signal reconstruction efficiency of 0.14.


\section{\label{sec:results}Results}

\begin{figure*}[!ht]
\centering
\includegraphics[width=0.73\textwidth]{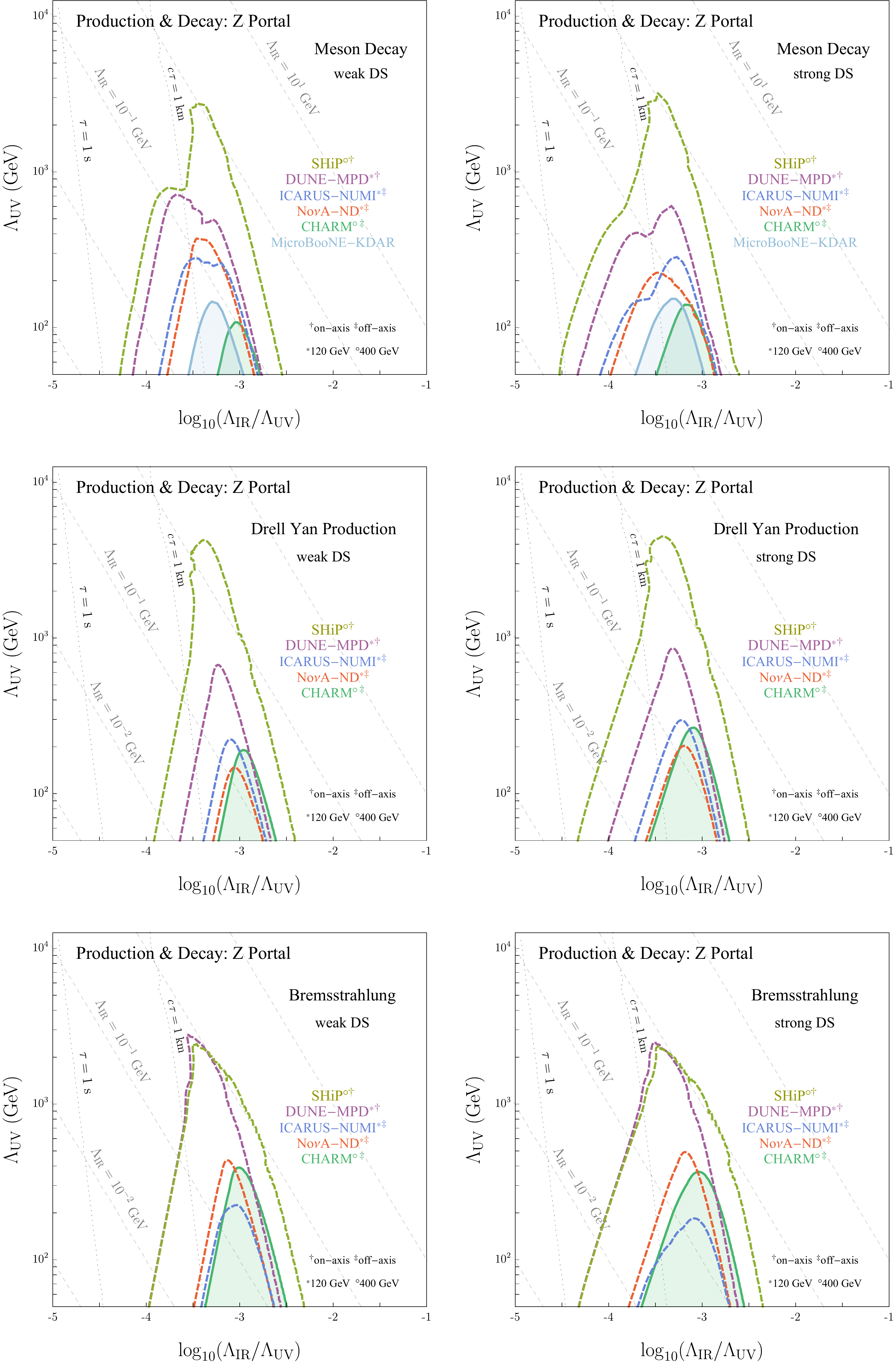}
\caption{\small{Constraints on DS production and decay from $D=6$ $Z$ portal for various production modes: Meson Decay (top row), Drell-Yan (mid row) and Dark Bremsstrahlung (bottom row), for weakly coupled (left) and strongly coupled (right) benchmark cases. Exclusions are shown in solid lines, while future projections are shown in dotted lines. The superscripts indicate the beam energy and the on/off axis nature of detectors for each of the experiments considered. The black dotted lines show the LDSP lifetime $\tau$ isocurves, while the gray dashed lines show $\Lir$ isocurves. The bounds assume $\pds^2/\Luv^2 < 0.1$ for EFT validity which is satisfied by restricting to $\Luv > 50$ GeV. All plots assume $\kappa_i ^2 c_i = 1$, where $i$ labels the portal, $\kappa$ is the portal coupling and $c$ is a measure of degrees of freedom of the DS.}}
\label{fig:ZPortal-IndividualProdModes}
\end{figure*}

In this section we present our bounds on the parameters $\Luv, \Lir$. As we have argued before, the dominant production mode is through the Z-portal, but the decay can proceed through either Z-portal or Higgs portal. As discussed in section~\ref{sec:DSprod}, there are three production modes for DS states, each of which has a different distribution in $\pds$ and therefore contributes differently depending on the energy and the detector geometries. In fig.~\ref{fig:ZPortal-IndividualProdModes} we show the exclusion regions individually for the three production modes, keeping to the $Z$ portal decay for simplicity, for both weakly and strongly coupled benchmark scenarios. We will assume that the LDSP has spin 1 in order to fix the decay parametrics. In fig.~\ref{fig:LHC_vs_nuexp_Zportal}, we show the combined bounds from this work, for both $Z$ and Higgs portal decays, and compare against bounds from LHC and LEP from ref.~\cite{Contino:2020tix}. The bounds for the Z-aligned $\JJ$ portal can be obtained from the $Z$ portal bounds simply by re-scaling the signal cross section as $\sigma_\text{S}^Z/\sigma_\text{S}^{JJ} \sim (\kappa_Z^2/\kappa_{JJ}^2)\,(v^2/m_Z^2)$. The case of generic $JJ$ portal is obtained from the Z-portal by an appropriate combination vector/axial parts of the current and a rescaling of the couplings. 

In all these exclusion plots, regions bounded by solid lines show the excluded parameter space from recasts of past and recent DS searches (CHARM~\cite{CHARM:1985anb} and MicroBooNE~\cite{MicroBooNE:2021usw}). 
The region bounded by dashed contours in our plots show the potential of current and future upcoming neutrino experiments: NO$\nu$A-ND and ICARUS (current), and DUNE-MPD (near future). To compare their potential with future DS experiments, we also show the projections coming from the future beam dump experiment SHiP. For these projected exclusions, we have shown the 10 signal events line assuming $100 \%$ reconstruction and detection efficiency. Following our discussion in section~\ref{sec:experiments}, for DUNE-MPD, we also show the 100 events line in fig.~\ref{fig:LHC_vs_nuexp_Zportal}.

There are several features of the bounds which make the neutrino experiments a very powerful probe for dark sectors, in the parametrization considered in this work. First of all, we find that the bounds from current and upcoming neutrino experiments are comparable to dedicated DS experiments, with a reach of $\Luv$ in multiple TeV range, for $\Lir$ in the MeV-GeV range. This is similar to the ranges probed in high energy experiments like LHC and LEP (as done in ref.~\cite{Contino:2020tix}). The typical scale $l$ at which the detectors are placed is much larger than the corresponding scale in the LHC DV searches, and the typical boosts involved are also different, which together select a somewhat larger $\tau$ and hence a smaller $\Lir$ region compared to LHC.  Importantly, neutrino experiments fill the gaps in the parameter space coming from trigger and event selection requirements at LHC and LEP, since they are sensitive to much lower energy activity in the detector. Even more importantly, in portals which are not enhanced by a resonant production, the EFT condition $(\pds^2)_\text{max} < \Luv^2$ makes LHC and LEP bounds inconsistent, an issue which is again alleviated at neutrino experiments due to a smaller $\sqrt{s}$ involved. All these features are seen in the bounds in fig.~\ref{fig:ZPortal-IndividualProdModes}, \ref{fig:LHC_vs_nuexp_Zportal}.

In the following subsections, we will discuss in detail how prospective DS searches at neutrino experiments can complement current bounds for different portals considered in this work. We will also emphasize the difference between the production modes, especially on how a  particular detector geometry can favor one mode over the other.

\subsection{Z Portal Production}
\label{subsection:result_zportal}
Consider first the MD production mode through the $Z$ portal (fig.~\ref{fig:ZPortal-IndividualProdModes} first row), where we show bounds from various experiments. We also show bounds coming from the MicroBooNE KDAR analysis~\cite{MicroBooNE:2021usw} which is only relevant for the MD mode. We find that for neutrino experiments based at 120 GeV proton beam, in general the strongest bounds come from radiative decays of K meson ($K \to \pi + \DS$) due to the large number of K mesons produced with respect to other mesons. However, for the strongly coupled DS case where the kinematic condition on $\pds^2$ is stronger due to a larger $n_\text{LDSP}$, we find that at DUNE and ICARUS, $\phi \to \text{DS}$ decays can give a stronger bound on $\LUV$ as compared to K meson bounds.
For experiments based on 400 GeV proton beam, heavier mesons like $B, \ J/\psi$ can be produced in large numbers and can contribute to bounds at CHARM and SHiP. These heavier mesons can in principle probe larger $\LIR$ due to the relaxed kinematic condition $n_\text{LDSP} \LIR \lesssim M$, where $M$ is the mass of the decaying meson. The $\Lir$ reach is correlated with the $\Luv$ reach, for fixed lifetime. At SHiP, we indeed find that $J/\psi$ decays probe the highest $\LUV$ scales as opposed to K meson decays which suffer from the kinematic condition. However, we find that K mesons can still improve reach on lower $\LIR$ values relative to $J/\psi, \, B$ mesons due to larger geometric acceptance for $K \to \pi + \DS$ and larger number of K mesons. At CHARM, we find that $J/\psi \to DS$ decays give the leading bounds which dominate those coming from $K \to \pi + \DS$ decays.

For the Drell-Yan (direct partonic) production mode (fig.~\ref{fig:ZPortal-IndividualProdModes} middle row), the best bounds come from SHiP, and the other experiments only give subleading bounds. Within them, due to the collinear nature of the produced DS beam, detectors of the on-axis type are more sensitive to this mode. Note that compared to the DB mode, the average DS boost is smaller for the DY mode, so that the LDSP spread is more and the off-axis detectors are penalized less.

For the case of dark bremsstrahlung (fig. \ref{fig:ZPortal-IndividualProdModes} bottom row), the DS is produced very collimated along the beam line, favoring detector geometries closer to it. We find that the best bounds from neutrino experiments for this mode come from DUNE-MPD. These are comparable in $\LUV$ and only probe slightly smaller $\LIR$ values, as compared to the future beam dump experiment SHiP. This is because in bremsstrahlung the typical $\pds^2$ is cut roughly around QCD scales, and an increase in $\sqrt{s}$ at SHiP as compared to DUNE does not lead to a large increase in the production cross section. The other experiments shown, ICARUS-NUMI, NO$\nu$A and CHARM give subleading reach in both $\Luv$ and $\Lir$. Despite an increase in the number of POTs with respect to NO$\nu$A, ICARUS still has lower sensitivity due to a reduced angular coverage. Both ICARUS and CHARM, due to their off-axis nature, miss out signal events from the forward DS beam, characteristic of the dark bremsstrahlung mode.

\begin{figure*}[t]
\includegraphics[width=0.99\textwidth]{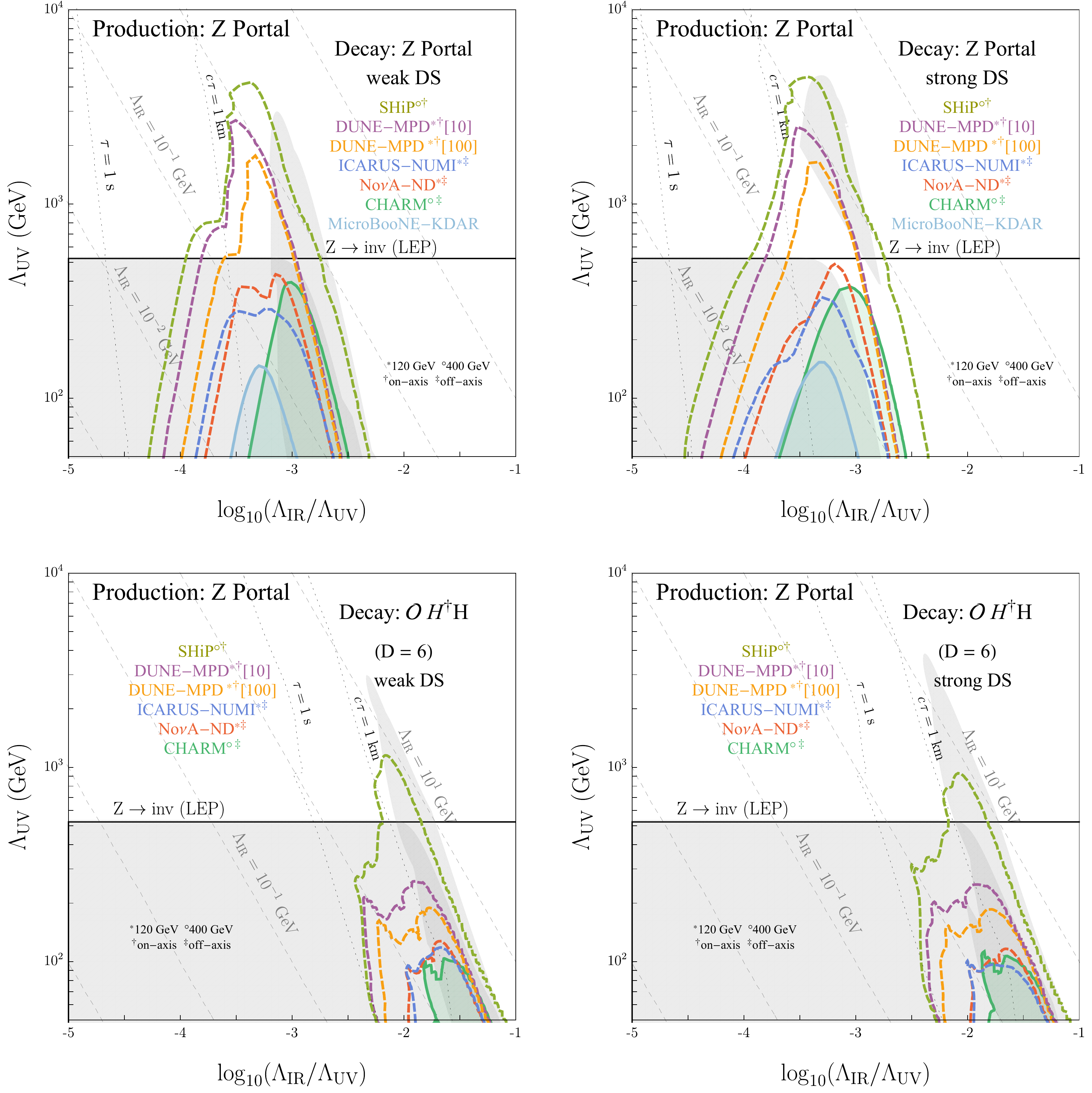}
\caption{
\small{Constraints on DS production through the $D=6$ $Z$ portal, and decay through the same $Z$ portal (top), or through $D=6$ $\mathcal{O}H^{\dagger} H$ portal (bottom), at various neutrino experiments. We have shown both the 10 event and the 100 event lines for DUNE. For comparison, bounds from high-energy colliders (obtained in ref.~\cite{Contino:2020tix}) are also shown in gray. The left (right) plots assume weakly coupled (strongly coupled) dark dynamics. The exclusion from the $Z$ invisible width measurement at LEP is shown by the horizontal solid black line. We restrict to $\Luv > 50$ GeV for EFT validity. All plots assume $\kappa_i ^2 c_i = 1$, where $i$ labels the portal, $\kappa$ is the portal coupling and $c$ is a measure of degrees of freedom of the DS.
}
}
\label{fig:LHC_vs_nuexp_Zportal}
\end{figure*}

Combining all the production modes, in fig. \ref{fig:LHC_vs_nuexp_Zportal} we show the final excluded parameter space, for both weakly and strongly coupled benchmarks. We also show the results from~\cite{Contino:2020tix}
which studied resonant DS production through $Z$ portals at high energy colliders and presented exclusion regions from ATLAS monojet search~\cite{ATLAS:2017bfj}, displaced vertex search~\cite{ATLAS:2018tup,ATLAS:2019jcm}, and from total $Z$ width bounds from LEP~\cite{ALEPH:2005ab}. The bounds presented here probe different parts of the parameter space, in particular in $\Lir$, even if the $\Luv$ reach is comparable to before, and also probe gaps in parameter space in earlier work which came from trigger requirements. Further, for portals which do not have a resonant production, the EFT condition at LHC invalidates the bounds, which is not an issue for neutrino experiments. The complementarity of the bounds at neutrino experiments, as compared to missing energy and displaced searches at LHC is due to the peculiar position of the near detectors of neutrino experiments, placed at $\mathcal{O}(10^2)$ meters. We also find that the bounds are stronger than past beam dump searches like E137 and NA64 (whose results can be found in~\cite{Contino:2020tix, Darme:2020ral}) due to a larger $\npot$, and in some cases, a larger beam energy.

From fig.~\ref{fig:LHC_vs_nuexp_Zportal} we see that one of the \textit{current} strongest bounds for $Z$ portal DS still comes from the indirect $Z$ width measurement at LEP, $ \LUV > 525 (k_J^2 c_J)^{1/4} \; \mathrm{GeV}$~\cite{Contino:2020tix} which is independent of $\Lir$ till the kinematic threshold. This bound is stronger than CHARM and MicroBooNE. Additionally, CHARM and MicroBooNE are also weaker than the  LHC monojet and displaced vertex bounds except for the strongly coupled case (fig.~\ref{fig:LHC_vs_nuexp_Zportal} top right) where they probe slightly higher $\LIR$ values. Prospective DS searches at current Fermilab neutrino facilities, ICARUS and NO$\nu$A-ND, improve on CHARM and MicroBooNE, but they are still weaker than the LEP bound.

Most importantly however, we find that future neutrino detector DUNE-MPD will be sensitive to $\LIR$ in the range $\mathcal{O}(0.1-1)$ GeV for $\LUV$ of few TeVs, a region not covered by LHC exclusions. Future LLP experiment SHiP based on 400 GeV proton beam would further improve sensitivity with respect to DUNE-MPD and LHC searches. These improvements are either due to a higher geometric acceptance from being on-axis or from having a wider detector, or due to a higher beam energy.

Emphasizing the complementarity of neutrino experiments with respect to the LHC searches, we note that for the strongly coupled case (fig.~\ref{fig:LHC_vs_nuexp_Zportal} top right), the LHC (ATLAS) searches are not sensitive in a gap of parameter space values close to $\LUV \sim 500 \ \mathrm{GeV}  - 2 \ \mathrm{TeV}$ and $\LIR \sim 0.1-0.5 \ \mathrm{GeV}$. We find that both future neutrino experiment DUNE-MPD and dedicated LLP experiment SHiP would remarkably fill this gap in the $Z$ portal DS parameter space. These gaps were due to trigger and event selection requirements.

So far we have only considered decay through the Z-portal, but the decay can also proceed through the Higgs portal. Before proceeding with that, a couple of comments about the interplay of the quantum numbers of LDSPs and the relevant decay portals is in order. According to the Landau-Yang theorem~\cite{PhysRev.77.242,Landau:1948kw}, a massive spin-1 particle can not decay into two massless spin-1 particles. This implies that if the LDSP is a spin-1 particle then for values of $\Lir < 2 m_e$, it will not decay into any visible SM particles, so that the only signal is a missing energy. If instead the LDSP is a spin-0 particle, there is no such condition. However for such small values of $\Lir$, the LDSPs are too long lived and cannot be efficiently constrained at the experiments considered here. For this reason it's crucial to realize that while production from the Higgs portal is very suppressed, it might be relevant for decays if its the only available decay mode. We should clarify that if multiple portals are available for decay, one has to consider the dominant one. When considering the LDSP decay through the Higgs portal, we are assuming it to be dominant compared to other portals. Note that a spin-1 LDSP can not decay through the Higgs portal due to quantum numbers.

In the bottom row of fig.~\ref{fig:LHC_vs_nuexp_Zportal} we show the bounds where the LDSP decay occurs through the $\Delta=4$ Higgs portal. The longer lifetime of the LDSPs, due to a very small coupling to leptons and the extra $\LIR^2/m_h^2$ suppression factor, effectively shifts the exclusion regions to higher $\LIR$ regions. However, bounds from such values of $\Lir$ can be suppressed due to being too close to the edge of allowed phase space, effectively chopping off the bounded region. This makes these bounds typically weaker than colliders, in their $\Luv$ reach, although they still cover regions unconstrained by $Z$ portal decays at higher $(\LIR/\LUV)$ ratio.

Finally, for completeness, we will now tabulate constraints coming from invisible meson decays where the LDSP is long-lived enough to escape detectors. Overall, we find that these bounds are weaker in their $\Luv$ reach than the ones coming from both LHC and neutrino detectors.

For the $Z$ portal (both production and decay), the strongest constraints from invisible meson decay come from flavour changing decays of $B$ and $K$ mesons (updated w.r.t ref.~\cite{Contino:2020tix}). We take the BaBAR upper limit for $B^+ \to K^+ $ decays~\cite{BaBar:2013npw}: $\mathcal{B}(B^+ \to K^+ \bar{\nu} \nu) < 1.6 \times 10^{-5}$, which gives
\begin{align}
\frac{\LUV}{\text{GeV}}  &>  60\:(\kappa_J^2\,c_J)^{1/4},\nonumber \\
\frac{\LIR}{\text{MeV}}  &\ll  108 \ (\kappa_J^2c_J)^{-0.2}\:\text{(weak)}\:,
65 \ (\kappa_J^2 c_J)^{-0.2}\:\text{(strong)}
\end{align}

For the case of $K \to \pi + \DS$, we take the upper limit from the NA62 Collaboration~\cite{NA62:2021zjw}:
$\mathcal{B}(K^{+} \to \pi^{+} + \bar{\nu} \nu) < 1.06 \times 10^{-10}$ which gives
\begin{align}
\frac{\LUV}{\text{GeV}}  &>  68.8\:(\kappa_J^2\,c_J)^{1/4},\nonumber \\
\frac{\LIR}{\text{MeV}}  &\ll  83\: (\kappa_J^2 c_J)^{-0.2}\:\text{(strong)}
\end{align}
The bounds for a weakly coupled DS are similar.
For the case in which the LDSP decay is via $\Delta_{\mathcal{O}}$ = 4 Higgs portal instead, the bounds for the B and K meson decays respectively are: 
\begin{align}
\frac{\LUV}{\text{GeV}}  &>  60\:(\kappa_J^2\,c_J)^{1/4}, \, \text{for} \,
\frac{\LIR}{\text{GeV}} \ll  1.6 \ (\kappa_J^2 c_J)^{-0.14} \\ \nonumber
\frac{\LUV}{\text{GeV}}  &>  68.8\:(\kappa_J^2\,c_J)^{1/4}, \, \text{for} \,
\frac{\LIR}{\text{MeV}} \ll  77 \ (\kappa_J^2 c_J)^{-0.14} \\
\end{align}
In the above, the condition on $\LIR$ has been calculated assuming a strongly coupled DS, and they do not change significantly for the weakly coupled case.

The invisible decays from $J/\psi$ which have been searched for by the BES Collaboration~\cite{BES:2007sxr} set an upper limit on $\mathcal{B}(J/\psi \to \bar{\nu} \nu ) < 7.2 \times 10^{-4} $. However, we found the resulting bound on $\LUV$ to be weaker than those coming from the BaBar and NA62 limits on $B, K$ decays, and we do not report it here. 

\subsection{$JJ$ Portal (Z-aligned) Production}

Even though the Z-aligned $JJ$ portal and Z-portal are equivalent at neutrino experiments after an appropriate rescaling of the $\kappa$, there is a distinction between them at high energy experiments that can produce a $Z$ on-shell. Contrary to Z-portal case, for the Z-aligned $JJ$ portal, the LHC bounds are generally weaker due to the lack of resonant production and EFT consistency condition on $\LUV$. For the same reason, there is no bound coming from  Z-width. In this scenario, the bounds on $\LUV$ come only from LEP missing energy searches (see fig. 8 in ref.~\cite{Contino:2020tix}). 
Therefore regions in the parameter space with too short lifetimes are not tested due to the requirement for the LDSP to decay outside the detector. On the other hand, the bounds coming from high-intensity experiments such as neutrino experiments are essentially unchanged with respect to the Z-portal case, so that all the discussion from before applies: they are able to exclude a larger portion of $\LUV$ by roughly one order of magnitude in the large lifetime region (low $\LIR$), compared to the LEP/LHC detector size, while it excludes a completely unexplored region at small lifetime (or large $\LIR$).

\subsection{$JJ$ Portal (generic) Production}

The previous sections can give us an insight on how high-intensity experiments can put a bound on a generic $J_{\mu}^\mathrm{SM} J^\mu_\DS$ portal, where $J_\mu^\mathrm{SM}$ is a generic flavor-conserving SM current.
Missing-energy bounds coming from LHC will still hold provided $\mathcal{O}(1)$ couplings to light quarks. If these are absent, (e.g. for $\bar{e} \gamma^{\mu} e\,J_\mu^{\text{DS}}$ portal) electron beam dump experiments like E137 and missing energy searches at NA64 put weaker bounds ($\LUV\gtrsim 10^2 \; \mathrm{GeV}$, see ref.~\cite{Contino:2020tix}).

Proton-beam based neutrino experiments cannot probe hadrophobic current interactions given that couplings to quarks are essential for all production modes. Since neutrino experiments typically exclude LDSP masses for $\Lir \lesssim 2 m_\pi$, if the decay proceeds through generic $JJ$ portal, we need non-zero couplings to electrons. If that is small, the Higgs portal may be relevant depending on couplings. This feature is not present in missing energy searches at high energy colliders and high intensity experiments, which only probe the production mode. This problem can be circumvented if instead of looking at displaced vertex signatures (where LDSP decays inside the detector), in which both DS production and DS decay into SM are required, scattering events are also considered. As mentioned in previous sections, we do not look at such signatures due to the extra assumptions needed with respect to LDSP decays.

We remark that no big difference is expected from changing the axial or vector nature of the SM current as long as their coupling is of the same order. While for $Z$ portal the axial contribution to bremsstrahlung is larger than the vector counterpart, due to the accidentally small coupling of the vector component, this is not necessary for a generic case. A similar argument also hold for DY mode, while for MD mode the quantum numbers of the SM current select the relevant meson processes (see App~\ref{app:MesonDecay} for details). To conclude, as long as DS has a coupling to proton and electrons in $J^\mu_\mathrm{SM}$, we expect the results to not change dramatically at fixed magnitude of the couplings: the bounds presented in sec.~\ref{subsection:result_zportal} apply.

\subsection{Higgs Portal Production}\label{sec:result_higgs}
The bounds at neutrino experiment for production through $\Delta_\mathcal{O} = 4$ Higgs portal $\mathcal{O} H^\dagger H$ are very weak:  $\LUV\ll 10^2$ GeV for DB and DY modes, since these modes are suppressed by a small Higgs coupling. Only radiative meson decays happening through a top loop do not suffer from such a problem. The strongest bounds for this case then come from meson decay where for DUNE-MPD, we get $\LUV \lesssim 150 \text{ GeV}$. This exclusion is much weaker than the bounds coming from missing energy searches at LHC and Higgs coupling fits~\cite{Contino:2020tix}, $\LUV \gtrsim 450\text{ GeV}$. For $\Delta \ge 4$ the rate is suppressed with respect to the $Z$ portal as explained in Sec.~\ref{sec:meson_prod}. The situation is slightly improved for high-energy beam experiments like SHiP, 
but is still not competitive with the ones coming from Higgs resonant production at LHC. For this reason, we do not show any plots for production through the Higgs portal. For a $\Delta_{\mathcal{O}} = 3$ Higgs portal, at SHiP, we find that bounds from K meson decays can probe $\LUV \sim 3 \text{ TeV}$, while B meson decays can probe $\LUV \sim 10-30 \text{ TeV}$ for $\LIR \sim  1-5\text{ GeV}$. The former bounds are weaker than LHC missing energy searches (which exclude $\LUV \lesssim 8 \ \mathrm{TeV}$) while the B meson bounds are stronger. 

\section{\label{sec:summary}Summary and Discussion}
Secluded sectors that interact very feebly with the SM have the potential to be probed at the high-intensity frontier, particularly at neutrino experiments (as has been previously explored in refs.~\cite{Batell:2009di,Batell:2019nwo,Berryman:2019dme,Ballett:2019bgd}, see also refs.~\cite{Blumlein:2011mv,Blumlein:2013cua}). Most of the past work has focused on the case of relevant portals, while the case of irrelevant portal DS scenario has only recently been explored~\cite{Contino:2020tix,Darme:2020ral,Cheng:2021kjg, Kelly:2020dda}. In this work, we have considered the sensitivity of DS that interacts with SM through a dimension 6 irrelevant portal, at past and current neutrino experiments, and its prospective discovery in both existing and future neutrino experiments based on proton beams. 

We have performed a detailed study of the possible production mechanisms of DS through non-renormalizable portals: meson decays ($M \to m + DS$, $V \to \DS$), direct partonic production ($\bar{q}q \to \DS$, $gg \to \DS$), and dark bremsstrahlung ($pp \to \DS + X$). The interplay between the various production mechanisms as a function of the DS invariant mass squared $\pds^2$ can be summarized in the plot shown in fig.~\ref{fig:prod modes}. Compared to previous works on irrelevant portals, we have added production details, and also considered strongly coupled dark sectors, and done so in a model agnostic framework.
Further, we have constrained such dark sectors using past and current analyses at beam dump/neutrino experiments, also showing projections for prospective searches at existing and future neutrino experiments. In order to emphasize the importance of these bounds, we have also compared our results with previous bounds on such portals. 

In an earlier work of this scenario~\cite{Contino:2020tix}, the most stringent bounds on DS excitations produced from the decay of $Z$ bosons was set by LHC monojet searches~\cite{ATLAS:2017bfj} and LHC displaced vertex search~\cite{ATLAS:2018tup,ATLAS:2019jcm}, in a range of $(\Luv, \Lir)$ values dictated by various factors such as the energy of the experiment and the lifetime of the DS etc. In the present work, we have tried to address the question if neutrino experiments, being placed farther from the interaction point (as compared to say the ATLAS detector at LHC) could probe a lower $\LIR$ range, thereby testing a complementary parameter space with respect to high energy colliders for such elusive dark sectors. Our main summary plots can be found in fig. \ref{fig:LHC_vs_nuexp_Zportal}.

While the present work focuses on the utility of neutrino experiments for probing dark sectors, we would like to mention the status of other probes of the dark sectors considered here, for completeness. Colliders and beam-dump probes produce the DS states directly. Other setups that also produce DS states directly result in astrophysical bounds from Supernova cooling, from lifetime of horizontal branch stars and from positronium lifetime. Complementary to those are indirect probes where the initial and final states are SM states, and DS degrees of freedom propagate internally. Examples of such probes are electroweak precision tests (EWPT), fifth-force constraints, torsion balance experiments, molecular spectroscopy, etc. Depending on the process, these indirect probes are UV sensitive (and in that case they do not probe the dark dynamics directly) or give weaker constraints. A careful analysis of all these direct and indirect effects was already carried out in ref.~\cite{Contino:2020tix} and we refer the reader to there.

In this work, for the case of $Z$ portal DS production, we find that past analyses and prospective DS searches at current neutrino experiments give weaker bounds when compared with the current bounds from LHC and LEP in resonant production scenario. However, future neutrino experiments such as DUNE-MPD would improve on this, and will be sensitive to $\LIR$ in the range $0.1 - 1$ GeV for $\LUV \sim 1 \ \mathrm{TeV}$. The current displaced vertex searches at LHC are already probing $\LUV$ as high as few TeVs, but only in the $\LIR$ range $\sim 0.6 - 2.5 \ \mathrm{GeV}$ for the strongly coupled DS case. The ATLAS DV searches lose sensitivity in the range of $\LIR \sim 0.1 - 0.6$ due to trigger requirements (as can be seen from the gap on the right plot in fig.~\ref{fig:LHC_vs_nuexp_Zportal}).\footnote{The trigger requirements imposed in~\cite{Contino:2020tix} depend on the $n_\ldsp$ distribution. Events where the number of LDSPs produced has a downward fluctuation can loosen the cut, but will also affect the total cross-section, the decay probability and the geometric efficiency. Including this effect systematically will reduce the un-probed region but not entirely.} Future neutrino experiment DUNE-MPD will have a unique sensitivity to access this gap in the parameter space for a range of $\LIR \sim 0.1 - 1 \ \mathrm{GeV}$ for $\LUV$ of a few TeVs.

We have also compared these results with projections from the proposed experiment SHiP which serves as a benchmark for LLP experiments. As can be seen from fig. \ref{fig:LHC_vs_nuexp_Zportal}, SHiP will improve on the reach of DUNE-MPD  by probing  $\LUV$ of few TeVs for a range of $\LIR \sim 0.1 - 2 \ \mathrm{GeV}$. This is mainly due to its higher proton beam energy of 400 GeV and larger geometric acceptance ($ \egeo \sim 1$ for the production modes bremsstrahlung and DY and $ \egeo \sim 0.1-0.9$ for K and $J/\psi$ decays).

As we have described in the previous section, we can recycle our bounds at neutrino experiments on $Z$ portal also for the case of $\JJ$ portal where $J_\mu^\text{SM} = \bar{f}\gamma_\mu f, \bar{f}\gamma_\mu \gamma^5 f, f = l, q$. The earlier work in~\cite{Contino:2020tix} found no bounds from LHC for this portal, once EFT considerations were taken into account. The only constraint presented is the one for $J_\mu^\text{SM} = \bar{e}\gamma_\mu\,e$ from monophoton searches at LEP (see fig. 8 in ref.~\cite{Contino:2020tix}) where the excluded space is restricted to $\LUV \lesssim 200 \ \mathrm{GeV}$ for $\LIR \lesssim 0.1 \ \mathrm{GeV}$. These bounds are much weaker than the bounds we get at neutrino experiments. Therefore neutrino experiments are a useful tool to study DS that do not directly mix with the $Z$ and that are not enhanced by resonant production at colliders.
We have already explained how our $Z$ portal bounds can be recycled for a $JJ$ portal, since a $JJ$ portal can be obtained from $Z$ portal after integrating out $Z$ mediator.

For the case of the Higgs portal $\mathcal{O} H^\dagger H$ production (for $\Delta_{\mathcal{O}} = 4$) the bounds from neutrino experiments are very weak, and are limited to values of $\LUV$ much below the electroweak scale. Whereas, in comparison, bounds from Higgs resonant production derived in ref.~\cite{Contino:2020tix} coming from missing energy and displaced vertex searches at high energy colliders are much stronger.

The bounds presented here are derived under a model agnostic approach, and are applicable to a large class of DS models (see~\cite{Contino:2020tix} for explicit examples). Knowledge of the underlying dark dynamics can be used to study other possible signatures like DS scatterings with SM particles, but will need to be done on a case-by-case basis, and hence is out of the scope of this work. We have discussed in detail our assumptions and limitations of our approach in section~\ref{sec:darkcft}. Our results are conservative and can be improved if the full theory is defined explicitly. Despite this, we claim that our approach can be very useful in giving a qualitative picture.

The point of this work is to convey the usefulness of a model agnostic approach to exploring dark sectors, and the potential of neutrino experiments (both current and future) as unique probes of irrelevant DS-SM portals. Future proposed LLP experiments at LHC interaction points like MATHUSLA, CODEX-b, ANUBIS are designed to improve reach on $\LIR$ scales for such elusive DS. However, future neutrino experiment DUNE Multi-Purpose Detector (MPD)~\cite{DUNE:2020ypp} running at the LNBF (Long Baseline Neutrino Facility) would probe low $\LIR$ scales in a shorter timescale. The forward LHC detectors like FASER and SND (see~\cite{Feng:2022inv} for a recent status report), built for searching feebly interacting particles would be taking data during the LHC Run 3, and could also give useful bounds for our DS. We hope our study would motivate analyses of neutrino-detector data for the search of such elusive dark sectors.

\section*{Acknowledgments}
We would like to thank R. Contino for initial discussions and for useful input at various stages of this project. We would additionally like to thank P. Asadi, B. Batell, D. Buttazzo, N. Foppiani, S. Homiller, W. Jang, K. Max, F. Poppi, D. Redigolo, M. Reece, M. Scodeggio, M. Strassler and L. Vittorio for reading the draft, for useful discussions, and for pointing us to relevant references. We also thank Adam Ritz and Saeid Foroughi-Abari for clarifications on their work on proton bremsstrahlung. MC is partially supported by PRIN 2017L5W2PT. RKM is supported by the National Science Foundation under Grant No. NSF PHY-1748958 and NSF PHY-1915071. SV is supported in part by the MIUR under contract 2017FMJFMW (PRIN2017). 

\appendix
\section{Time-like form factors}\label{app:FormFactors}
In the extended Vector Meson Dominance (eVMD) formalism, form factors are modelled as sum over meson states $\mathfrak{m}$ sharing the same quantum numbers of the SM operator $X$ involved in the matrix element.
Typically the formalism is applied in the space-like region, and the form factors are essentially the sum of the propagators of the virtual mesons. However given that the exchanged momentum $\pds$ in our case is time-like, the form factors are modelled as Breit-Wigners, to allow the virtual exchanged meson to go on-shell and resonantly mix with the DS system:
\begin{align}\label{eq:ff_bw}
    F_\mathfrak{m}(\pds^2)=\sum_i f_{\mathfrak{m}_i}\frac{m_{\mathfrak{m}_i}^2}{\pds^2-m_{\mathfrak{m}_i}^2+i m_{\mathfrak{m}_i}\Gamma_{\mathfrak{m}_i}}\:,
\end{align}
where $m_\mathfrak{m}, \Gamma_\mathfrak{m}, f_\mathfrak{m}$ are the mass, the decay width and the ``couplings'' for the meson $\mathfrak{m}$ respectively.
The number of resonances in the meson tower sharing the $\mathfrak{m}$ quantum numbers is such that it allows to enforce the correct asymptotic behaviour in $q^2$ of the form factor coming from sum rules. The couplings $f_{\mathfrak{m}_i}$ are fitted from data and by the overall coupling normalization.
In order to get the total form factor $F_X$, we need to also take into account the cut-off for too high virtuality of eq.~\eqref{eq:cutoff_FF}:
\begin{align}
    F_X = F_D (Q^2) F_\mathfrak{m}(\pds ^2) \;.
\end{align}
We next discuss the Higgs and the $Z$ portal case separately.

\subsection{Higgs portal}\label{app:ff_h}
The Higgs coupling to the protons is of the form
\begin{align}
g_{hNN} F_H(q^2)\,h\,\bar{u}_p\,u_p\:,
\end{align}
where $u_p$ is the proton spinor.
The form factor $F_H$ also includes the virtuality cut-off of eq.~\eqref{eq:brem_thresh}:
\begin{equation}
    F_H = F_D F_S
\end{equation}
The form factor $F_S(\pds ^2)$ is estimated using eq.~\eqref{eq:ff_bw}, and we include the first three CP-even, scalar resonances. The values used for masses, width and $f$ used in the FF are given in table~\ref{tab:ff}.
\begin{table*}
    \centering
    \begin{ruledtabular}
    \begin{tabular}{c|ccc||ccc|ccc|cc}
      & \multicolumn{3}{c}{$f_0$ (Scalar)} & \multicolumn{3}{c}{ $\omega$ (Z, vector iso-singlet)} & \multicolumn{3}{c}{$\rho$ (Z,  vector iso-triplet)} & \multicolumn{2}{c}{$a_1$ (Z, axial iso-triplet)}\\
     \hline
    $m_\mathfrak{m}$ (GeV) & 0.5 & 0.980 & 1.37 & 0.782 & 1.42 & 1.67 & 0.775 & 1.45 & 1.72 & 1.23 & 1.647  \\
    $\Gamma_\mathfrak{m}$ (GeV) & 0.275 & 0.5 & 0.35 & $8 \times 10^{-3}$ & 0.2 & 0.3 & 0.149 & 0.4 & 0.25 & 0.4 & 0.254  \\
    $f_\mathfrak{m}$ & 0.28 & 1.8 & -0.99 & 1.011 & -0.881 & 0.369 & 0.616 &   0.223 & -0.339 & 2.26 & -1.26 \\
    \end{tabular}
    \end{ruledtabular}
    \caption{Masses $m_\mathfrak{m}$, width $\Gamma_\mathfrak{m}$ and coupling $f_\mathfrak{m}$ of the mesons $\mathfrak{m}$ for the scalar, vector iso-singlet, vector iso-triplet, and axial iso-triplet proton Form Factors used in the eVMD approach.}
    \label{tab:ff}
\end{table*}
\subsection{Z portal}
The effective $Z$ vertex for the proton is modelled as:
\begin{align}
     & \frac{g_\mathrm{EW}}{\cos \theta_W}\bar{u}_p \gamma^{\mu}\left( F^\rho (q^2)\left(\frac{1}{2}-\sin^2 \theta_W\right)-\sin^2 \theta_W F^\omega(q^2)\right) u_p \nonumber \\
     & + \frac{g_\mathrm{EW}}{4\cos \theta_W} g_A F_A(q^2)\bar{u}_p \gamma^{\mu}\gamma^5 u_p\:,
     \label{eq:Z_ff}
\end{align}
where $q$ is the exchanged momentum between the virtual and real proton ($\pds$ in our case) and $u_p$ is the proton spinor.
The prefactors in front of the iso-singlet and iso-triplet vector form factors ($F^\omega$ and $F^\rho$ respectively) come from the decomposition of the vector piece of the quark $Z$-current in the singlet and triplet component in isospin space, using the approximate isospin symmetry of proton and neutron. The axial-vector form factor has a slightly different normalization than the vector one. In particular, there is an extra $g_A\simeq 1.2$ multiplying the overall form factor in eq.~\eqref{eq:ff_bw}, as shown in eq.~\eqref{eq:Z_ff}. We can see that the vector part is subleading with respect to the axial current due to the absence of $\sin^2 \theta_W$ suppressing factor and a $\mathcal{O}(1)$ coupling $g_A$. Notice that in principle, along with the Dirac-like form factors appearing in eq.~\eqref{eq:Z_ff}, there should be the non-renormalizable Pauli-like terms~\cite{PhysRevD.87.014005}:
\begin{align}
    G_V(q^2)  i\frac{\left[\gamma^\mu,\gamma^\nu \right]}{m_p}q_\nu + G_A(q^2) \gamma^5 \frac{1}{m_p}q^\mu\:,
\end{align}
where the second piece (the axial one) is mediated by pion exchange. It turns out that this contribution vanishes when contracted with the $JJ$ correlator, due to current conservation. The vector piece, corresponding to the anomalous magnetic proton contribution, does not vanish. Given that it's numerically subleading, and that there are large uncertainties to extrapolate such form factor in the time-like region~\cite{ritz_comm, Gorbunov:2014wqa}, we will neglect this contribution.

The form factors appearing in eq.~\eqref{eq:Z_ff} are again computed in the eVMD formalism, as a sum of Breit-Wigners of $\rho, \omega, a_1$ meson for $F^\rho, F^\omega, F_A$ respectively. The coefficients used in the form factors are found in table~\ref{tab:ff}.

The final form factors appearing in eq.~\eqref{eq:z_brem_vec},\eqref{eq:z_brem_ax} are defined respectively as:
\begin{align}
    &F^\text{V}_Z = F_D \left(-\sin^2 \theta_W F^\omega+ F^\rho \left(\frac{1}{2}-\sin^2 \theta_W\right)\right) \;, \nonumber \\
    &F^\text{A}_Z = F_D \frac{g_A}{4} F_A \;.
\end{align}
We can generalize the formalism for any combination of iso-singlet and iso-triplet axial and axial-vector currents, meaning that we can get the form factor for all the possible flavor-conserving quark coupling structures of generic $JJ$ portals. 

Notice that in table~\ref{tab:ff} there are no axial-vector iso-singlets form factors, but they can be obtained from lattice computations. 

\section{Meson decay matrix elements}
\label{app:MesonDecay}
Here we summarise the QCD matrix elements and form factor parametrizations we use for computing decays of mesons.
For the case of annihilation decays, the meson decays to DS states entirely, while for radiative decays, a heavier meson decays to lighter mesons, along with DS. The matrix element for the process can be factorized into a short distance contribution (and only involves DS matrix elements) and a long distance (QCD) contribution. For the process $\textbf{M} \to \text{DS}$, where $\textbf{M}$ is the decaying meson, the full amplitude $\left<\text{DS}| \mathcal{O}_\text{SM}\,\mathcal{O}_\text{DS}|\textbf{M}\right>$ factors into $\left<0| \mathcal{O}_\text{SM}|\text{M}\right>\times \left<\text{DS}|\mathcal{O}_\text{DS}|0\right>$
while for the process $\textbf{H} \to \textbf{L} + \text{DS}$, where $\textbf{H}$ ($\textbf{L}$) are the heavy (light) SM mesons, the amplitude $ \langle \textbf{L}, \text{DS} |\mathcal{O}_\text{SM}\,\mathcal{O}_\text{DS} | \textbf{H} \rangle$ factors as $\langle \textbf{L} | \mathcal{O}_\text{SM} | \textbf{H} \rangle \times \langle \text{DS} | \mathcal{O}_\text{DS} | 0 \rangle$. For the annihilation case, we only consider vector mesons, denoted by V\footnote{For pseudoscalar mesons, the matrix element is proportional to $p_\mu$ and vanishes when contracted with the DS current, by current conservation.}. The SM matrix element for annihilation decay is simple, and is generically given as
\begin{align}
    \left< 0 | \bar{u}\gamma_\mu u| V(p)\right> &= i f_V m_V \epsilon_\mu (p)\:,
\end{align}
where $u$ are the quark spinors, $\epsilon_\mu(p)$ is the polarization vector of the vector meson V, $f_V$ is the decay constant, and $m_V$ is the mass of the meson.

For the radiative decay case, the SM contribution to the amplitude is the same as SM semileptonic meson decays, and is less straightforward than the annihilation case. We now give details of the matrix elements and form factors used to compute width of decays of $\text{H} \to \text{L} + \text{DS}$ where H can be $B, K$ and L can either be a light pseudoscalar P (e.g. $K, \pi$) or vector $V$ (e.g., $K^*, \rho,\phi$).

\subsection{Decay to Pseudoscalars} \label{appendix:mesontoP}
For the decay of mesons to pseudoscalars P (e.g., $K, \pi$) we use the usual matrix element definitions  (see \cite{Ball:2004ye} for example):

\begin{align}\label{eqn:BtoP_meSM}
    \bra{P(p_P)} V^{\mu}  \ket{H(p_H)}  = f_+(q^2) p^{\mu}
     + \left(f_0(q^2) -f_+(q^2)\right)\,D\,q^{\mu}\:,
\end{align}
where $V^{\mu} = \bar{u}_L \gamma^{\mu} u_H$, $p^{\mu} = p_H^\mu + p_{P}^{\mu}$, $D=(M_H^2 - M_P^2)/\pds^2$, and $q^{\mu} \equiv p_{\text{DS}}^\mu = p_H^{\mu} - p_P^{\mu}$. Here, $u_L$ ($u_H$) denotes a light (heavy) quark field, $p_H (p_P)$ is the 4-momentum of the decaying heavy (light) meson with mass $m_H (m_P)$, and $f_0(q^2), f_+(q^2) $ are dimensionless form factors which encode the strong interaction effects. For the case of $\mathcal{O}_\text{DS} = J_\mu^\text{DS}$, a conserved current, terms proportional to $q^\mu\,J_\mu^\text{DS}(q)$ vanish, so that we only need to specify the $f_+(q^2)$ form factors.

For $K^+ \to \pi^+ + \DS$ decay we use the explicit form factor data points defined at each $q^2$ in table IV of ref.~\cite{Carrasco:2016kpy}. For $B \to \pi, K$ decays, we use the form factor definitions and values from ref.~\cite{Ball:2004ye}:
\begin{align}
    f_{+}^{B\to \pi}(q^2) & = \frac{A}{\big(1 - q^2/D^2\big)\big(1 - q^2/E^2\big)}\:, \nonumber \\
     f_{+}^{B \to K}(q^2) & = \frac{B}{1 - q^2/F^2} + \frac{C}{\big(1 - q^2/F^2\big)^2}\:,
\end{align}
where $A=0.258, B=0.173, C=0.162$ and $D=5.32 \text{ GeV}$, $E= 6.38\text{ GeV}$, and $F = 5.41 \text{ GeV}$.

\subsection{Decay to Vector Mesons}\label{appendix:mesontoV}
For the case of $B$ meson decaying to light vector mesons V  (e.g., $K^{*}, \rho, \phi$), the QCD matrix element is defined as (see ref.~\cite{Ball:2004rg}):
\begin{align}
\label{eq:M_decay_vec}
&\bra{V(p_V)} J_{\mu}  \ket{B (p_B)}  = - i \epsilon_{\mu*}(p_V) (m_B + m_V)\,A_1(q^2) 
\nonumber \\
& \:\:+ i p_{\mu} (\epsilon^{*}(p_V)\cdot q)\,\frac{A_2(q^2)}{m_B+m_V}
\nonumber \\
& \:\:+ i\,q_{\mu} (\epsilon^{*}(p_V)\cdot q)\,\frac{2 m_V}{q^2} (A_3(q^2) - A_0(q^2)) \nonumber \\
& \:\:+ \epsilon_{\mu \nu \rho \sigma} \epsilon^{\nu *}(p_V) p_{B}^\rho p_{V}^\sigma\,\frac{2 V (q^2)}{m_B+m_V}\:,
\end{align}
where $J^{\mu} = \bar q_L \gamma^{\mu} (1- \gamma_5) q_H$, $p^{\mu} = (p_B + p_V)^{\mu}$, and $q^{\mu} \equiv \pds^{\mu} = p_B^{\mu} - p_V^{\mu}$. Here $p_B$ ($p_V$) are the 4-momentum of the $B$ (vector) meson, $m_B$ ($m_V$) is the mass of the B (V) meson, $\epsilon^{\mu}$ is the polarization vector of V, and $A_i (q^2), V(q^2)$ are the dimensionless form factors encoding strong interaction effects. Note that, unlike the pseudoscalar case, in the case of vector meson, the $\gamma_5 \gamma^{\mu}$ part of the current does not vanish. Further, the third term in the above does not contribute since it is zero by current conservation, as in the pseudoscalar case. 

For the form factors, we use the parametrizations as given in ref.~\cite{Ball:2004rg}:
\begin{align}
    \label{eq:ff_vector1}
    V(q^2) & = \frac{r_1}{1 -q^2/m_R^2} + \frac{r_2}{1 - q^2/m_{fit}^2}\:, \\
    \label{eq:ff_vector2}
    A_1(q^2) & = \frac{r_2}{1 - q^2/m_{fit}^2}\:, \\
    \label{eq:ff_vector3}
    A_2 (q^2) & = \frac{r_1}{1 - q^2/m_{fit}^2} + \frac{r_2}{(1 - q^2/m_{fit}^2)^2}\:.
\end{align}
We give the values of the various fit parameters for the different decays in table~\ref{tab:fit_BtoV}. To evaluate further the squared matrix element, we make the calculations in the rest frame of $B$ meson, where only the longitudinal polarization of $\epsilon_{\mu}^{*}(p_V)$ contributes. Thus, using $\epsilon_{\mu}^{*}(p_V) = \bigg( \frac{|\vec{p}_V|}{m_V} , \frac{\vec{p}_V}{|\vec{p}_V|}\frac{E_V}{m_V} \bigg)$ and $\pds = p_B - p_V$, we can find the production width of DS from radiative $B \to V$ decays.

\begin{table}[t]
    \centering
    \begin{tabular}{c|c|c|c|c}
    & $r_1$ & $r_2$ & $m_R^2 \ [\text{GeV}^2]$  & $m_{fit}^2 \ [\text{GeV}^2]$ \\[0.1cm]
    \hline
      $V^{B \to K^*}$ &  0.923 & -0.511 & 28.30 & 49.4\\[0.1cm]
      $ A_1^{B \to K^*}$  & - & 0.290 & - & 40.38 \\[0.1cm]
      $A_2^{B \to K^*}$ & -0.084 & 0.342 & - & 52.00 \\[0.1cm]
      $V^{B \to \rho}$ & 1.045 & -0.721 & 28.30 & 38.34 \\[0.1cm]
      $ A_1^{B \to \rho}$ & - & 0.240 & - & 37.51 \\[0.1cm]
      $ A_2^{B \to \rho}$ & 0.009 & 0.212 & - & 40.82 \\[0.1cm]
      $V^{B_s \to \phi}$ & 1.484 & -1.049 & 29.38 & 39.52 \\[0.1cm]
      $ A_1^{B_s \to \phi}$ & - & 0.308 & - & 36.54 \\[0.1cm]
      $ A_2^{B_s \to \phi}$ & -0.054 & 0.288 & - & 48.94 \\[0.1cm]
    \end{tabular}
    \caption{\small{Fit parameters for the form factors defined in eq.~ \eqref{eq:ff_vector1}, \eqref{eq:ff_vector2}, \eqref{eq:ff_vector3} for various $B \to V$ transitions where $V = K^{*}, \rho, \phi$ (taken from~\cite{Ball:2004rg}).}}
    \label{tab:fit_BtoV}
\end{table}

\subsection{Sensitivity to non-conformal contributions in $K, B$ meson decays}
For the case of DS production from irrelevant portals, we expect the production cross-section to grow with $\pds^2$ which is necessary to make the contribution away from the IR threshold more important, and is needed for usefulness of our model agnostic approach. For radiative meson decays however, the DS production cross-section does not grow as $\pds^2$, but is rather flat, upto the kinematic threshold. For annihilation decays, $\pds^2$ is fixed and equals the parent meson mass-squared. To justify our model agnostic approach we need to ensure we are away from the thresholds, which must be imposed as a self-consistent criteria.

The overall signal is obtained by an integral over the range of allowed $\pds$. The kinematic condition $\pds/(n_\text{LDSP}\,\Lir) \geq 1$ is always stronger than the condition to be in the conformal regime $\pds/\Lir \gtrsim 1$. We also need to make sure that the relevant $\Lir$ probed in the experiment under consideration (which depends on the portal and the lifetime) is away from the kinematic threshold $M-m$. For $\Lir$ close to $M-m $, a small change in the lower limit of $\pds$ integration would have a bigger impact, but this does not happen in the cases we consider. For $B \to K + \DS$, $(m_B - m_K) \sim 5$ GeV is larger than typical $\Lir$ probed which is $0.001-1\text{ GeV}$, while for $K \to \pi + \DS$, $(m_K - m_{\pi}) \sim 0.4$ GeV, while the typical $\Lir$ probed is $0.001-0.1\text{ GeV}$. 

\section{Probability of Decay}
\label{app:DecayProb}
To compute the number of signal events, we need to calculate the number of LDSPs that decay inside the detector. To correctly compute this quantity, the differential cross section (in both energy and angle) of LDSP production must be convoluted with the probability $P$ for at least an LDSP to decay inside the detector. The probability $P_1$ for a particle to decay inside the detector can be roughly estimated as the probability $P_{1, \mathrm{dec}}$ to decay within the radial distance at which the detector is, multiplied with $\egeo$, the geometric acceptance accounting for the particles flying in the detector direction:
\begin{align}
    P_1 &\approx \egeo P_{1, \mathrm{dec}}\;,
\end{align}
In a simplified setup (see fig.~\ref{fig:typical_exp}), $P_{1,\mathrm{dec}}$ can be estimated to be
\begin{align}
    P_{1, \mathrm{dec}} = \exp \left(-\frac{l}{c \tau (\gamma\,\beta)_\text{LDSP}} \right) - \exp \left(-\frac{l+d}{c \tau (\gamma\,\beta)_\text{LDSP}} \right) \;,
    \label{eq:Pdecay}
\end{align}
where $l$ and $d$ are the distance of the detector from the target location and the length of the detector respectively
\footnote{Notice that the distance should be a function of the direction of the LDSP for generic geometry of the detector. We will work under the assumption of spherical detectors, where the distances appearing in $P$ are independent of the line of flight. This is a good approximation for neutrino experiments, since they consist of small boxes (of size $\mathcal{O}(10)$ m) positioned far ($\mathcal{O}(10^3)$ m) from the target. The error due to the geometrical approximation is therefore of order $1 \%$ (the ratio of the typical distances).}, $\gamma_\text{LDSP}$ is the (energy dependent) Lorentz factor of the decaying particle, $\beta_\text{LDSP}$ is its velocity and $\tau$ is its proper lifetime.
In general, for more than 1 LDSP, a slightly more refined procedure accounting for the presence of multiple particles that can go inside the detector is needed. We now discuss our procedure for multiple particle events.
We consider both the weakly coupled case, where $n_\text{LDSP}=2$, and a generic strongly coupled case in which $n_\text{LDSP}\sim \mathcal{O}(10)$ are produced.

For the weakly coupled case, the directions of the two LDSPs are fully correlated, at fixed $\pds$, by momentum conservation. 
If the direction of the DS sytem (\emph{i.e.} $\vec{p}_\text{DS}$) does not intersect the detector, at most one of the two LDSPs can have the correct direction to hit the detector. Indeed, the two particles have opposite azimuthal angle $\phi$ (computed with respect to the DS momentum in the lab frame), so that only one particle at most can travel to the correct side. 
Therefore the probability to have at least one particle decaying inside the detector is (in the notation of eq.~\eqref{eq:factorized_xsec}):
\begin{align*}
    \egeo\,P_\text{decay} &=2 \egeo P_{1, \mathrm{dec}}\;,
\end{align*}
where the 2 reflects the fact that $\egeo$ has been computed for a single particle only, and there are two possible particles that can decay in the detector (or in other words, we only considered one of the particles to be aimed toward the detector, but the opposite direction is also a valid choice given the presence of the other LDSP).

If instead the DS direction intersects the detector there is the possibility for both LDSPs to fall inside the detector. This can happen only if the LDSP velocity in the DS frame is slower than the velocity of the DS system in the lab frame, so that the LDSP traveling in the direction $\theta>\pi/2$ (the backward direction) in the DS frame is boosted in the forward direction. The events in which the backward LDSP gets boosted forward in the lab frame are only significant if the DS system velocity is much larger than the velocity of the LDSP in the DS frame. Given that typically $n_\text{LDSP}\Lir \ll \sqrt{\pds^2}$, and that the bulk of the cross section is dominated by high $\pds^2$ events, the events will have fast LDSPs in the DS frame. Therefore events with two particles boosted forward are negligible, and the formula to have a particle decaying inside the detector holds unchanged.

In the strongly coupled case, where $n_\text{LDSP}$ is large, we can neglect the fact that the momenta are correlated by momentum conservation. Therefore we will assume that the $n_\text{LDSP}$ particles share democratically the energy in the rest frame, and that the directions are independent samples from an isotropic distribution in such a frame. In this scenario, multiple particles can in principle have the correct direction to get inside the detector in the lab frame.
Using the probability for a single particle to decay inside the detector from eq.~\eqref{eq:Pdecay}, and recalling that this probability is very small (both $P_{1,\mathrm{dec}}$ and typical $\egeo$ are small), it follows that events in which multiple particles simultaneously decay inside the detector are very rare. Therefore we can approximate the probability for a single event to contain at least one particle decaying inside the detector as
\begin{align*}
    \egeo\,P_\text{decay} &= n_\text{LDSP}\,\egeo\, P_{1, \mathrm{dec}}\:.
\end{align*}
Notice that in the end (although for slightly different reason) this formula matches the one in the weakly coupled case for $n_\text{LDSP}=2$.

\section{$\egeo$ estimates}\label{app:egeo}
In order to compute the signal events, we have to compute the fraction of events that contain at least one LDSP  with the direction intersecting the detector.
Given the assumptions outlined in Section~\ref{app:DecayProb}, we define $\egeo$ as the probability for a single particle to have such a direction.
We work under the assumption that LDSPs are produced isotropically in the DS rest frame
, the frame in which only the time component of $\pds$ is non-zero. We will also assume that in this frame, all the LDSPs produced share the same energy $p_\DS^0/n_\text{LDSP}$. We take the beam direction to be aligned to the $z$-axis ($\theta=0$ in polar coordinates), and $\phi$ as the azimuthal angle (measured in the plane orthogonal to the beam line). 

In order to get an estimate for the angular coverage of the neutrino detector, we work under the simplifying assumption that the detector surface lies on a 2D plane orthogonal to the beam (for the off-axis case, the angle is very small) and all points on the detector are at the same distance from the interaction point. This is a good approximation for neutrino detectors since corrections are of order $(d/l)^2 \ll 1$, where $l$ is the distance at which the detector is placed, and $d$ is the typical size of the detector. There are two relevant cases where this approximation fails. 

For a closer experiment like SHiP, $l$ and $d$ are of same order, and the proposed shape is a rectangle of dimensions $(a, 2a)$, $a = 5$ m.
We will compute the angular acceptance by taking the largest side of the rectangle and requiring the trajectory to intersect the first layer, which gives $\theta_\mathrm{acc}=\operatorname{ArcTan}\left(2a/2l \right)$ . Being more specific about the shape of the detector does not change the estimate significantly since the DS is produced very boosted, with a small opening angle, and is fully covered by the extent of the detector. The geometric acceptances are close to unity, see table~\ref{tab:avg_boosts}.

For MicroBooNE, the produced Kaons are at rest in the lab frame, and not collimated, so that the DS is produced isotropic in the lab frame. In this case, the above approximation again does not hold. We calculate $\egeo$ without this approximation for MicroBooNE.

To calculate $\egeo$, we need to find the overlap of the LDSPs with the detector, which is easier to do in the $*$ frame. By the assumption of isotropic decay, the LDSP directions are distributed uniformly in the $(\cos \theta^*, \phi^*)$ plane. In this plane $\egeo$ is the area (normalized by $1/4\pi$) that corresponds to lab frame configurations in which the LDSP falls inside the detector. This in general depends on $\pds$ and $n_\text{LDSP}$ through the boost factor.

To be precise, we need to introduce some notation. There are two frames to consider, the lab frame, and the DS frame (the one where $\pds$ only has a time component). We refer to the DS frame as the * frame, in what follows. The angles and the boost factors with a * superscript are defined in the DS frame, and without a * superscript are in the lab frame.

\begin{figure*}[t]
\includegraphics[width=0.85\textwidth]{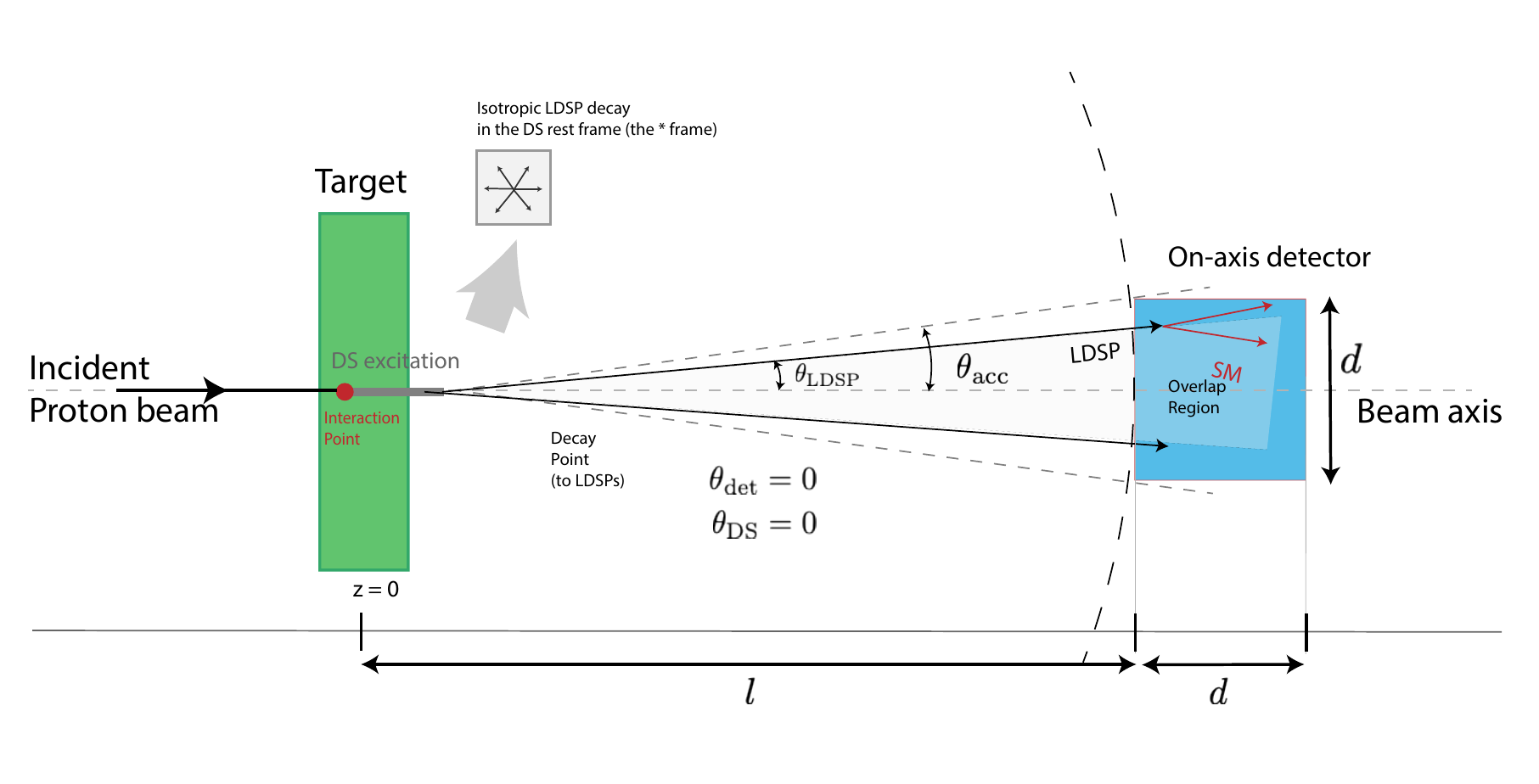}
\includegraphics[width=0.85\textwidth]{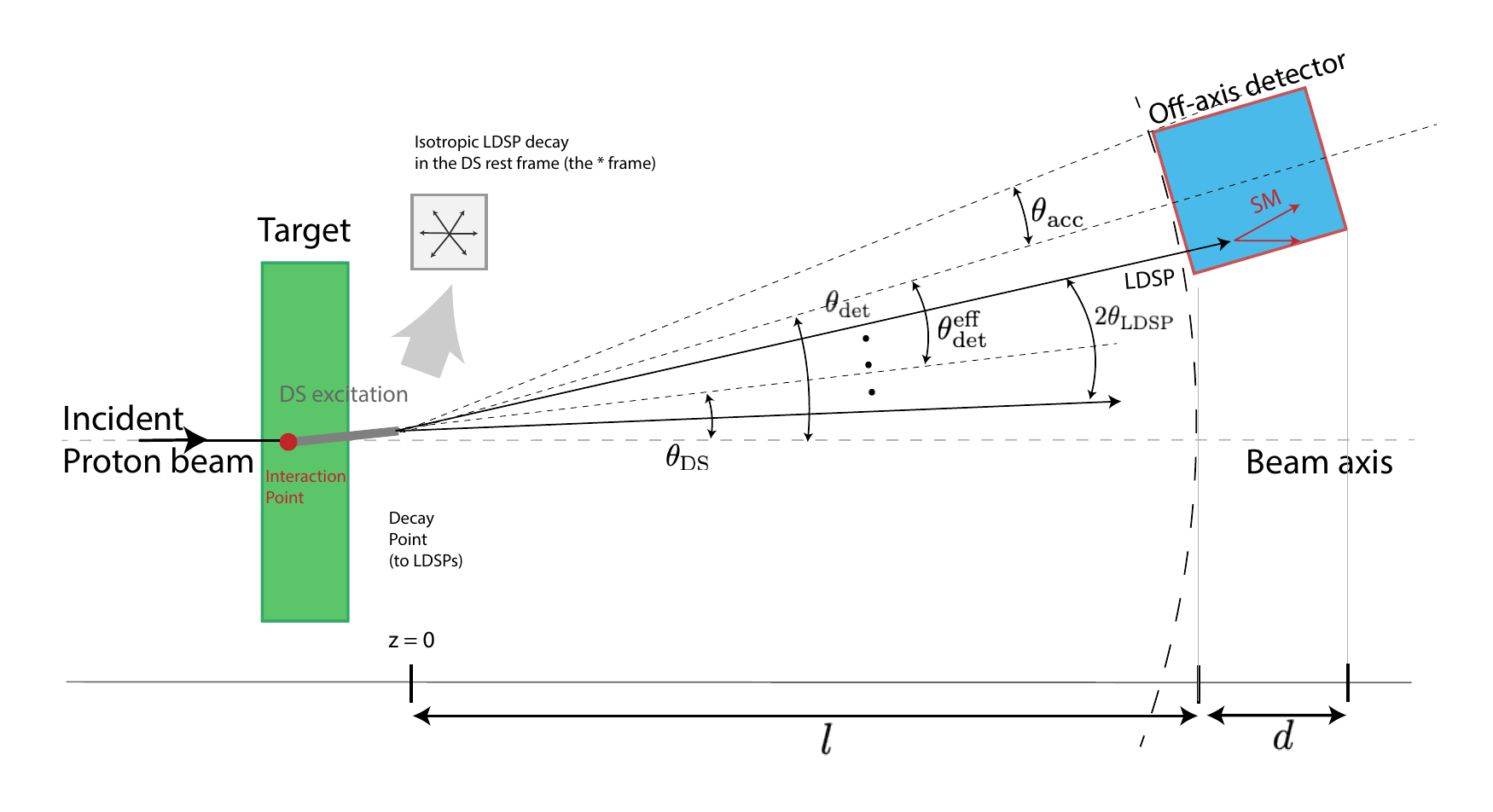}
\caption{\small{The geometry relevant for the estimation of $\egeo$ in the on- and off-axis case (top and bottom respectively). Various quantities defined in the text, and involved in the estimation of $\egeo$ are shown, not to scale. In particular, the DS line (gray thick) is drawn exaggerated for clarity.}}
\label{fig:egeom_OnAndOffAxis}
\end{figure*}

The detector direction, the DS direction, and the LDSP direction are all defined by a $\theta$, and the last two depend in general on kinematical quantities such as $\pds$ and other process-dependent quantities \emph{e.g.} $z$ in bremsstrahlung. We take the detector to be positioned at $\theta=\theta_\text{det}$. The accepted directions form a cone around it with an angle $\theta_\text{acc}$ around it. The DS system may be along the beam line or at an angle from it. We define the DS system to be at $\theta = \theta_\text{DS}$. Finally, the LDSP is at an angle $\theta_\text{LDSP}$ w.r.t. the DS system, so that the DS direction has a cone of angle $\theta_\text{LDSP}$ around it. It is useful to also define the DS direction w.r.t. the detector, which is denoted by $\theta_{\mathrm{det}}^{\mathrm{eff}}\approx \left(\theta_{\mathrm{DS}}^2+\theta_{\mathrm{det}}^2-2\theta_{\mathrm{det}}\theta_{\mathrm{DS}}\cos \phi_{\mathrm{DS}}\right)^{1/2}$. All these quantities are shown in fig~\ref{fig:egeom_OnAndOffAxis} (drawn not to scale), and are defined in the lab frame.

We additionally denote the LDSP in the * frame to be at $(\theta^*, \phi^*)$. For a given $\pds$ and $\theta^*$, $\theta_\mathrm{LDSP}$ is given as
\begin{align}\label{eq:thetastar}
    \tan\theta_\mathrm{LDSP}
    &=\frac{\sin \theta^*}
    {\gamma_\DS \left(\beta_\DS/\beta_\text{LDSP}^*+\cos \theta^*\right)} \,,
\end{align}
where $\beta_\text{DS},\gamma_\text{DS}$ are the boost factors to go from $*$ to the lab frame and $\beta_\text{LDSP}^*$ is the velocity of the LDSP with respect to the $*$ frame. Note that the boosts along the DS direction do not change $\phi$. Given the width $\theta_\text{LDSP}$ of the lab cone there is only a range of $\phi_*=\phi_{\mathrm{lab}}$ for which the LDSP direction intersects the detector. $\egeo$ is then the area (computed in the $*$ frame) of the overlap between the circular detector and all the possible LDSP cones in the lab frame, in general a function of $z$ and of $\pds$. To compute such area, we will make a \emph{linearization approximation}: we will consider that in the $*$ frame the shape of allowed $\cos \theta^*-\phi^*$ is bound by straight lines and not curved ones.
This approximation is expected to hold at the $10\%$ level for the relevant boosts. 

We now discuss the details specific to the three production modes considered in this work.

\subsection{$\egeo$ for meson decay production}\label{appendix:ego_mesons}

In order to find $\epsilon_\mathrm{geo}$ for LDSPs in the meson case, we use the prescription explained in the section before. For this, we must boost the DS kinematic variables twice: once from the meson rest frame to the lab frame, then from the lab frame to the DS rest frame (or $*$ frame). We will denote all quantities in the meson rest frame with a subscript 0.

The energy and 3-momentum of the DS can be easily computed  using momentum conservation in the parent meson rest frame as shown in eq.~(\ref{eq:pds0_rad}). Using the parent meson boost $\gamma_{M}$  and velocity $v_{M}$ (obtained from simulation), we can find the DS lab frame variables and express them as a function of $\cos \theta_0$ (angle of DS with meson flight direction in the meson rest frame, DS is isotropic in this) and $\pds^2$ i.e. $E_\text{DS}^\text{lab}(\cos \theta_0, \pds^2)$, $\left|\vec{p}_\text{DS}^\text{\ lab}|\right (\cos \theta_0, \pds^2)$.

In order to compute $\epsilon_\mathrm{geo}$ we must find the $\theta^*$ values corresponding to the angular coverage of the detector. For this, we must boost the lab frame DS variables to the DS rest frame (or $*$ frame) using the DS boost given by $\gamma_\text{DS}(\cos \theta_0, \pds^2) = E_{\text{DS}}^{\text{lab}}/ \sqrt{\pds^2}$ and DS velocity $\beta_{\text{DS}}(\cos \theta_0, \pds^2) = \left|\vec{p}_{\DS}^\text{\  lab}|\right/E_{\DS}^\text{lab}$ (which will be close to 1). 
Note that for the annihilation decay mode $M \to \DS$, $ E_{\DS}^{\text{lab}}$ gets fixed by momentum conservation to $\sqrt{M^2 + \left|\vec{p}_{\text{M}}\right|}$ where $M$ is the parent meson mass, and $\left|\vec{p}_{\text{M}}\right|$ is the parent meson momentum in the lab frame. Thus, $\gamma_{\text{DS}}$ is fixed to $\sqrt{M^2 + \left|\vec{p}_{\text{M}}\right|}/M$.

We can now solve for the $\theta^*$ angle using eq.~\eqref{eq:thetastar}, plugging $\beta^*_{\ldsp} = \left|\vec{p}_{\ldsp}^{\ *} \right| / E_{\ldsp}^{\ *}$, $E_{\ldsp}^{\ *} = \sqrt{\pds^2}/n_\ldsp$ and using $\LIR = \sqrt{(E_{\ldsp}^{*})^2 - \left|\vec{p}_{\ldsp}^{\ *} \right|^2}$ to get $\left|\vec{p}_{\ldsp}^{\ *} \right|$. 

We also need $\theta_\DS$ for estimating $\egeo$. Note that in the annihilation decay case, $\theta_{\DS}$ is 0 due to momentum conservation, Whereas for the radiative decay case, $\theta_{\DS}$ can be expressed in terms of meson rest frame variables as $\tan \theta_{\DS} = \sin \theta_0 \left|\vec{p}\right|_{\text{DS}, 0}/\left(\gamma_M\,( \left|\vec{p}\right|_{\text{DS}, 0} \cos \theta_0 + v_M E_{\DS, 0})\right)$.

Finally, we calculate $\expval{\egeo}$ by a weighted average over the differential decay width $d \Gamma/ d \pds^2\,d\lambda_i$, which is a function of $\Lir$ only (the $\Luv$ dependence factors out):
\begin{align}
\expval{\egeo} & = 
\frac{\int \dd\pds^2\,\dd\lambda_i \:\egeo(\lambda_i, \pds^2) \: \frac{d \Gamma}{d \pds^2 d \lambda_i}}
{\int \dd\pds^2\,\dd\lambda_i \: \frac{d \Gamma}{d \pds^2 d \lambda_i}}\:,
\label{eq:avg_egeo_meson}
\end{align}
where $\lambda_i$ are the angular variables $\theta_0, \phi_0$ integrated over the full range and $\pds^2$ is integrated in the allowed kinematic range. Note that for the annihilation decay case, $\pds^2 = M^2$ and $\theta_{\text{DS}} = 0$ by momentum conservation, which simplifies the equation above.

\subsection{$\egeo$ for DY}
In DY mode the DS system is by construction directed along the original beam line (the z direction).
The boost of the DS system is $\gamma_\DS=E_{\mathrm{DS}}/\sqrt{\pds^2}$, where $E_{\mathrm{DS}}$ is the energy of the DS system in the lab frame. While $\pds$ is already one of the variables used in DY production, $E_{\mathrm{DS}}$ is computed by first getting the DS energy in the DY CM frame, $E_{\mathrm{DS,CM}}=\sqrt{s}(x+\pds^2/(s x))/2$ using the DY variables introduced in \ref{sec:dy_prod}, and then boosting it to the lab frame with Lorentz parameter $\gamma_{\mathrm{CM}}=E_{\mathrm{beam}}/\sqrt{s}$. This gives $E_{\mathrm{DS}}=x E_{\mathrm{beam}}$. Putting everything together, the boost factor to go from the lab frame to the DS system is $\gamma_\DS=x E_\mathrm{beam}/\sqrt{\pds^2}$.
To get the average value of $\egeo$, we compute the integral as in eq.~(\ref{eq:avg_egeo_meson}) with the appropriate distribution where now we average over the kinematic variables $x$ and $\pds^2$.

\subsection{$\egeo$ for DB}
DB mode is different from the DY mode because in general the DS system will be produced at an angle with the beam-line. We will call $\theta_\mathrm{DS}, \phi_\mathrm{DS}$ the pair of angles indicating the direction of the radiated DS system relative to the beam line.
Using the kinematic variables introduced in \ref{sec:brem_prod}, we have $\theta_\mathrm{DS}=\text{tan}^{-1}(p_T/z E_\mathrm{beam})$. The boost factor to go from the lab to the DS frame is $\gamma_\DS \approx z E_{\mathrm{beam}}/\sqrt{\pds^2}$.
As before, to get the average value of $\egeo$, we average over the kinematic quantities $z, p_T^2$, and $\pds^2$. 
we have checked that these values are in good agreement with the $\egeo$ for the average DS angle, as expected for very collimated DS excitations.

\begin{figure*}[t]
\includegraphics[width=0.31\textwidth]{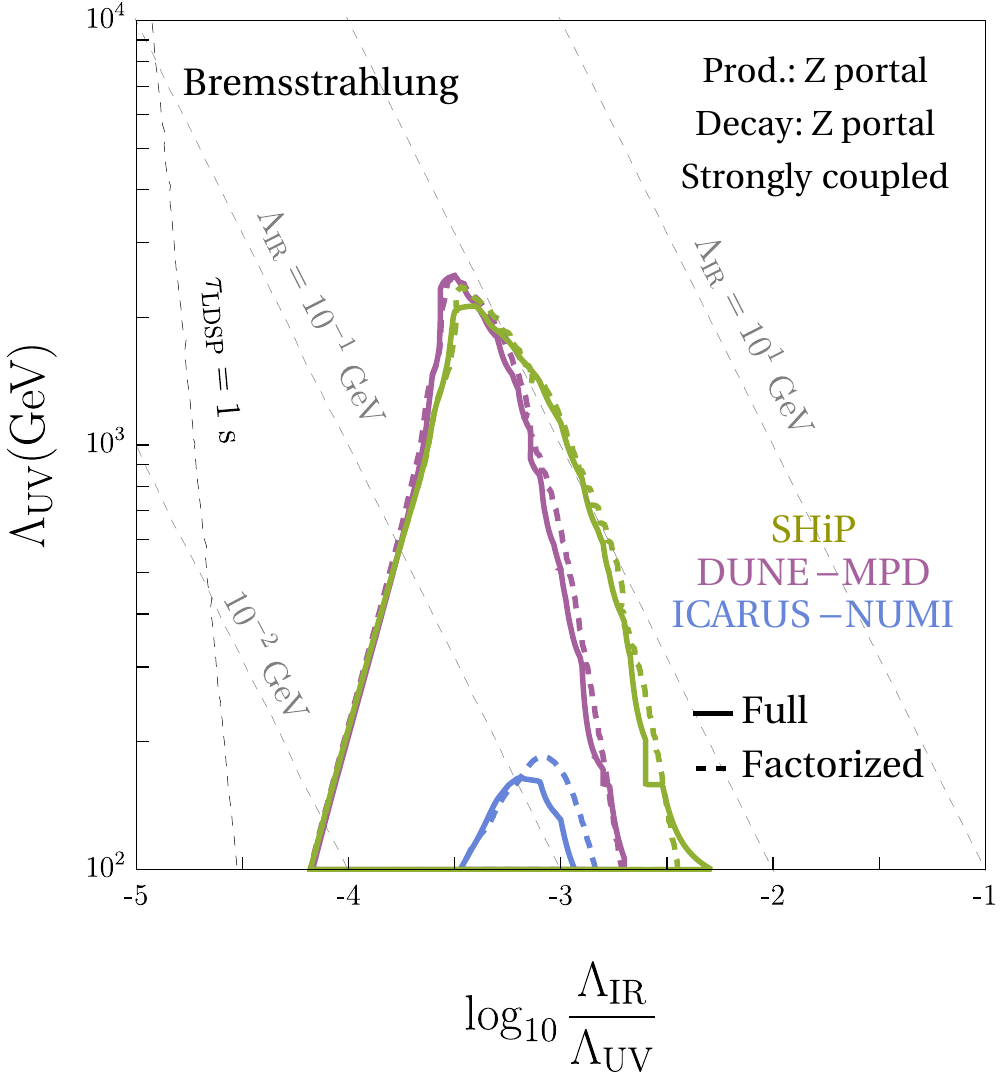}
\includegraphics[width=0.31\textwidth]{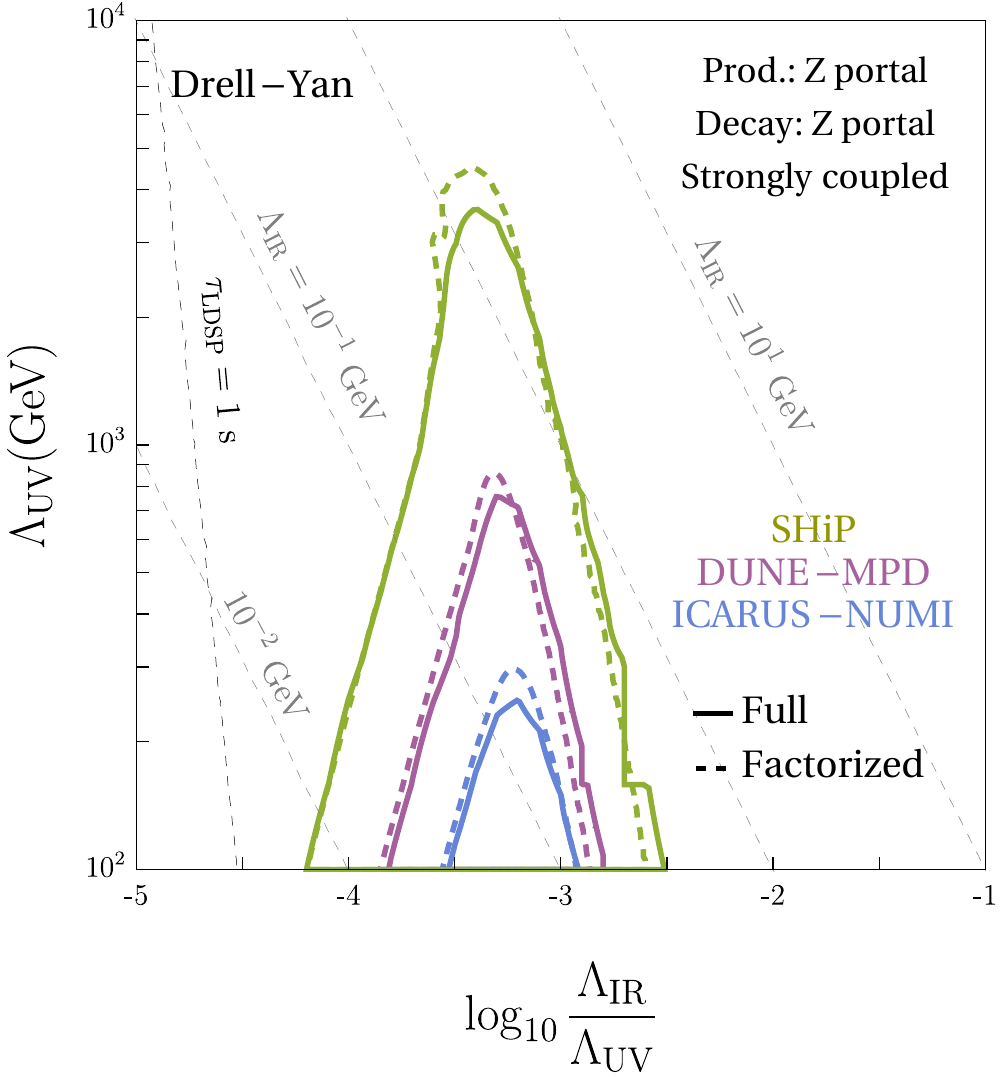}
\includegraphics[width=0.31\textwidth]{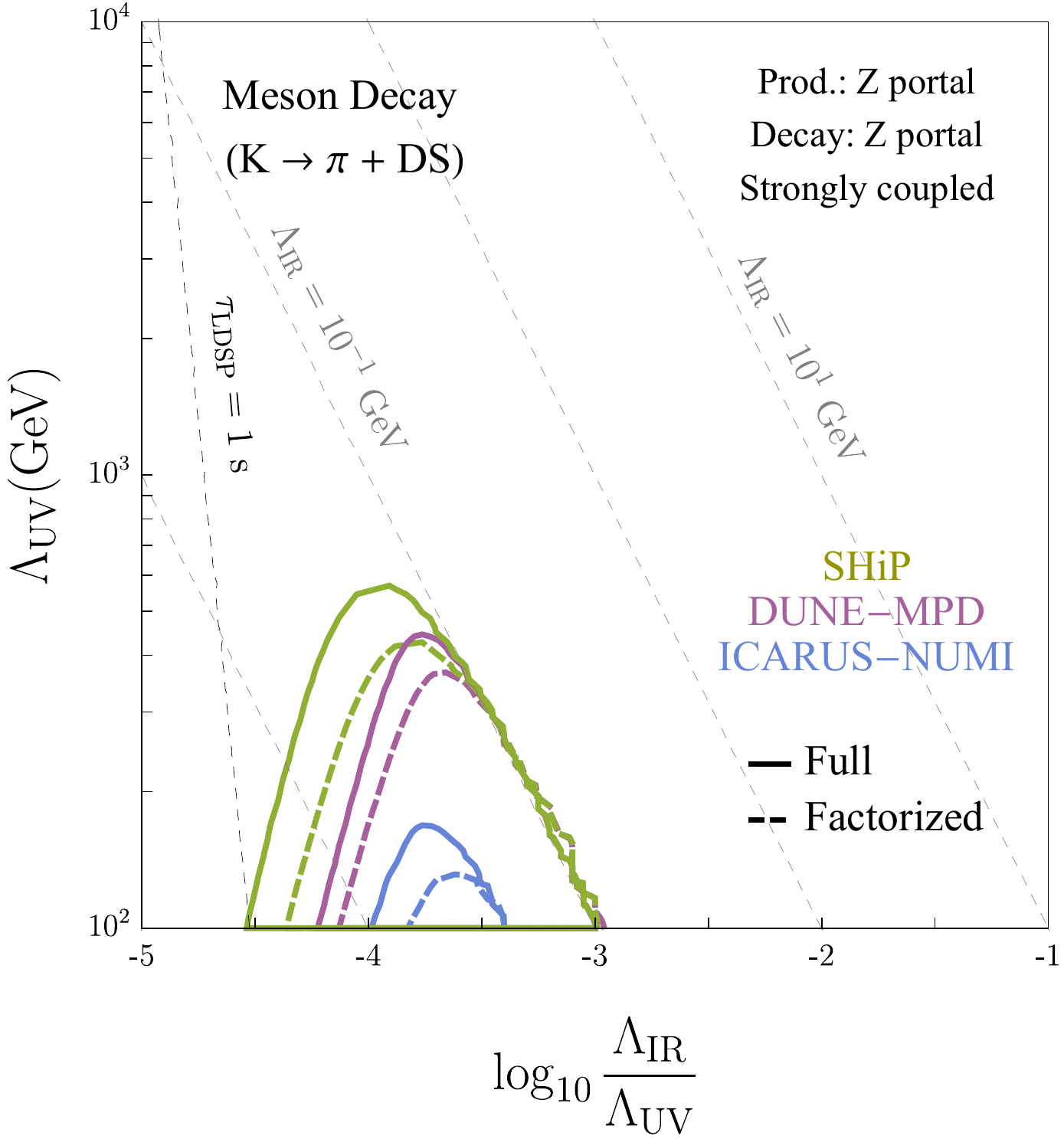}
\caption{
\small{Comparison of the factorized (dashed) vs full (solid) approach for production and decay through $Z$ portal of a strongly interacting DS, for various production modes: DB (left), DY (mid) and MD (right). The experiments considered are ICARUS-NuMI (blue), DUNE-MPD (purple) and SHiP (green).}}
\label{fig:ZZ_fullvsfac}
\end{figure*}

\section{Factorization approximation}
\label{app:factorization}
In this appendix we compare the approximate method we delineated in the main text with the correct procedure of doing the integral of the product of the differential quantities ($\egeo$, cross-section, decay probability) without factorizing them. 

In order to get the correct number of signal events, the differential cross section $\dd \sigma / \dd \lambda_i$ must be folded with $n_\text{LDSP} \,\egeo\,P_\mathrm{1,dec}$, all of which are a function of kinematic variables $\lambda_i$:
\begin{align}
    N_S (\Lir, \Luv) &= \frac{N_\text{POT}}{\sigma_{pN}}
    \int \dd \lambda_i \, 
    \frac{\dd\sigma}{\dd\lambda_i}\, n_\text{LDSP}\,\egeo\, P_{1, \text{dec}}\:.
\end{align}
We call this procedure the \textit{full} approach. We compute the integrals numerically, using the CUBA integration tools \cite{Hahn:2004fe}.
The method used in the main text is done instead by replacing the full integral with the average $\egeo$ and using average kinematic quantities to estimate the decay probability and the average $n_\mathrm{LDSP}$\footnote{We have checked that this procedure agrees very well (percent level) to a true average of $n_\ldsp$.}: 
\begin{align}
    N_S &\approx  N_\text{POT}\,\frac{\sigma_S}{\sigma_{pN}}\,
    n_\text{LDSP}(\langle \lambda_i \rangle) \,
    P_{1,\mathrm{dec}}(\langle \lambda_i \rangle)\,
    \langle \egeo \rangle\, .
\end{align}
where the average is defined by a weighted integral over the differential cross-section/decay-width (e.g. see eq.~\eqref{eq:avg_egeo_meson}).
We call this approach the \textit{factorized} approach. The advantage of this approach is that the production integral must be done only once, and not repeated for each $\left(\LIR, \LUV \right)$ pair: the dependence on them is essentially factorized. In particular, for a fixed production portal, this allows changing the decay portal, without having to redo the production integral from scratch. This is particularly useful when exploring all the various combinations of production and decay portals.

To compare the factorized and the full approach, and to show that the factorized approach is very efficient, in fig.~\ref{fig:ZZ_fullvsfac} we show the comparison for DB, DY and radiative meson decay $K \to \pi + \DS$, for the strongly coupled DS, for DUNE, SHiP and ICARUS. The two approaches are in very good agreement. The factorized approach is conservative at most, and can miss rare events appearing in the tails of distributions (see e.g. ref.~\cite{Egana-Ugrinovic:2019wzj}). However for the purposes of the present work, the factorized approach suffices.


\newpage
\bibliography{references}

\end{document}